\documentclass[journal]{IEEEtran}
\usepackage{pbox}
\usepackage{enumerate}
\usepackage{cite}
\usepackage{graphicx}
\usepackage{graphicx, subfigure}
\usepackage{alltt}
\usepackage{amsmath}
\usepackage{amsopn}
\usepackage{amssymb}

\newtheorem{lemma}{Lemma}
\newtheorem{proposition}{Proposition}
\newtheorem{corollary}{Corollary}
\newtheorem{remark}{Remark}
\newtheorem{definition}{Definition}
\newtheorem{claim}{Claim}

\DeclareMathOperator{\mincut}{mincut}

\DeclareMathOperator{\out}{out}
\DeclareMathOperator{\inp}{in}

\DeclareMathOperator{\DC}{DC}
\DeclareMathOperator{\MBR}{MBR}
\DeclareMathOperator{\MSR}{MSR}

\DeclareMathOperator{\GF}{GF}

\newcommand{\gf}{\mathcal{G}_F}
\newcommand{\gfp}{\mathcal{G}_{F^+}}

\title{When Can Helper Node Selection Improve Regenerating Codes? Part I: Graph-Based Analysis}

\author{Imad~Ahmad,~\IEEEmembership{Student Member,~IEEE,}
        and~Chih-Chun~Wang,~\IEEEmembership{Senior Member,~IEEE}
\thanks{This work was supported in parts by NSF grants CCF-0845968, CNS-0905331, and CCF-1422997. Part of the results was presented in the 2014 Allerton Conference on Communication, Control, and Computing.}

\thanks{I. Ahmad and C.-C. Wang are with the School of Electrical and Computer Engineering, Purdue University, West Lafayette,
IN, 47906 USA e-mail: \{ahmadi,chihw\}@purdue.edu.}

}

\begin{document}
\maketitle
\thispagestyle{plain}
\pagestyle{plain}

\begin{abstract}
Regenerating codes (RCs) can significantly reduce the repair-bandwidth of distributed storage networks. Initially, the analysis of RCs was based on the assumption that during the repair process, the newcomer does not distinguish (among all surviving nodes) which nodes to access, i.e., the newcomer is oblivious to the set of helpers being used. Such a scheme is termed the {\em blind helper selection (BHS)} scheme.  Nonetheless, it is intuitive in practice that the newcomer should choose to access only those ``good'' helpers. In this two-part paper, a new characterization of the effect of choosing the helper nodes in terms of the storage-bandwidth tradeoff is given. Specifically, the answer to the following fundamental question is provided: Under what condition does proactively choosing the helper nodes improve the storage-bandwidth tradeoff?

\par
Through a graph-based analysis, this Part~I paper answers this question by providing a necessary and sufficient condition under which optimally choosing good helpers strictly improves the storage-bandwidth tradeoff. A low-complexity helper selection solution, termed the \emph{family helper selection (FHS)} scheme, is proposed and the corresponding storage/repair-bandwidth curve is characterized. This Part~I paper also proves that under some design parameters, the FHS scheme is indeed optimal among all helper selection schemes. In the Part~II paper, an explicit construction of an exact-repair code is proposed that achieves the minimum-bandwidth-regenerating (MBR) point of the FHS scheme. The new exact-repair code can be viewed as a generalization of the existing \emph{fractional repetition} code.
\end{abstract}


\section{Introduction} \label{sec:intro}
\IEEEPARstart{T}{he} need for storing very large amounts of data reliably is one of the major reasons that has pushed for distributed storage systems. Examples of distributed storage systems include data centers \cite{ghemawat2003google} and peer-to-peer systems \cite{rhea2001maintenance,bhagwan2004total}.  One way to protect against data loss is by replication coding, i.e, if a disk in the network fails, it can be replaced and its data can be recovered from a replica disk. Another way is to use maximum distance separable (MDS) codes. Recently, regenerating codes (RCs) \cite{dimakis2010network} have been proposed to further reduce the repair-bandwidth of MDS codes.

One possible mode of operation is to let the \emph{newcomer}, the node that replaces the failed node, {\em always} access/connect to all the remaining nodes. On the other hand, under some practical constraints we may be interested in letting the newcomer communicate with only a subset of the remaining nodes, termed the {\em helpers}. For example, reducing the number of helpers decreases I/O overhead during repair and thus mitigates one of the performance bottlenecks in cloud storage systems. In the original storage versus repair-bandwidth analysis of RCs \cite{dimakis2010network}, it is assumed that the newcomer does not distinguish/choose its helpers. We term such a solution the {\em blind helper selection (BHS) scheme.} Nonetheless, it is intuitive that the newcomer should choose to access only those ``good'' helpers of the remaining nodes.

The idea of choosing good helpers in RCs has already been used in constructing exact-repair codes as in \cite{el2010fractional,papailiopoulos2012simple} that are capable of outperforming RCs with BHS in \cite{dimakis2010network} in some instances.\footnote{Reference \cite{el2010fractional} observes that choosing good helpers
can strictly outperform BHS at the minimum-bandwidth point by giving a code example \cite[Section~VI]{el2010fractional} for parameters (using the notation of RCs) $(n,k,d,\alpha,\beta)=(6,3,3,3,1)$. As will be seen later, the helper selection scheme and the associated code construction in \cite{el2010fractional} can be viewed as a special case of the helper selection schemes proposed in this Part~I paper and the new code construction in the companying
Part~II \cite{part2}.} Under the subject of {\em locally repairable codes (LRCs)} some additional progress has been done with the goal of minimizing the storage \cite{gopalan2012locality,kamath2014codes,papailiopoulos2012locally}. In the literature of LRCs, helper selection is not blindly decided but is judiciously chosen/fixed (see Section~\ref{sec:comparison} for an in-depth comparison with these references). We note that there are at least two classes of helper selection schemes:\footnote{SHS and DHS will be formally defined in Section~\ref{sec:hs}.} The {\em stationary helper selection (SHS)} schemes, which are more practical and are currently used in all existing literatures \cite{el2010fractional,gopalan2012locality,papailiopoulos2012locally,kamath2014codes}; and the {\em dynamic helper selection} (DHS) schemes, which are the most general form of helper selection. 

\par However, a complete characterization of the effect of choosing the helper nodes in RCs, including SHS and DHS, on the storage-bandwidth tradeoff is still lacking. This motivates the following open questions: Under what condition is it beneficial to proactively choose the helper nodes (including SHS and DHS)? Is it possible to analytically quantify the benefits of choosing the good helpers? Specifically, the answers to the aforementioned fundamental questions were still unknown.

\par In this paper, we answer the first question by providing a necessary and sufficient condition under which optimally choosing the helpers strictly improves the storage-bandwidth tradeoff. By answering such a fundamental information-theoretic question, our answers will provide a rigorous benchmark/guideline when designing the next-generation smart helper selection solutions.

\par The main contribution of this work is two-fold. Firstly, we prove that, under a certain condition, even the best optimal helper selection can do no better than the simplest BHS scheme. Using information-theoretic terminology, this answers the {\em converse} part of the problem. Secondly, we prove that when those conditions are not satisfied, we can always design a helper selection scheme that strictly outperforms BHS, the {\em achievability} part of the problem. For the achievability part, we propose a new low-complexity solution, termed the {\em family helper selection (FHS) scheme}, that is guaranteed to harvest the benefits of (careful) helper selection when compared to a BHS solution. We then characterize analytically the storage-bandwidth tradeoff of the FHS scheme and its extension, the family-plus helper selection scheme, and prove that they are \emph{optimal} (as good as any helper selection one can envision) in some cases and {\em weakly optimal} in general, see Sections~\ref{sec:achievability_result} and \ref{sec:optimality_result}). We also note that even though the purpose of introducing FHS and its extension is to prove the achievability part in theory, the FHS schemes have the same complexity as the existing BHS solution \cite{dimakis2010network} and demonstrate superior performance for practical system parameters. 

\par In this Part~I, we focus exclusively on the graph-based analysis of helper selection. In Part~II \cite{part2}, we provide an explicit construction of an exact-repair code that can achieve the minimum-bandwidth-regenerating (MBR) points of the family and family-plus helper selection schemes predicted by the graph-based analysis. The new MBR-point code in Part~II is termed the \emph{generalized fractional repetition} code, which can be viewed as a generalization of the existing fractional repetition codes \cite{el2010fractional}.

\par The rest of this paper is organized as follows. Section~\ref{sec:prob_stat} motivates the problem and introduces key definitions and notation. Section~\ref{sec:comparison} compares our setup to existing code setups. Section~\ref{sec:preview} gives a preview of our main results in this paper. Section~\ref{sec:results} states the main results of this paper. Section~\ref{sec:converse} states and proves the converse part of our main results. Section~\ref{sec:achievability} states and proves the achievability part by proposing the FHS scheme and its extension and analyzing their performance. Section~\ref{sec:conc} concludes this paper.

\section{Problem Statement} \label{sec:prob_stat}
\subsection{The Parameters of a Distributed Storage Network}
{\bf Parameters $n$ and $k$:} 
We denote the total number of nodes in a storage network by $n$. For any $1\leq k\leq n-1$, we say that a code can satisfy the reconstruction requirement if any $k$ nodes can be used to reconstruct the original data/file. For example, consider a network of 7 nodes. A $(7,4)$ Hamming code can be used to protect the data. We say that the Hamming code can satisfy the reconstruction requirement for $k=6$. Specifically, any 6 nodes can construct the original file. By the same definition, the Hamming code can also satisfy the reconstruction requirement for $k=5$ and $k=4$, but cannot satisfy the reconstruction requirement for $k=3$. The smallest $k$ of the $(7,4)$ Hamming code is thus $k^*=4$. In general, the value of $k$ is related to the {\em desired} protection level of the system while the value of $k^*$ is related to the {\em actual} protection level offered by the specific distributed storage code implementation. 

\par For example, suppose the design requirement is $k=6$. We can still opt for using the $(7,4)$ Hamming code to provide the desired level of protection. However, using the $(7,4)$ Hamming code may be an overkill since the $(7,4)$ Hamming code has $k^*=4$ and it is possible to just use a single-parity bit to achieve $k=6$. For any valid code construction, we always have $k\geq k^*$, meaning that  the design requirement must be met by the actual protection level. However, for some specific types of construction, it is possible to have $k>k^*$. Also see the subsequent discussion in the next paragraph. 

\par {\bf Parameter $d$:} We denote the number of nodes that a newcomer can access during repair by $d$. For example, \cite{dimakis2010network} proposes the concept of RCs that achieves the design goal $(n,k,d)=(10,7,9)$. Specifically, each newcomer can access $d=9$ helpers and any $k=7$ nodes can be used to reconstruct the original file. At the same time, \cite{dimakis2010network} also provides RCs to achieve the design goal when $(n,k,d)=(10,7,5)$. However, those RCs can be an overkill in this scenario of $(n,k,d)=(10,7,5)$ since any RC construction in \cite{dimakis2010network} that can achieve $(n,k,d)=(10,7,5)$ can always achieve $k^*= d=5$. As a result, even though the high-level design goal is to only protect against $10-7=3$ failures, the RC in \cite{dimakis2010network} cannot take advantage of this relatively low protection-level requirement since it always has $k^*\leq d=5$, which is strictly smaller than the design requirement $k=7$. 

\par Note that the above observation does not mean that the system designer should never use the RCs \cite{dimakis2010network} when the design goal is $(n,k,d)=(10,7,5)$. The reason is that these RCs with BHS have many other advantages that may be very appealing in practice, e.g., some very efficient algebraic code construction methods \cite{shah2012interference}, allowing repair with $n-d$ simultaneous failures, and admitting efficient collaborative repair when more than one node fails \cite{shum2013cooperative}. The fact that $k^*\leq d$ for any RCs in \cite{dimakis2010network} simply means that when the requirement is $(n,k,d)=(10,7,5)$, the system designer should be aware that the RCs with BHS in \cite{dimakis2010network} do not take full advantage of the relatively loose required protection level since we have in this scenario $k>d\geq k^*$. 

\par In this work, we focus on the design target $k$ instead of the actual performance parameter $k^*$, since given the same $k$, the actual $k^*$ value may depend on how we implement the codes. For example, when locally repairable codes \cite{gopalan2012locality} are used, it is possible to design a system with $k=k^*>d$. However, when RCs are used together with BHS, we may have $k>d\geq k^*$. As we will see later, when RCs are used together with some carefully designed helper selection schemes, we may again achieve $k=k^*>d$. For any given $(n,k,d)$ values, the goal of this paper is to compare the best performance of any possible helper selection scheme that can still satisfy the desired $(n,k,d)$ values regardless whether they offer over-protection ($k> k^*$) or not.

\par {\bf The parameter tuple $(n,k,d)$ and other notation:} From the above definitions, the $n$, $k$, and $d$ values must satisfy
\begin{align}
2\leq n,\quad 1\leq k\leq n-1,\quad\text{and}\quad 1\leq d\leq n-1.\label{eq:ccw1}
\end{align}
In all the results in this work, we assume {\em implicitly} that the $n$, $k$, and $d$ values satisfy \eqref{eq:ccw1}. The overall file size is denoted by $\mathcal{M}$. The storage size for each node is $\alpha$, and during the repair process, the newcomer requests $\beta$ amount of traffic from each of the helpers. The total repair-bandwidth is thus $\gamma\stackrel{\Delta}{=}d\beta$.  We use the notation $(\cdot)^+$ to mean $(x)^+=\max(x,0)$. We also define the indicator function as follows
\begin{align}
1_{\{B\}}=
\begin{cases}1, \mbox{ if condition $B$ is true} \\
0, \mbox{ otherwise}.
\end{cases}
\end{align}

\par In this work, we consider exclusively single failure at any given time. The setting of multiple simultaneous failed nodes \cite{el2010fractional,shum2013cooperative,kamath2014codes} is beyond the scope of this work. We consider the multiple failures scenario in a separate work, see \cite{arxiv_multiple}.

\subsection{Dynamic Versus Stationary Helper Selection Schemes} \label{sec:hs}
\par In general, the helper selection at current time $\tau$ can depend on the history of the failure patterns and the helper choices for all the previous time slots 1 to $(\tau-1)$. We call such a general helper selection scheme {\em the dynamic helper selection (DHS)}. Mathematically, the helper set decision at time $\tau$ can be written in function form as $D_\tau(\{F_l\}_{l=1}^{\tau})$ that takes $F_l$, the failed node at time $l$, for all $l=1$ to $\tau$ and returns the set of helpers for the latest newcomer. The function $D_\tau(\cdot)$ can be designed independently for each time slot $\tau=1,2,\cdots$. One can see that the DHS schemes are the most general form of helper selection.

\par A sub-class of the DHS schemes is the set of \emph{stationary helper selection (SHS)} schemes that assign fixed helper sets of $d$ nodes to each node. The idea is that, for a given node failure, the same helper set is used at any time instant and thus the name stationary. Mathematically, in SHS, each node $i$ is associated with a set of indices $D_i$ where the size of $D_i$ is $d$. Whenever node $i$ fails, the newcomer (for node $i$) simply accesses those helpers $u\in D_i$ and requests $\beta$ amount of data from each helper. 

\par It is not hard to see that SHS is indeed a subset of DHS by observing that any SHS is a DHS with the helper set decision at time $\tau$ being 
\begin{align}
D_\tau(\{F_l\}_{l=1}^{\tau})=D_{i}~\text{if $F_\tau=i$}.\nonumber
\end{align}
Also note that while DHS allows different $D_\tau(\cdot)$ for different $\tau$, the helper set collection $\{D_i:\text{all nodes i}\}$ of SHS is fixed. 

\par Our FHS scheme described in Section~\ref{subsec:desc_fr} is an example of a SHS scheme. Since our FHS scheme, along with its extension, are sufficient to prove the achievability part of Proposition~\ref{prop:comparison}, we do not have to design a DHS scheme for that purpose. More specifically, we have proved that whenever there exists a DHS scheme that strictly outperforms BHS, there always exists another SHS  scheme that strictly outperforms BHS. As a result, at least when considering only single node failure, there is no clear advantage of DHS over SHS. However, for the multiple failures scenario, we have shown in a separate work \cite{arxiv_multiple} that it is possible to have DHS$\succ$SHS$=$BHS. Specifically, under some scenarios, only DHS can strictly outperform BHS while the best SHS  design is no better than the simple BHS solution.

\subsection{The Information-Flow Graph and the Corresponding Graph-Based Analysis}\label{subsec:ifg_existing}
As in \cite{dimakis2010network}, the performance of a distributed storage system can be characterized by the concept of information flow graphs (IFGs). IFGs depict the storage in the network and the communication that takes place during repair. For readers who are not familiar with IFGs, we provide its detailed description in Appendix~\ref{app:ifg}.

\par Intuitively, each IFG reflects one unique history of the failure patterns and the helper selection choices from time $1$ to $(\tau-1)$ \cite{dimakis2010network}. Consider any given helper selection scheme $A$ which can be either dynamic or stationary. Since there are infinitely many different failure patterns (since we consider $\tau=1$ to $\infty$), there are infinitely many IFGs corresponding to the same given helper selection scheme $A$. We denote the collection of all such IFGs by $\mathcal{G}_A(n,k,d,\alpha,\beta)$. We define $\mathcal{G}(n,k,d,\alpha,\beta)=\bigcup_{\forall A}\mathcal{G}_A(n,k,d,\alpha,\beta)$ as the union over all possible helper selection schemes $A$. We sometimes drop the input argument and use $\mathcal{G}_A$ and $\mathcal{G}$ as shorthands.

\par Given an IFG $G\in\mathcal{G}$ and a data collector $t\in \DC(G)$, we use $\mincut_G(s,t)$ to denote the {\em minimum cut value} \cite{west2001introduction} separating $s$, the root node (source node) of $G$, and $t$.

\par The key reason behind representing the repair problem by an IFG is that it casts the problem as a multicast scenario \cite{dimakis2010network}. This allows for invoking the results of network coding in \cite{ahlswede2000network}, \cite{ho2006random}. More specifically, for any helper scheme $A$ and given system parameters $(n,k,d,\alpha,\beta)$, the results in \cite{ahlswede2000network} prove that the following condition is \emph{necessary} for the RC with helper selection scheme $A$ to satisfy the reliability requirement
\begin{align}
\min_{G\in \mathcal{G}_A}\min_{t\in \DC(G)}\mincut_G(s,t)\geq \mathcal{M}. \label{eq:condition}
\end{align}
If we limit our focus to the BHS scheme, then the above necessary condition becomes
\begin{align}
\min_{G\in \mathcal{G}}\min_{t\in \DC(G)}\mincut_G(s,t)\geq \mathcal{M}.\label{eq:condition-BR}
\end{align}
An important contribution of \cite{dimakis2010network} is a closed-form expression of the left-hand side (LHS) of \eqref{eq:condition-BR}
\begin{align}
\min_{G\in \mathcal{G}}\min_{t\in \DC(G)}\mincut_G(s,t)=\sum_{i=0}^{k-1} \min((d-i)^+\beta,\alpha),\label{eq:ex_low_b}
\end{align}
which allows us to numerically check whether \eqref{eq:condition-BR} is true.

\par Reference \cite{wu2010existence} further proves that \eqref{eq:condition-BR} is not only necessary but also sufficient for the existence of a blind RC with some finite field $\GF(q)$ that satisfies the reliability requirement. Namely, as long as ``the right-hand side (RHS) of \eqref{eq:ex_low_b} $\geq \mathcal{M}$" is true, then there exists a RC that meets the system design parameters $(n,k,d,\alpha,\beta)$ even for the worst possible helper selection scheme (since we take the minimum over $\mathcal{G}$).

\par In contrast with the existing results on the BHS scheme, this work focuses on any given helper selection scheme $A$ and we are thus interested in the bandwidth-storage tradeoff specified in \eqref{eq:condition} instead of \eqref{eq:condition-BR}. For example, the Minimum Bandwidth Regenerating (MBR) and Minimum Storage Regenerating (MSR) points of a given helper selection scheme $A$ can be defined by 
\begin{definition} For any given $(n,k,d)$ values, the MBR point $(\alpha_{\text{MBR}}, \beta_{\text{MBR}})$ of a helper scheme $A$ is defined by 
\begin{align}
\beta_{\text{MBR}}\stackrel{\Delta}{=}\min \{\beta: (\alpha,\beta)\text{ satisfies \eqref{eq:condition} and } \alpha=\infty\}\label{eq:MBR-beta-def}\\
\alpha_{\text{MBR}}\stackrel{\Delta}{=}\min \{\alpha: (\alpha,\beta)\text{ satisfies \eqref{eq:condition} and } \beta=\beta_{\text{MBR}}\}.\label{eq:MBR-alpha-def}
\end{align}
\end{definition}

\begin{definition} For any given $(n,k,d)$ values, the MSR point $(\alpha_\text{MSR}, \beta_{\text{MSR}})$ of a helper scheme $A$ is defined by 
\begin{align}
\alpha_{\text{MSR}}\stackrel{\Delta}{=}\min \{\alpha: (\alpha,\beta)\text{ satisfies \eqref{eq:condition} and } \beta=\infty\}\label{eq:MSR-def}\\
\beta_{\text{MSR}}\stackrel{\Delta}{=}\min \{\beta: (\alpha,\beta)\text{ satisfies \eqref{eq:condition} and } \alpha=\alpha_{\text{MSR}}\}.\nonumber
\end{align}
\end{definition}
Specifically, the MBR and MSR are the two extreme ends\footnote{An alternative definition of the MSR point is when a scheme stores only $\alpha=\frac{\mathcal{M}}{k}$ packets, which is different from the definition we used in \eqref{eq:MSR-def}. For example, when $(n,k,d)=(5,3,2)$, one can prove that $\min_{\text{all codes}}\alpha_{\text{MSR}}=\frac{\mathcal{M}}{2}$ based on the definition in \eqref{eq:MSR-def}. We thus say that the MSR point of the best possible scheme is $\alpha^*_{\text{MSR}}=\frac{\mathcal{M}}{2}$ for $(n,k,d)=(5,3,2)$. In contrast, the alternative MSR definition will say that the MSR point does not exist for the parameter $(n,k,d)=(5,3,2)$ since no scheme can achieve
\begin{align} 
\alpha=\frac{\mathcal{M}}{k}=\frac{\mathcal{M}}{3}<\alpha^*_{\text{MSR}}=\frac{\mathcal{M}}{2}.\nonumber
\end{align}
} of the bandwidth-storage tradeoff curve in \eqref{eq:condition}.  

\par By comparing \eqref{eq:condition} and \eqref{eq:condition-BR}, we note that it is possible mathematically that when focusing on $\mathcal{G}_A$ ($\mathcal{G}_A$ is by definition a strict subset of $\mathcal{G}$) we may have
\begin{align}
\min_{G\in \mathcal{G}_A}\min_{t\in \DC(G)}\mincut_G(s,t)> \min_{G\in \mathcal{G}}\min_{t\in \DC(G)}\mincut_G(s,t). \label{eq:outperform}
\end{align}
If \eqref{eq:outperform} is true, then the given helper selection scheme $A$ strictly outperforms the BHS solution. Whether (or under what condition) \eqref{eq:outperform} is true is the main focus of this work.

\subsection{Optimality and Weak Optimality of a Helper Selection Scheme}\label{sec:optimality}
\par For future reference, we define the following optimality conditions.

\begin{definition} \label{def:scheme-optimal}
For any given $(n,k,d)$ value, a helper selection scheme $A$ is \emph{absolutely optimal}, or simply \emph{optimal}, if for any DHS scheme $B$ the following is true
\begin{align}
\min_{G\in \mathcal{G}_A}\min_{t\in \DC(G)}\mincut_G(s,t)\geq \min_{G\in \mathcal{G}_B}\min_{t\in \DC(G)}\mincut_G(s,t)\nonumber
\end{align}
for all $(\alpha,\beta)$ combinations. That is, scheme $A$ has the best $(\alpha,\beta)$ tradeoff curve among all DHS schemes and thus allows for the protection of the largest possible file size.
\end{definition}
\begin{definition} \label{def:collection-scheme-optimal}
A class/collection of helper selection schemes $\mathcal{A}=\{A_1,A_2,\cdots\}$ is optimal if for any $(n,k,d)$ values, we can always find one $A\in\mathcal{A}$ such that $A$ is optimal. 
\end{definition}

\par By the above definitions, it is thus of significant practical/theoretic interest to find an optimal helper selection scheme $A$ for a given $(n,k,d)$ value, and to characterize the smallest optimal helper scheme class $\mathcal{A}$. 

\par While we have been able to devise an optimal helper selection scheme $A$ for some $(n,k,d)$ combinations, see our results in Section~\ref{sec:results}, the problem of finding a small optimal helper scheme class $\mathcal{A}$ remains unsolved. Instead, we will characterize a small class of helper schemes that is {\em weakly optimal:}

\begin{definition} \label{def:scheme-weakly-optimal}
For any given $(n,k,d)$ value, a helper selection scheme $W$ is \emph{weakly optimal}, if the Boolean statement ``there exists a DHS scheme $A$ such that \begin{align}
\min_{G\in \mathcal{G}_A}\min_{t\in \DC(G)}\mincut_G(s,t)> \min_{G\in \mathcal{G}}\min_{t\in \DC(G)}\mincut_G(s,t)\nonumber
\end{align}
for some $(\alpha_1,\beta_1)$'' implies 
\begin{align}
\min_{G\in \mathcal{G}_W}\min_{t\in \DC(G)}\mincut_G(s,t)> \min_{G\in \mathcal{G}}\min_{t\in \DC(G)}\mincut_G(s,t)\nonumber
\end{align}
for some $(\alpha_2,\beta_2)$. 
\end{definition}

\par Comparing Definitions~\ref{def:scheme-optimal} and~\ref{def:scheme-weakly-optimal}, the difference is that the absolute optimality needs to be the best among all DHS schemes, while the weak optimality definition uses the BHS as a baseline and only requires that if the optimal scheme $A^*$ can strictly outperform the BHS scheme, then so can a weakly optimal scheme $W$. 

\par Following the same logic, we can define a weakly optimal collection of helper selection schemes:
\begin{definition}\label{def:collection-scheme-weakly-optimal}
A class/collection of helper selection schemes $\mathcal{W}=\{W_1,W_2,\cdots\}$ is weakly optimal if for any $(n,k,d)$ value, we can always find one $W\in\mathcal{W}$ such that $W$ is weakly optimal. 
\end{definition}

\subsection{From Graph-Based Analysis to Explicit Code Construction}
\par This Part~I of our work focuses exclusively on the graph-based analysis. As discussed in Section~\ref{subsec:ifg_existing}, the graph-based analysis only gives a necessary condition (cf.\ \cite{dimakis2010network}) while the sufficient condition needs to be proved separately through explicit code construction (cf.\ \cite{wu2010existence}). Although the graph-based analysis only gives a necessary condition, in the literature of distributed storage, there is not yet any example in which the min-cut-based characterization is provably not achievable by any finite field, which is an evidence of the power/benefits of graph-based analysis.  

\par To complement the necessary conditions derived by the graph-based min-cut analysis in this Part~I, we have proved the following (partial) sufficiency statement in Part~II \cite{part2} of this work.
\begin{quote} For any $(n,k,d)$ value, consider the two helper selection schemes proposed in this work, termed the family and the family plus helper selection schemes. With a sufficiently large finite field, we can explicitly construct an {\em exact-repair} code with $(\alpha,\beta)$ equal to the MBR point $(\alpha_{\text{MBR}},\beta_{\text{MBR}})$ computed by the min-cut-based analysis \eqref{eq:condition}, \eqref{eq:MBR-beta-def}, and \eqref{eq:MBR-alpha-def}. That is, {\em the necessary condition \eqref{eq:condition} is also sufficient for the MBR point $(\alpha_{\text{MBR}},\beta_{\text{MBR}})$ of the tradeoff curve.}
\end{quote}
As will be discussed in Section~\ref{subsec:fhs_analysis}, the MBR point is the point when good helper selection results in the largest improvement over the BHS scheme. Since our focus is on studying the benefits of helper selection, the above partial statement proved in Part~II is sufficient for our discussion.

\section{Comparison with Existing Codes}\label{sec:comparison}
\begin{table*}[ht!]
\caption{The comparison table among blind-repair regenerating codes, locally repairable codes, and the smart-repair regenerating codes.}
\begin{center}
\begin{tabular}{| p{2cm} || p{4.5cm} | p{4.5cm} | p{4.5cm} |}
\hline
    &Original RC \cite{dimakis2010network,wu2009reducing,rashmi2009explicit,shah2012interference,rashmi2011optimal}& Locally Repairable Codes \cite{gopalan2012locality,papailiopoulos2012locally,kamath2014codes,rawat2012optimal,kamath2013explicit}&
Dynamic Helper Selection\\
\hline\hline

Repair Mode& Functional/Exact-Repair & Exact-Repair & Functional-Repair\\ \hline

Helper Selection&Blind&Stationary (Fixed over time)& Dynamic (helper choices may depend on failure history)\\ \hline

$(n,k,d)$ range&

Designed for $k\leq d$.

&
Designed for $k>d$.

&Allow for arbitrary $(n,k,d)$ values\\ \hline

Contribution&Storage/repair-bandwidth tradeoff for the worst possible helper selection&Storage/repair-bandwidth characterization for the specific stationary helper selection of the proposed exact-repair local code, which may/may not be optimal&First exploration of the storage/repair-bandwidth tradeoff for the optimal dynamic helper selection\\

\hline
\end{tabular}
\label{tab:comparison}
\end{center}
\end{table*}

\par Recall that RCs are distributed storage codes that minimize the repair-bandwidth (given a storage constraint). In comparison, codes with local repair or (when with all-symbol locality) {\em locally repairable codes (LRC)}, recently introduced in \cite{gopalan2012locality}, are codes that minimize the number of helpers participating in the repair of a failed node. LRCs were proposed to address the disk I/O overhead problem that the repair process may entail on a storage network since the number of helpers participating in the repair of a failed node is proportional to the amount of disk I/O needed during repair. Subsequent development has been done on LRCs in \cite{papailiopoulos2012locally,kamath2014codes,rawat2012optimal,kamath2013explicit}.

\par In Table~\ref{tab:comparison}, we compare the setting of the original RCs, LRCs, and the DHS considered in this work. As first introduced in \cite{dimakis2010network}, original RCs were proposed under the functional-repair scenario, i.e., nodes of the storage network are allowed to store any combination of the original packets as long as the reliability requirement is satisfied. In subsequent works \cite{wu2009reducing,rashmi2009explicit,shah2012interference,rashmi2011optimal,
cadambe2013asymptotic,shah2012distributed}, RCs were considered under the exact-repair scenario in which nodes have to store the same original packets at any given time. In contrast, LRCs are almost always considered under the exact-repair scenario. However, in this work, for RCs with DHS, we consider functional-repair as the mode of repair as we aim at understanding the absolute benefits/limits of helper selection in RCs. Albeit our setting is under functional-repair, in Part~II, we are able to present an explicit construction of exact-repair codes that achieve the optimal or weakly optimal MBR point of the functional-repair. 

\par Table~\ref{tab:comparison} also compares the three scenarios in terms of the helper selection mechanisms. The original RCs are codes that do not perform helper selection at all, i.e., BHS, while LRCs are codes that can perform SHS only. In this work, we consider the most general setting in which codes are allowed to have DHS.

\par Moreover, as shown in Table~\ref{tab:comparison}, the $(n,k,d)$ range of operation of each of the three code settings is different. The original RCs were designed for storage networks with large $d$ values, whereas LRCs are designed for small $d$ values. In contrast, this work allows for arbitrary $(n,k,d)$ values and studies the benefits of helper selection under different $(n,k,d)$ values.

\par The comparison above illustrates the main differences in the setup and contributions between the three scenarios. The original RCs are concerned with the storage/repair-bandwidth tradeoff for the worst possible helper selection. LRCs, on the other hand, are concerned with only data storage (ignoring repair-bandwidth) of the codes when restricting to SHS and exact-repair. Some recent developments \cite{kamath2014codes,kamath2013explicit} in LRCs consider using RCs in the construction of the codes therein (as local codes) in an attempt to examine the repair-bandwidth performance of LRCs. This approach, however, is not guaranteed to be optimal in terms of storage/repair-bandwidth tradeoff. 

\par In this work, we present the first exploration of the optimal storage-bandwidth tradeoff for RCs that allow {\em dynamic helper selection (DHS)} for arbitrary $(n,k,d)$ values. The closest setting in the existing literature is in \cite{hollmann2014minimum}. That work finds upper bounds on the file size $\mathcal{M}$ when $\alpha=d\beta$ and $\alpha=\beta$ for functional-repair with DHS. However, \cite{hollmann2014minimum} considers the case of $k=n-1$ only. Also, it is not clear whether the provided upper bounds for $k=n-1$ are tight or not. A byproduct of the results of this work shows that the upper bounds in \cite{hollmann2014minimum} are tight in some cases and loose in others, see Corollary~\ref{cor:existing_loose} and Propositions~\ref{prop:optimal_2} and~\ref{prop:family-plus_optimal}. 

\section{Preview Of The Results} \label{sec:preview}

In the following, we give a brief preview of our results through concrete examples to illustrate the main contributions of this work. Although we only present here specific examples as a preview, the main results in Section~\ref{sec:results} are for general $(n,k,d)$ values.

\par\emph{Result~1:} For $(n,k,d)=(6,3,4)$, RCs with BHS are absolutely optimal, i.e., there exists no RCs with DHS that can outperform BHS. 

\par
\begin{figure}[h!]
\centering
\includegraphics[width=0.475\textwidth]{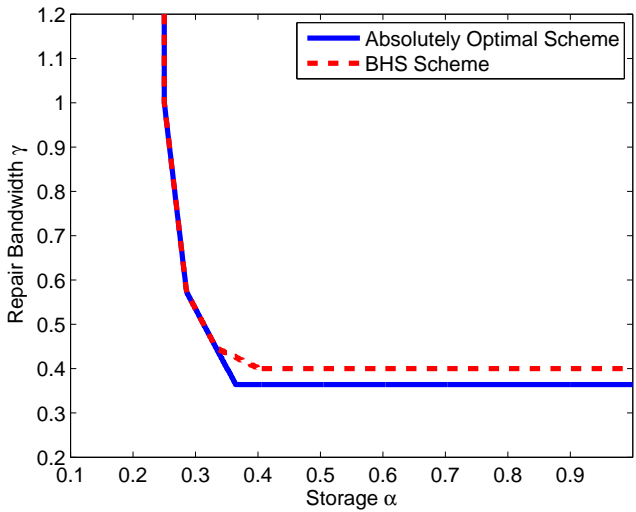}
\caption{Storage-bandwidth tradeoff curves of RCs with BHS versus RCs with the absolutely optimal scheme (FHS) for $(n,k,d)=(6,4,4)$ and file size $\mathcal{M}=1$.}
\label{fig:storage_vs_bandwidth_(6-4-4)}
\end{figure}

\emph{Result~2:} For $(n,k,d)=(6,4,4)$, the RCs with the new family helper selection (FHS) scheme proposed in this paper are absolutely optimal in terms of the storage-bandwidth tradeoff among all RCs with DHS, also see Definition~\ref{def:scheme-optimal}. In Fig.~\ref{fig:storage_vs_bandwidth_(6-4-4)}, the storage-bandwidth tradeoff curve of the FHS scheme, the optimal helper selection scheme, is plotted against the BHS scheme with file size $\mathcal{M}=1$. In Part~II, we provide an explicit construction of an exact-repair code that can achieve $(\alpha,\gamma)=(\frac{4}{11},\frac{4}{11})$, the MBR point of the storage-bandwidth tradeoff curve of the FHS scheme in Fig.~\ref{fig:storage_vs_bandwidth_(6-4-4)}. If we take a closer look at Fig.~\ref{fig:storage_vs_bandwidth_(6-4-4)}, there are 3 corner points on the FHS scheme curve and they are $(\alpha,\gamma)=(0.25,1)$, $(\frac{2}{7},\frac{4}{7})$, and $(\frac{4}{11},\frac{4}{11})$. Since the two corners  $(\alpha,\gamma)=(0.25,1)$ and $(\frac{2}{7},\frac{4}{7})$ can be achieved by the scheme in \cite{wu2010existence} and the new corner point $(\alpha,\gamma)=(\frac{4}{11},\frac{4}{11})$ is proved to be achievable in Part~II, we can thus achieve the entire optimal tradeoff curve in Fig.~\ref{fig:storage_vs_bandwidth_(6-4-4)} by space-sharing while no other scheme can do better, as stated in Proposition~\ref{prop:optimal_main}.\footnote{If we analyze the LRCs proposed in \cite{kamath2014codes,kamath2013explicit,papailiopoulos2012locally} for $(n,k,d)=(6,4,4)$, we can show that those codes/schemes cannot do better than the BHS curve at the MSR point. As a result, the LRCs in \cite{kamath2014codes,kamath2013explicit,papailiopoulos2012locally} are no better than the absolutely optimal scheme curve in Fig.~\ref{fig:storage_vs_bandwidth_(6-4-4)}, as predicted by Proposition~\ref{prop:optimal_main}.}

\par
\begin{figure}[h!]
\centering
\includegraphics[width=0.475\textwidth]{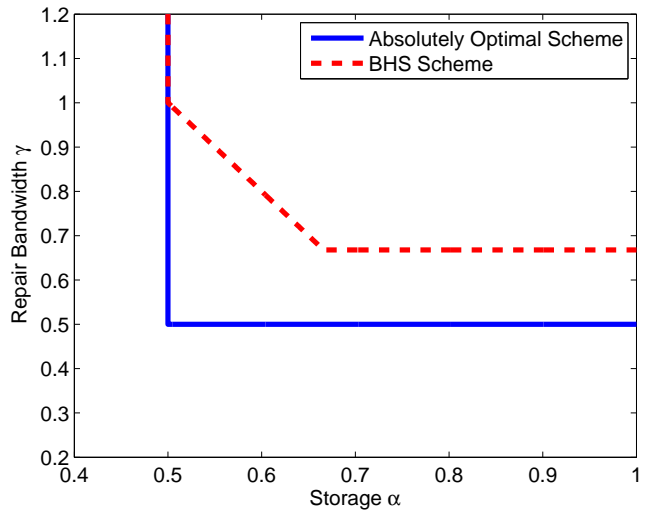}
\caption{Storage-bandwidth tradeoff curves of RCs with BHS versus RCs with the absolutely optimal scheme (FHS) for $(n,k,d)=(5,3,2)$ and file size $\mathcal{M}=1$.}
\label{fig:storage_vs_bandwidth_(5-3-2)}
\end{figure}

\emph{Result~3:} For $(n,k,d)=(5,3,2)$, the proposed FHS scheme again outperforms the BHS scheme, and is provably optimal.\footnote{Using Proposition~\ref{prop:low_b}, we have that the tradeoff of FHS is characterized by $2\min(2\beta,\alpha)\geq\mathcal{M}$ for $(n,k,d)=(5,3,2)$. It is not hard to prove, in a similar way to the proof of Proposition~\ref{prop:optimal}, that any arbitrary DHS scheme is bound to do no better than this tradeoff.} We note that BHS is inherently inefficient in this example since BHS always has $k^*\leq d$ and thus overprotects the data when $d<k$. However, for this particular $(n,k,d)$ combination we do not have any other existing scheme that can be used as a baseline. For that reason, we still compare to BHS in this example for the sake of illustration. Fig.~\ref{fig:storage_vs_bandwidth_(5-3-2)} shows a tradeoff curve comparison between the FHS scheme and the BHS scheme. An interesting phenomenon is that the tradeoff curve of the FHS scheme has only one corner point $(\alpha,\gamma)=(0.5,0.5)$ and we can achieve this point by an exact-repair scheme, see Part~II \cite{part2}. Note that this exact-repair scheme for $(\alpha,\gamma)=(0.5,0.5)$ has the same storage consumption as the MSR point of the original RC ($(\alpha,\gamma)=(0.5,1)$) while using strictly less than the bandwidth of the MBR point of the original RC ($(\alpha,\gamma)=(\frac{2}{3},\frac{2}{3})$). Since the provably optimal FHS scheme has only a single corner point, it means that we can achieve minimum-storage (the MSR point) and minimum-bandwidth (the MBR point) simultaneously.

\par
\begin{figure}[h!]
\centering
\includegraphics[width=0.475\textwidth]{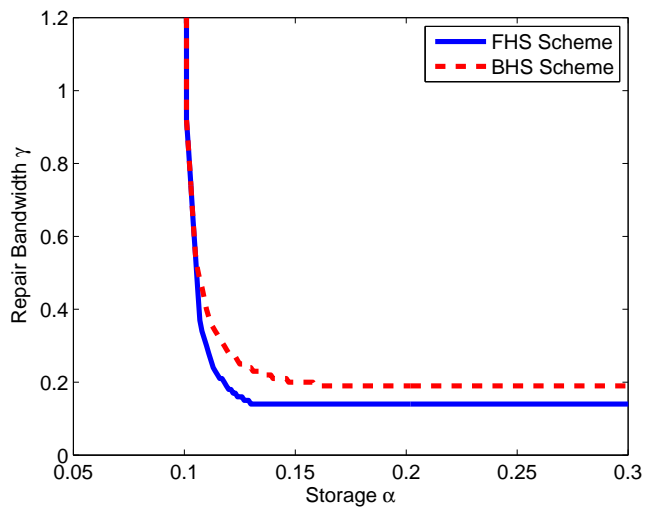}
\caption{Storage-bandwidth tradeoff curves of RCs with BHS versus RCs with FHS for $(n,k,d)=(20,10,10)$ and file size $\mathcal{M}=1$.}
\label{fig:storage_vs_bandwidth_(20-10-10)}
\end{figure}

\emph{Result~4:} For $(n,k,d)=(20,10,10)$, we do not know what is the absolutely optimal DHS scheme. On the other hand, the FHS scheme again outperforms the BHS scheme.  Fig.~\ref{fig:storage_vs_bandwidth_(20-10-10)} shows a tradeoff curve comparison between the FHS scheme and the BHS scheme.

\par
\begin{figure}[h!]
\centering
\includegraphics[width=0.475\textwidth]{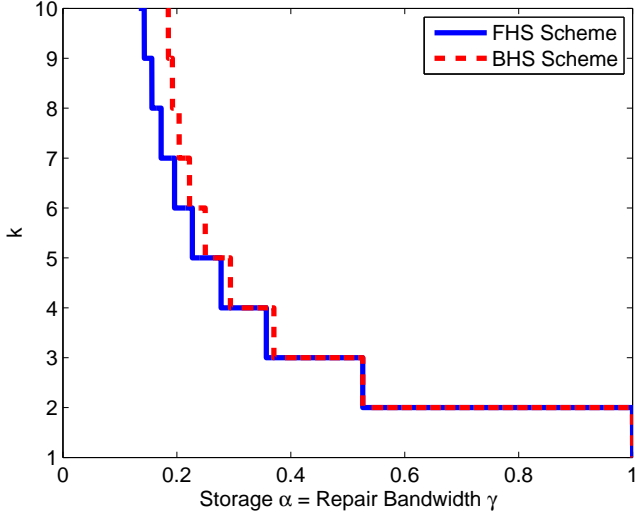}
\caption{The $k$ value versus repair-bandwidth $\gamma$ curve comparison at the MBR point for $(n,d)=(60,10)$ and file size $\mathcal{M}=1$.}
\label{fig:k_vs_gamma_(60-10)}
\end{figure}

\emph{Result~5:} For $(n,d)=(60,10)$, we do not know what is the absolutely optimal DHS scheme. However, in Fig.~\ref{fig:k_vs_gamma_(60-10)}, we plot a $k$ versus repair-bandwidth curve to compare the BHS scheme to the FHS scheme while restricting to the MBR point. Examining Fig.~\ref{fig:k_vs_gamma_(60-10)}, we can see that the BHS scheme performs poorly compared to FHS as $k$ grows larger. When $k=d=10$, the FHS scheme only uses $73.33\%$ of the bandwidth of the BHS scheme.

\par\emph{Result~6:} Although the main focus of this work is to investigate the benefits of helper selection, a byproduct of our results is a new explicit construction of exact-repair codes for arbitrary $(n,k,d,\alpha,\beta)$ values satisfying $\alpha=d\beta$. This code construction is presented in Part~II of this work. Numerically, the proposed codes demonstrate good performance in all $(n,k,d)$ cases. Analytically, it achieves the absolutely optimal MBR points, among all DHS schemes, for all $(n,k,d,\alpha,\beta)$ values satisfying (i) $n\neq 5$, $k=n-1$, and $d=2$; (ii) $n$ is even, $k=n-1$, and $d=3$; (iii) $n\notin \{7,9\}$, $k=n-1$, and $d=4$; (iv) $n$ is even, $n\notin \{8,14\}$, $k=n-1$, and $d=5$; and (v) $n\notin \{10,11,13\}$, $k=n-1$, and $d=6$. This result is the combination of Proposition~\ref{prop:family-plus_optimal} and the explicit code construction in Part~II.

\section{The Main Results} \label{sec:results}
\par The main result in this paper is the answer to the question ``When is it beneficial to choose the good helpers?''. This is stated as a necessary and sufficient condition in the following proposition. 
\begin{proposition}  \label{prop:comparison}
(\emph{The converse:}) If at least one of the following two conditions is true:  (i) $d=1$, $k=3$, and $n$ is odd; and (ii) $k\leq \left\lceil \frac{n}{n-d}\right\rceil$, then BHS is absolutely optimal, see Definition~\ref{def:scheme-optimal}. That is, even the best DHS scheme has identical performance to the BHS. \par (\emph{The achievability:}) For any $(n,k,d)$ values that satisfy neither (i) nor (ii), there exists a DHS scheme and a pair of $(\alpha,\beta)$ values such that we can protect a file of size strictly larger than that of BHS.
\end{proposition}
\par The converse and the achievability of the above proposition are formally stated and proved in Sections~\ref{sec:converse} and \ref{sec:achievability}, respectively. The converse is proved by a new min-cut based analysis. The achievability is proved by analyzing a new scheme termed the family helper selection (FHS) scheme, along with its extension, described in Sections~\ref{subsec:desc_fr} and~\ref{sec:family-plus}.

\par We have two other major results that state the optimality of our new FHS schemes.
\begin{proposition} \label{prop:optimal_main} 
For any $(n,k,d)$ values satisfying simultaneously the following three conditions (i) $d$ is even, (ii) $n=d+2$, and (iii) $k=\frac{n}{2}+1$; the FHS scheme is absolutely optimal. 
\end{proposition}

\begin{proposition}  \label{prop:optimal_2_main} 
For any $(n,k,d,\alpha,\beta)$ values satisfying simultaneously the following two conditions (i) $k=n-1$, (ii) we can rewrite $n=\sum_{b=1}^B n_b$ for positive integers $n_b$ satisfying $n_b\bmod(n_b-d)=0$ for all $b=1,\cdots, B$, the extension of the proposed FHS scheme, see Section~\ref{sec:family-plus}, achieves the minimum repair bandwidth among all DHS schemes. More explicitly, our proposed scheme has the $(\alpha,\beta)$ value satisfying $\beta=\min_{\text{all possible codes }} \beta_{\text{MBR}}$.
\end{proposition}

Propositions~\ref{prop:optimal_main} and \ref{prop:optimal_2_main} will be restated and proved in Propositions~\ref{prop:optimal} and \ref{prop:optimal_2}, respectively, in Section~\ref{sec:achievability}. 

\section{The Converse}\label{sec:converse}

\par Before proving the converse result, we introduce the following definition and lemma.
\begin{definition}\label{def:m_set}
A set of $m$ active storage nodes (input-output pairs) of an IFG is called an $m$-set if the following conditions are satisfied simultaneously.  (i) Each of the $m$ active nodes has been repaired at least once; and (ii) jointly the $m$ nodes satisfy the following property: consider any two distinct active nodes $x$ and $y$ in the $m$-set and, without loss of generality, assume that $x$ was repaired before $y$. Then there exists an edge in the IFG that connects $x_\text{out}$ and $y_\text{in}$.
\end{definition}

\begin{lemma} \label{lem:set} Fix a helper selection scheme $A$. Consider an arbitrary $G\in \mathcal{G}_A(n,k,d,\alpha,\beta)$ such that each active node in $G$ has been repaired at least once. Then there exists a $\left\lceil \frac{n}{n-d}\right\rceil$-set in $G$.
\end{lemma}

\begin{IEEEproof}
We prove this lemma by proving the following stronger claim: Consider any integer value $m\geq 1$. There exists an $m$-set in every group of $(m-1)(n-d)+1$ active nodes that have been repaired at least once in the past. Since the $G$ we consider has $n$ active nodes and each of them has been repaired at least once, the above claim implies that $G$ must contain a $\left\lceil \frac{n}{n-d}\right\rceil$-set.

\par We prove this claim by induction on the value of $m$. When $m=1$, by the definition of the $m$-set, any group of 1 active node in $G$ forms a 1-set. The claim thus holds naturally.

\par Suppose the claim is true for all $m<m_0$, we now claim that in every group of $(m_0-1)(n-d)+1$ active nodes of $G$ there exists an $m_0$-set. The reason is as follows. Given an arbitrary, but fixed group of $(m_0-1)(n-d)+1$ active nodes, we use $y$ to denote the youngest active node in this group  (the one which was repaired last). Obviously, there are $(m_0-1)(n-d)$ active nodes in this group other than $y$. On the other hand, since any newcomer accesses $d$ helpers out of $n-1$ surviving nodes during its repair, node $y$ was able to ``avoid'' connecting to at most $(n-1)-d$ surviving nodes (the remaining active nodes). Therefore, out of the remaining $(m_0-1)(n-d)$ active nodes in this group, node $y$ must be connected to at least $(m_0-1)(n-d)-(n-1-d)=(m_0-2)(n-d)+1$ of them. By induction, among those $\geq (m_0-2)(n-d)+1$ nodes, there exists an $(m_0-1)$-set. Since, by our construction, $y$ is connected to {\em all} nodes in this $(m_0-1)$-set, node $y$ and this $(m_0-1)$-set jointly form an $m_0$-set. The proof of this claim is complete and hence the proof of Lemma~\ref{lem:set}.
\end{IEEEproof}

In the following proposition, we restate the converse  part of Proposition~\ref{prop:comparison} and prove it.
\begin{proposition}  \label{prop:converse}
If at least one of the following two conditions is true:  (i) $d=1$, $k=3$, and $n$ is odd; and (ii) $k\leq \left\lceil \frac{n}{n-d}\right\rceil$, then for any arbitrary DHS scheme $A$ and any arbitrary $(\alpha,\beta)$ values, we have
\begin{align} \label{eq:neg}
\min_{G\in\mathcal{G}_A}\min_{t\in \DC(G) } \mincut_G(s,t)=\sum_{i=0}^{k-1}\min ((d-i)^+\beta,\alpha),
\end{align}
that is, BHS is absolutely optimal.
\end{proposition}
\begin{IEEEproof}
Assume condition (ii) holds and consider an IFG $G\in \mathcal{G}_A$ in which every active node has been repaired at least once. By Lemma~\ref{lem:set}, there exists a $\left\lceil \frac{n}{n-d} \right\rceil$-set in $G$. Since condition (ii) holds, we can consider a data collector of $G$ that connects to $k$ nodes out of this $\left\lceil \frac{n}{n-d} \right\rceil$-set. Call this data collector $t$. If we focus on the edge cut that separates source $s$ and the $k$ node pairs connected to $t$, one can use the same analysis as in \cite[Lemma 2]{dimakis2010network} and derive ``$\mincut(s,t)\leq \sum_{i=0}^{k-1}\min ((d-i)^+\beta,\alpha)$''  for the given $G\in G_A$ and the specific choice of $t$. By further taking the minimum over all $t\in \DC(G)$ and all $G\in \mathcal{G}_A$, we have
\begin{align}\label{eq:worse_br}
\min_{G\in\mathcal{G}_A}\min_{t\in \DC(G) } \mincut_G(s,t)\leq \sum_{i=0}^{k-1}\min((d-i)^+\beta,\alpha).
\end{align}
On the other hand, since by definition $\mathcal{G}_A\subseteq \mathcal{G}$, we have
\begin{align}
\min_{G\in\mathcal{G}_A}\min_{t\in \DC(G) } \mincut_G(s,t)\geq \min_{G\in\mathcal{G}}\min_{t\in \DC(G) } \mincut_G(s,t). \label{eq:new2_better}
\end{align}
Then by \eqref{eq:worse_br}, \eqref{eq:new2_better}, and \eqref{eq:ex_low_b}, we have proved that whenever condition (ii) holds, the equality \eqref{eq:neg} is true.

Now, assume condition (i) holds. We first state the following claim and use it to prove \eqref{eq:neg}.

\begin{claim} \label{clm:vertex-cut}
For any given DHS scheme $A$ and the corresponding collection of IFGs $\mathcal{G}_A$, we can always find a $G^*\in \mathcal{G}_A$ that has a set of 3 active nodes, denoted by $x$, $y$, and $z$, such that the following three properties hold simultaneously: (a) $x$ is repaired before $y$ and $y$ is repaired before $z$; (b) $(x_\text{out},y_\text{in})$ is an edge in $G^*$; and (c) either $(x_\text{out}, z_\text{in})$ is an edge in $G^*$ or $(y_\text{out},z_\text{in})$ is an edge in $G^*$.
\end{claim}

\par Suppose the above claim is true. We let $t^*$ denote the data collector that is connected to $\{x,y,z\}$. By properties (a) to (c) we can see that node $x$ is a vertex-cut separating source $s$ and the data collector $t^*$. The min-cut value separating $s$ and $t^*$ thus satisfies $\mincut_{G^*}(s,t^*)\leq\min(d\beta,\alpha)=\sum_{i=0}^{k-1}\min((d-i)^+\beta,\alpha)$, where  the inequality follows from $x$ being a vertex-cut separating $s$ and $t^*$ and the equality follows from that condition (i) being true implies $d=1$ and $k=3$. By the same arguments as used in proving the case of condition (ii), we thus have \eqref{eq:neg} when condition (i) holds.

\par We prove Claim~\ref{clm:vertex-cut} by explicit construction. Start from any $G\in \mathcal{G}_A$ with all $n$ nodes having been repaired at least once. We choose one arbitrary active node in $G$ and denote it by $w^{(1)}$. We let $w^{(1)}$ fail and denote the newcomer that replaces $w^{(1)}$ by $y^{(1)}$. The helper selection scheme $A$ will choose a helper node (since $d=1$) and we denote that helper node as $x^{(1)}$. The new IFG after this failure and repair process is denoted by $G^{(1)}$. By our construction $x^{(1)}$, as an existing active node, is repaired before the newcomer $y^{(1)}$ and there is an edge $(x^{(1)}_\text{out},y^{(1)}_\text{in})$ in $G^{(1)}$.

\par Starting now from $G^{(1)}$, we choose another $w^{(2)}$ which is not one of $x^{(1)}$ and $y^{(1)}$ and let this node fail. Such $w^{(2)}$ always exists since $n$ is odd by condition (i). We use $y^{(2)}$ to denote the newcomer that replaces $w^{(2)}$. The helper selection scheme $A$ will again choose a helper node based on the history of the failure pattern. We denote the new IFG (after the helper selection chosen by scheme $A$) as $G^{(2)}$. If the helper node of $y^{(2)}$ is $x^{(1)}$, then the three nodes $(x^{(1)},y^{(1)}, y^{(2)})$ are the $(x,y,z)$ nodes satisfying properties (a), (b) and the first half of (c). If the helper node of $y^{(2)}$ is $y^{(1)}$, then the three nodes $(x^{(1)},y^{(1)}, y^{(2)})$ are the $(x,y,z)$ nodes satisfying properties (a), (b) and the second half of (c).  In both cases, we can stop our construction and let $G^*=G^{(2)}$ and we say that the construction is complete in the second round. 

\par Suppose neither of the above two is true, i.e., the helper of $y^{(2)}$ is neither $x^{(1)}$ nor $y^{(1)}$. Then, we denote the helper of $y^{(2)}$ by $x^{(2)}$. Note that after this step, $G^{(2)}$ contains two disjoint pairs of active nodes such that there is an edge $(x^{(m)}_\text{out}, y^{(m)}_\text{in})$ in $G^{(2)}$ for $m=1,2$.

\par We can repeat this process for the third time by failing a node $w^{(3)}$ that is none of $\{x^{(m)},y^{(m)}:\forall m=1,2\}$. We can always find such a node $w^{(3)}$ since $n$ is odd when condition (i) holds. Again, let $y^{(3)}$ denote the newcomer that replaces $w^{(3)}$ and the scheme $A$ will choose a helper for $y^{(3)}$. The new IFG after this failure and repair process is denoted by $G^{(3)}$.  If the helper of $y^{(3)}$ is $x^{(m)}$ for some $m=1,2$, then the three nodes $(x^{(m)},y^{(m)}, y^{(3)})$ are the $(x,y,z)$ nodes satisfying properties (a), (b) and the first half of (c).  If the helper node of $y^{(3)}$ is $y^{(m)}$ for some $m=1,2$, then the three nodes $(x^{(m)},y^{(m)}, y^{(3)})$ are the $(x,y,z)$ nodes satisfying properties (a), (b) and the second half of (c).  In both cases, we can stop our construction and let $G^*=G^{(3)}$ and we say that the construction is complete in the third round. If neither of the above two is true, then we denote the helper of $y^{(3)}$ by $x^{(3)}$, and repeat this process for the fourth time and so on.

\par We now observe that since $n$ is odd, if the construction is not complete in the $m_0$-th round, we can always start the $(m_0+1)$-th round since we can always find a node $w^{(m_0+1)}$ that is none of $\{x^{(m)},y^{(m)}:\forall m=1,2,\cdots, m_0\}$.   On the other hand, we cannot repeat this process indefinitely since we only have a finite number of $n$ active nodes in the network. Therefore, the construction must be complete in the $\tilde{m}$-th round for some finite $\tilde{m}$. If the helper of $y^{(\tilde{m})}$ is $x^{(m)}$ for some $m=1,2,\cdots \tilde{m}-1$,  then the three nodes $(x^{(m)},y^{(m)}, y^{(\tilde{m})})$ are the $(x,y,z)$ nodes satisfying properties (a), (b) and the first half of (c).  If the helper node of $y^{(\tilde{m})}$ is $y^{(m)}$ for some $m=1,2,\cdots, \tilde{m}-1$, then the three nodes $(x^{(m)},y^{(m)}, y^{(\tilde{m})})$ are the $(x,y,z)$ nodes satisfying properties (a), (b) and the second half of (c). Let $G^*=G^{(\tilde{m})}$  denote the final IFG. The explicit construction of $G^*$ and the corresponding $(x,y,z)$ nodes is thus complete.
\end{IEEEproof}

\par To illustrate Proposition~\ref{prop:converse}, consider $(n,k,d)=(6,3,4)$. We have that $k=3\leq \left\lceil \frac{n}{n-d}\right\rceil=3$, i.e., condition (ii) of Proposition~\ref{prop:converse} is satisfied, implying Result~1 in Section~\ref{sec:preview} that BHS is absolutely optimal.


\section{The Achievability} \label{sec:achievability}
In this section, we restate the achievability result of Proposition~\ref{prop:comparison} and prove it. Before we do that, we first describe and analyze  our low-complexity schemes, the family and the family-plus helper selection schemes, that will be used later to prove the achievability.
\subsection{The Family Helper Selection Scheme and Its Notation} \label{subsec:desc_fr}
\begin{figure*}
\centering
\includegraphics[width=7.1in,keepaspectratio]{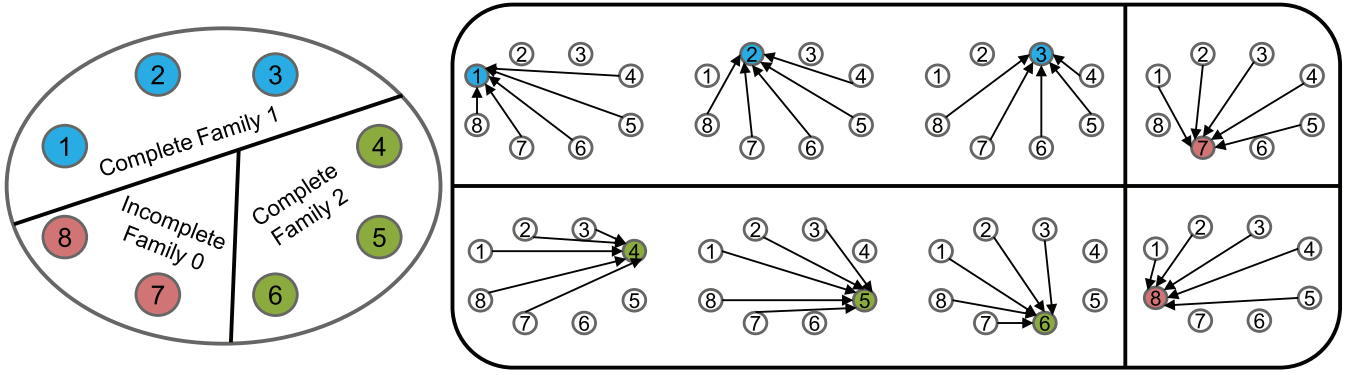} 
\caption{The FHS scheme for $(n,d)=(8,5)$ and the illustration of the repair process of each of the 8 nodes.}
\label{fig:fhs}
\end{figure*}

{\bf The description of the family helper selection (FHS) scheme:} We propose a new helper selection scheme, which is termed the \emph{family helper selection (FHS) scheme} and is a sub-class of SHS schemes. To describe the FHS scheme, we first arbitrarily sort all storage nodes and denote them by $1$ to $n$. We then define a {\em complete family} as a group of $(n-d)$ physical nodes. The first $(n-d)$ nodes are grouped as the first complete family and the second $(n-d)$ nodes are grouped as the second complete family and so on. In total, there are $\left\lfloor \frac{n}{n-d}\right\rfloor$ complete families. The remaining $n\bmod(n-d)$ nodes are grouped as an {\em incomplete family}. The helper set $D_i$ of any node $i$ in a complete family contains all the nodes {\em not} in the same family of node $i$. That is, a newcomer only seeks help from {\em outside} its family. The intuition is that we would like each family to preserve as much information (or equivalently as diverse information) as possible. To that end, we design the helper selection sets such that each newcomer refrains from requesting help from its own family. For any node in the incomplete family,\footnote{\label{footnote:incomplete}All the concepts and intuition are based on complete families. The incomplete family is used to make the scheme consistent and applicable to the case when $n\bmod(n-d)\neq 0$. } we set the corresponding $D_i=\{1,\cdots, d\}$. The description of the FHS scheme is complete. 

\par For example, suppose that $(n,d)=(8,5)$. There are $2$ complete families, $\{1,2,3\}$ and $\{4,5,6\}$, and $1$ incomplete family, $\{7,8\}$. See Fig.~\ref{fig:fhs} for illustration. The FHS scheme for this example is illustrated in Fig.~\ref{fig:fhs}. Let us say node $4$ fails. The corresponding newcomer will access nodes $\{1,2,3,7,8\}$ for repair since nodes 1, 2, 3, 7, and 8 are outside the family of node 4. If node $7$ (a member of the incomplete family) fails, then the newcomer will access nodes $1$ to $5$ for repair. 

{\bf Notation that is useful when analyzing the FHS scheme:} The above description of the FHS is quite simple. On the other hand, to facilitate further analysis, we need the following notation as well. By the above definitions, we have in total $\left\lceil\frac{n}{n-d}\right\rceil$ number of families, which are indexed from $1$ to $\left\lceil\frac{n}{n-d}\right\rceil$. However, since the incomplete family has different properties from the complete families, we replace the index of the incomplete family with $0$. Therefore, the family indices become from $1$ to $c\stackrel{\Delta}{=}\left\lfloor\frac{n}{n-d}\right\rfloor$ and then $0$, where $c$ is the index of the last Complete family. If there is no incomplete family, we simply omit the index $0$. Moreover, by our construction, any member of the incomplete family has $D_i=\{1,\cdots, d\}$. That is, it will request help from {\em all} the members of the first $(c-1)$ complete families, {\em but only from} the first $d-(n-d) (c-1)=n\bmod(n-d)$ members of the last complete family. Among the $(n-d)$ members in the last complete family, we thus need to distinguish those members who will be helpers for incomplete family members, and those who will not. Therefore, {\em we add a negative sign to the family indices of those who will ``not'' be helpers for the incomplete family.}

\par From the above discussion, we can now list the family indices of the $n$ nodes as an $n$-dimensional {\em family index vector}. Consider the same example as above where $(n,d)=(8,5)$. There are two complete families, nodes 1 to 3 and nodes 4 to 6. Nodes 7 and 8 belong to the incomplete family and thus have family index 0. The third member of the second complete family, node $6$, is not a helper for the incomplete family members, nodes $7$ and $8$, since $D_7=D_8=\{1,\cdots, d\}=\{1,2,\cdots, 5\}$. Therefore, we replace the family index of node 6 by $-2$. In sum, the {\em family index vector} of this $(n,d)=(8,5)$ example becomes $(1,1,1,2,2,-2, 0,0)$. Mathematically, we can write the family index vector as
\begin{align}
\left(\overbrace{1,\cdots, 1}^{n-d}, \right. \overbrace{2,\cdots, 2}^{n-d}&, \cdots, \overbrace{ c, \cdots, c}^{ n\bmod(n-d)},\nonumber\\
&\left.\overbrace{ -c, \cdots, -c}^{n-d-(n\bmod(n-d))} , \overbrace{0,\cdots, 0}^{n\bmod (n-d)}\right).\label{eq:ccw5}
\end{align}

\par
\begin{figure}[h!]
\centering
\includegraphics[width=0.475\textwidth]{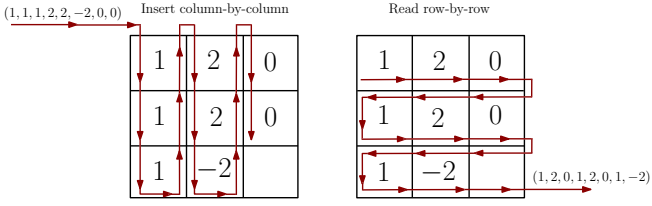}
\caption{The construction of the RFIP for $(n,d)=(8,5)$.}
\label{fig:rfip}
\end{figure}

 A {\em family index permutation} is a permutation of the family index vector defined in \eqref{eq:ccw5}, which we denote by $\pi_f$. Continuing from the previous example, one instance of family index permutations is $\pi_f=(1,1,0,2,0,-2,1,2)$. A \emph{rotating family index permutation (RFIP)} $\pi_f^*$ is a special family index permutation that puts the family indices of \eqref{eq:ccw5} in an $(n-d)\times \left\lceil \frac{n}{n-d}\right\rceil$ table column-by-column and then reads it row-by-row. Fig.~\ref{fig:rfip} illustrates the construction of the RFIP for $(n,d)=(8,5)$. The input is the family index vector $(1,1,1,2,2,-2,0,0)$ and the output is the RFIP $\pi_f^*=(1,2,0, 1,2,0,1,-2)$.

\subsection{Analysis of the Family Helper Selection Scheme} \label{subsec:fhs_analysis}

\par We analyze in this section the performance of the FHS scheme. Recall that FHS is a special example of the SHS. In the following, we first provide a lower bound on the performance of any given SHS scheme that will later be used in the analysis of FHS.
\begin{proposition} \label{prop:low_b_gen}
Consider any SHS scheme $A$ and denote its collection of helper sets by $\{D_1,D_2,\dots,D_n\}$. We have
\begin{align}
\min_{G\in \mathcal{G}_A}\min_{t\in \DC(G)}\mincut(s,t)\geq \min_{\mathbf{r}\in R}\sum_{i=1}^{k}\min ((d-z_i(\mathbf{r}))\beta,\alpha) \label{eq:low_b_gen},
\end{align}
where $\mathbf{r}$ is a $k$-dimensional integer-valued vector, $R=\{(r_1,r_2,\cdots,r_k):\forall i \in\{1,\cdots,k\}, 1\leq r_i\leq n\}$, and $z_i(\cdot)$ is a function $z_i:\{1,\cdots, n\}^k\mapsto \mathbb{N}$ defined as $z_i(\mathbf{r})=|\{a\in D_{r_i}:\exists j<i, a=r_j\}|$,  where $\mathbb{N}$ is the set of all positive integers and $D_{r_i}$ is the helper set of node $r_i$. For example, suppose $n=6$,  $k=4$, $D_3=\{1,4\}$, and $\mathbf{r}=(1,2,1,3)$, then we have $r_4=3$ and $z_4(\mathbf{r})=|\{a\in D_3:\exists j<4, a=r_j\}|= 1$.
\end{proposition}

\par The proof of Proposition~\ref{prop:low_b_gen} is relegated to Appendix~\ref{app:low_b_gen}.

\par Proposition~\ref{prop:low_b_gen} above establishes a lower bound on the cut capacity of any SHS  scheme. Therefore, when designing any SHS scheme, one simply needs to choose $(n,k,d,\alpha,\beta)$ values and the helper sets $D_i$ so that the RHS of \eqref{eq:low_b_gen} is no less than the file size $\mathcal{M}$. However, since we do not have equality in \eqref{eq:low_b_gen}, the above construction is sufficient but not necessary. That is, we may be able to use smaller $\alpha$ and $\beta$ values while still guaranteeing that the resulting regenerating code with the given SHS meets the reliability requirement.

\par When we focus on the FHS scheme introduced in Section~\ref{subsec:desc_fr}, a special SHS scheme, the inequality \eqref{eq:low_b_gen} can be further sharpened to the following equality.
\begin{proposition} \label{prop:low_b}
Consider any given FHS scheme $F$ with the corresponding IFGs denoted by $\mathcal{G}_F(n,k,d,\alpha,\beta)$. We have that
\begin{align}
\min_{G\in\mathcal{G}_F} \min_{t\in\DC(G)}&\mincut_G(s,t) = \nonumber\\
&\min_{\forall \pi_f} \sum_{i=1}^{k}\min \left(\left(d-y_i(\pi_f)\right)\beta,\alpha\right),\label{eq:low_b}
\end{align}
where $\pi_f$ can be any family index permutation and $y_i(\pi_f)$ is computed as follows. If the $i$-th coordinate of $\pi_f$ is $0$, then $y_i(\pi_f)$ returns the number of $j$ satisfying both (i) $j<i$ and (ii) the $j$-th coordinate $>0$. If the $i$-th coordinate of $\pi_f$ is not $0$, then $y_i(\pi_f)$ returns the number of $j$ satisfying both (i) $j<i$ and (ii) the absolute value of the $j$-th coordinate of $\pi_f$ and the absolute value of the $i$-th coordinate of $\pi_f$ are different.  For example, if $\pi_f=(1,2,-2,1,0,0,1,2)$, then $y_6(\pi_f)= 3$ and $y_8(\pi_f)= 5$.
\end{proposition}

\par The proof of Proposition~\ref{prop:low_b} is relegated to Appendix~\ref{app:low_b_proof}.

\begin{remark} In general, the minimum cut of an IFG may exist in the interior of the graph. When computing the min-cut value in the LHS of \eqref{eq:low_b_gen}, we generally need to exhaustively consider all possible cuts for any $G\in {\mathcal{G}}_A$, which is why we have to choose $\mathbf{r}\in R$ in \eqref{eq:low_b_gen} that allows for repeated values in the coordinates of $\mathbf{r}$ and we can only prove the inequality (lower bound) in \eqref{eq:low_b_gen}.
\end{remark}

\par Recall that the family index permutation $\pi_f$ is based on the family index vector of all ``currently active nodes.'' Proposition~\ref{prop:low_b} thus implies that when focusing on the FHS scheme $F$, we can reduce the search scope and consider only those cuts that directly separate $k$ currently active nodes from the rest of the IFG (see \eqref{eq:low_b}). This allows us to compute the corresponding min-cut value with equality.

\par Combining Proposition~\ref{prop:low_b} and \eqref{eq:condition}, we can derive the new storage-bandwidth tradeoff ($\alpha$ vs.\ $\beta$) for the FHS scheme. For example, Fig.~\ref{fig:storage_vs_bandwidth_(20-10-10)} plots $\alpha$ versus $\gamma\stackrel{\Delta}{=}d\beta$ for the $(n,k,d)$ values $(20,10,10)$ with file size $\mathcal{M}=1$. As can be seen in Fig.~\ref{fig:storage_vs_bandwidth_(20-10-10)}, the MBR point (the smallest $\gamma$ value) of the FHS scheme uses only $73.33\%$ (a ratio of $\frac{11}{15}$) of the repair-bandwidth of the MBR point of the BHS scheme ($\gamma_{\MBR}=\frac{2}{15}$ vs.\ $\frac{2}{11}$). It turns out that for any $(n,k,d)$ values, the biggest improvement of FHS over BHS always happens at the MBR point.\footnote{If we compare the min-cut value of FHS in \eqref{eq:low_b} with the min-cut value of BHS in \eqref{eq:ex_low_b}, we can see that the greatest improvement happens when the new term $(d-y_i(\pi_f))\beta\leq \alpha$ for all $i$. These are the mathematical reasons why the MBR point sees the largest improvement.} The intuition is that choosing the good helpers is most beneficial when the per-node storage $\alpha$ is no longer a bottleneck (thus the MBR point).

\par The RHS of \eqref{eq:low_b} involves taking the minimum over a set of $\mathcal{O}\left(\left(\frac{n}{n-d}\right)^k\right)$ entries. As a result, computing the entire storage-bandwidth tradeoff is of complexity $\mathcal{O}\left(\left(\frac{n}{n-d}\right)^k\right)$. The following proposition shows that if we are interested in the most beneficial point, the MBR point, then we can compute the corresponding $\alpha$ and $\beta$ values in polynomial time.

\begin{proposition} \label{prop:mbr}
For the MBR point of \eqref{eq:low_b}, i.e., when $\alpha$ is sufficiently large, the minimizing family index permutation is the RFIP $\pi_f^*$ defined in Section~\ref{subsec:desc_fr}. That is, the $\alpha$, $\beta$, and
$\gamma$ values of the MBR point can be computed by
\begin{align} \label{eq:gamma}
\alpha_{\MBR}=\gamma_{\MBR}=d\beta_{\MBR}=\frac{d\mathcal{M}}{\sum_{i=1}^{k} (d-y_i(\pi_f^*)) }.
\end{align}
\end{proposition}

\par The proof of Proposition~\ref{prop:mbr} is relegated to Appendix~\ref{app:mbr_proof}.

\par Using Proposition~\ref{prop:mbr} above, we can find the MBR point of the FHS tradeoff curve in Fig.~\ref{fig:storage_vs_bandwidth_(20-10-10)}. This is done by first finding the RFIP $\pi_f^*=(1,2,1,2,\dots,1,2)$, and then finding ${\sum_{i=1}^{k} (d-y_i(\pi_f^*))}=75$. Recall that $\mathcal{M}$ is assumed to be 1 in Fig.~\ref{fig:storage_vs_bandwidth_(20-10-10)}. Using \eqref{eq:gamma}, we thus get that $\gamma_{\MBR}=\frac{2}{15}$.

\par Unfortunately, we do not have a general formula for the least beneficial point, the MSR point, of the FHS scheme. Our best knowledge for computing the MSR point is the following

\begin{proposition}\label{prop:msr}
For arbitrary $(n,k,d)$ values, the minimum-storage of \eqref{eq:low_b} is $\alpha_{\MSR}= \frac{\mathcal{M}}{\min(d,k)}$. If the $(n,k,d)$ values also satisfy $d\geq k$, then the corresponding $\beta_{\MSR}=\frac{\mathcal{M}}{k(d-k+1)}$. If $d<k$, then the corresponding $\beta_{\text{MSR}}$ can be upper bounded by $\beta_{\MSR}\leq \frac{\mathcal{M}}{d}$.
\end{proposition}
\par The proof of Proposition~\ref{prop:msr} is relegated to Appendix~\ref{app:msr_proof}.
\par By Proposition~\ref{prop:msr}, we can quickly compute $\alpha_{\MSR}$ and $\beta_{\MSR}$ when $d\geq k$. If $d<k$, then we still have $\alpha_{\MSR}=\frac{\mathcal{M}}{\min(d,k)}$ but we do not know how to compute the exact value of $\beta_{\MSR}$ other than directly applying the formula in Proposition~\ref{prop:low_b}.

\begin{remark}
If we compare the expressions of Proposition~\ref{prop:msr} and the MSR point of the BHS scheme\footnote{Recall from \cite{dimakis2010network} that for BHS we have $\alpha_{\MSR}=\frac{\mathcal{M}}{\min(d,k)}$ and $\gamma_{\MSR}=\frac{d\mathcal{M}}{\min(d,k)(d-\min(d,k)+1)}$.}, Proposition~\ref{prop:msr} implies that the FHS scheme does not do better than the BHS scheme at the MSR point when $d\geq k$. However, it is still possible that the FHS scheme can do better than the BHS scheme at the MSR point when $d<k$. One such example is the example we considered in Section~\ref{sec:preview} when $(n,k,d)=(5,3,2)$. For this example, we have $\alpha_{\MSR}=\frac{\mathcal{M}}{2}$, $\beta_{\MSR}=\frac{\mathcal{M}}{4}$, and  $\gamma_{\MSR}=\frac{\mathcal{M}}{2}$ for the FHS scheme where $\beta_{\MSR}=\frac{\mathcal{M}}{4}$ is derived by searching over all family index permutations $\pi_f$ in \eqref{eq:low_b}. For comparison, the BHS scheme has $\alpha_{\MSR}=\frac{\mathcal{M}}{2}$, $\beta_{\MSR}=\frac{\mathcal{M}}{2}$, and  $\gamma_{\MSR}=\mathcal{M}$. This shows that the FHS scheme can indeed do better at the MSR point when $d<k$ in terms of the repair-bandwidth although we do not have a closed-form expression for this case.
\end{remark}

\subsection{The Family-plus Helper Selection Scheme} \label{sec:family-plus}
In the FHS scheme, there are $\left\lfloor\frac{n}{n-d}\right\rfloor$ complete families and $1$ incomplete family 
(if $n\bmod (n-d)\neq 0$). For the scenario in which the $n$ and $d$ values are comparable, we have many complete families and the FHS solution harvests almost all of the benefits of choosing good helpers, also see Proposition~\ref{prop:optimal_main} for the case of $n=d+2$. However, when $n$ is large but $d$ is small, we have only one complete family and one incomplete family. Therefore, even though the FHS scheme can still outperform the BHS scheme, the performance of the FHS scheme is far from optimal due to having only $1$ complete family. In this section, we propose the {\em family-plus helper selection} scheme that further improves the storage-bandwidth tradeoff when $n$ is large but $d$ is small.

The main idea is as follows. We first partition the $n$ nodes into several disjoint groups of $2d$ nodes and one disjoint group of $n_{\text{remain}}$ nodes. The first type of groups is termed the regular group while the second group is termed the remaining group. If we have to have one remaining group (when $n\bmod (2d)\neq 0$), then we enforce the size of the remaining group to be as small as possible but still satisfying $n_\text{remain}\geq 2d+1$. For example, if $d=2$ and $n=8$, then we will have 2 regular groups and no remaining group since $n\bmod (2d)=0$. If $d=2$ and $n=9$, then we choose $1$ regular group $\{1,2,3,4\}$ and $1$ remaining group $\{5,6,7,8,9\}$ since we need to enforce $n_\text{remain}\geq 2d+1$.

After the partitioning, we apply the FHS scheme to the individual groups. For example, if $d=2$ and $n=8$, then we have two regular groups $\{1,2,3,4\}$ and $\{5,6,7,8\}$. Applying the FHS scheme to the first group means that nodes $1$ and $2$ form a family and nodes $3$ and $4$ form another family. Whenever node $1$ fails, it will access helpers from outside its family, which means that it will access nodes $3$ and $4$. Node $1$ will never request help from any of nodes $5$ to $8$ as these nodes are not in the same group as node $1$. Similarly, we apply the FHS scheme to the second group $\{5,6,7,8\}$. All the FHS operations are always performed within the same group.

\par Another example is when $d=2$ and $n=9$. In this case, we have 1 regular group $\{1,2,3,4\}$ and 1 remaining group $\{5,6,7,8,9\}$. In the remaining group, $\{5,6,7\}$ will form a complete family and $\{8,9\}$ will form an incomplete family. If node 6 fails, it will request help from both nodes 8 and 9. If node 9 fails, it will request help from nodes $\{5,6\}$, the first $d=2$ nodes of this group. Again, all the repair operations for nodes 5 to 9 are completely separated from the operations of nodes 1 to 4. The above scheme is termed the \emph{family-plus helper selection scheme}.

One can easily see that when $n\leq 2d$, there is only one group and the family-plus helper selection scheme collapses to the FHS scheme. When $n>2d$, there are approximately $\frac{n}{2d}$ regular groups, each of which contains two complete families. Therefore, the construction of the family-plus helper selection scheme ensures that there are many complete families even for the scenario of $n\gg d$.

\subsection{Analysis of the Family-plus Scheme}
In the following proposition, we characterize the performance of the family-plus helper selection scheme.

\begin{proposition} \label{prop:low_b_plus}
Consider any given $(n,k,d)$ values and the family-plus helper selection scheme $F^+$. Suppose we have $B$ groups in total (including both regular and remaining groups) and each group has $n_b$ number of nodes for $b=1$ to $B$. Specifically, if the $b$-th group is a regular group, then $n_b=2d$. If the $b$-th group is a remaining group (when $n\bmod (2d)\neq 0$), then $n_b=n-2d(B-1)$. We use ${\mathcal{G}}_{F^+}(n,k,d,\alpha,\beta)$ to denote the collection of IFGs generated by the family-plus helper selection scheme. We have that
\begin{align}\label{eq:low_b_plus}
\min_{G\in\gfp} &\min_{t\in\DC(G)} \mincut(s,t) = \nonumber \\
& \min_{\mathbf{k}\in K} \sum_{b=1}^{B} \min_{H_b\in \gf(n_b, k_b,d,\alpha,\beta)} \min_{t_b\in \DC(H_b)} \mincut_{H_b}(s, t_b),
\end{align}
where $\mathbf{k}$ is a $B$-dimensional integer-valued vector, $K=\{(k_1,k_2,\cdots, k_B): \forall b\in\{1,\cdots, B\}, 0\leq k_b\leq n_b, \sum_{b=1}^{B}k_b=k\}$. Note that for any given $\bf{k}$, the RHS of \eqref{eq:low_b_plus} can be evaluated by Proposition~\ref{prop:low_b}.
\end{proposition}

\begin{IEEEproof} Observe that any IFG $G\in\gfp$ is a union of $B$ parallel IFGs that are in $\gf(n_b,\cdot,d,r,\alpha,\beta)$ where ``$\cdot$'' means that we temporarily ignore the placement of the data collectors. For any data collector $t$ in $G_{F^+}$, we use $k_b$ to denote the number of active nodes that $t$ accesses in group $b$. Therefore, the $\mincut_G(s,t)$ is simply the summation of the $\mincut_{H_b}(s,t_b)$ for all $b\in \{1,\cdots, B\}$ where $t_b$ corresponds to the ``sub-data-collector'' of group $b$ and $H_b$ is the $b$-th parallel IFG. Since we run the original FHS scheme in each of the $b$-th group, $H_b$ is a member of $\mathcal{G}_F(n_b,k_b,d,\alpha,\beta)$. By further minimizing over all possible data collectors $t$ (thus minimizing over $\{k_b\}$), we get \eqref{eq:low_b_plus}.
\end{IEEEproof}

To evaluate the RHS of \eqref{eq:low_b_plus}, we have to try all possible $\mathbf{k}$ vectors and for each $\mathbf{k}$, we need to evaluate each of the $B$ summands by Proposition~\ref{prop:low_b}, which requires checking all $n_b!$ different family index permutations. Fortunately, for the MBR point of the family-plus helper selection scheme, we can further simplify the computation complexity following similar arguments as used in Proposition~\ref{prop:mbr}.
\begin{corollary}\label{cor:mbr_plus}
The MBR point of the family-plus helper selection scheme is
\begin{align}
\alpha_{\MBR}=\gamma_{\MBR}=d\beta_{\MBR}\nonumber
\end{align}
and $\beta_{\MBR}$ can be computed by solving the following equation
\begin{align}\label{eq:gamma_plus}
\Bigg(&1_{\{n\bmod(2d)\neq 0\}}\cdot \sum_{i=0}^{\min(k,2d-1)-1}\left(d-i+\left\lfloor\frac{i}{2}\right\rfloor\right)+\nonumber\\
&d^2\left\lfloor \frac{(k-n_l)^+}{2d}\right\rfloor+ \sum_{i=0}^{q}\left(d-i+\left\lfloor\frac{i}{2}\right\rfloor\right)\Bigg) \beta_{\MBR}=\mathcal{M},
\end{align}
where $\mathcal{M}$ is the file size,
\begin{align}
&q=((k-n_l)^+\bmod(2d))-1, \text{ and}\nonumber\\
&n_l=
\begin{cases}
n_{\text{remain}},& \text{ if } n\bmod(2d)\neq 0\\
0,& \text{ otherwise}.
\end{cases}\nonumber
\end{align}
\end{corollary}
\par The proof of Corollary~\ref{cor:mbr_plus} is relegated to Appendix~\ref{app:mbr_plus_proof}.

\begin{figure}[h!]
\centering
\includegraphics[width=0.475\textwidth]{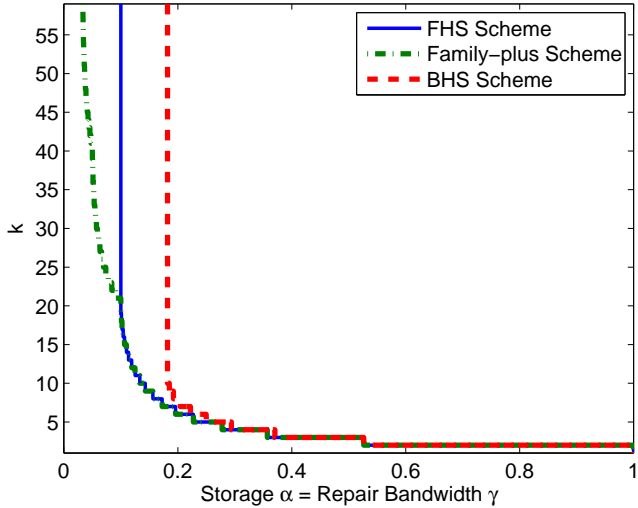}
\caption{The $k$ value versus repair-bandwidth $\gamma$ curve comparison between FHS, family-plus, and BHS at the MBR point for $(n,d)=(60,10)$ and file size $\mathcal{M}=1$.}
\label{fig:k_vs_gamma_(60-10)_plus}
\end{figure}

\par In Fig.~\ref{fig:k_vs_gamma_(60-10)_plus}, we plot the $k$ vs. $\gamma$ curves for the BHS, the FHS, and the family-plus helper selection schemes for the case of $(n,d)=(60,10)$ using Proposition~\ref{prop:mbr}, and Corollary~\ref{cor:mbr_plus}, respectively. As can be seen in Fig.~\ref{fig:k_vs_gamma_(60-10)_plus}, when $k>d$, the BHS scheme stops improving any further since RCs with BHS always have $k^*\leq d$ and thus overprotect the data when the protection-level requirement $k>d$. Therefore, BHS is not able to take advantage of the looser protection-level requirement when $k>d$. In contrast, the bandwidth consumption of FHS continues to decrease until the improvement stops when $k>2d$. The reason is that, for $(n,d)=(60,10)$, FHS only has two families. The family-plus scheme, on the other hand, divides $n=60$ nodes into 3 groups and each group has 2 complete families (6 families in total). As a result, the family-plus scheme can continue harvesting the looser and looser protection-level requirement even when $k>2d$ and the bandwidth consumption keeps decreasing continuously. 

\par For example, when $k=40$, the repair-bandwidth of the family-plus helper selection scheme is only $28\%$ of the repair-bandwidth of the BHS scheme (cf.\ the repair-bandwidth of the FHS scheme is $58\%$ of the repair-bandwidth of the BHS scheme). This demonstrates the benefits of the family-plus helper selection scheme, which creates as many complete families as possible by further partitioning the nodes into several disjoint groups.

\subsection{The Achievability Result and the Corresponding Proof}\label{sec:achievability_result}

We are now ready to use the FHS scheme and the family-plus helper selection scheme to prove the achievability result of Proposition~\ref{prop:comparison}. 
\begin{proposition}\label{prop:weak} Consider a family-plus helper selection scheme denoted by $F^+$ and its corresponding collection of IFGs ${\mathcal{G}}_{F^+}(n,k,d,\alpha,\beta)$. For any $(n,k,d)$ values satisfying neither of the (i) and (ii) conditions in Proposition~\ref{prop:comparison}, there exists a pair $(\alpha,\beta)$ such that
\begin{align}\label{eq:weak}
\min_{G\in\mathcal{G}_{F^+}}\min_{t\in \DC (G) } \mincut_G(s,t)>\sum_{i=0}^{k-1}\min((d-i)^+\beta,\alpha).
\end{align}
\end{proposition}
Since the family-plus scheme is strictly better than the BHS scheme when $(n,k,d)$ satisfies neither of the (i) and (ii) conditions in Proposition~\ref{prop:comparison}, the achievability result of Proposition~\ref{prop:comparison} is thus proved. Also, Proposition~\ref{prop:weak} immediately implies that the collection of family-plus helper selection schemes is weakly optimal, also see Definition~\ref{def:collection-scheme-weakly-optimal}.

\begin{IEEEproof}
The first step in our proof is to show that whenever $\alpha=d\beta$, we have
\begin{align} \label{eq:fpr_vs_fr}
\min_{G\in\mathcal{G}_{F^+}}\min_{t\in \DC (G)} \mincut_G(s,t)&\geq\nonumber\\
\min_{G\in\mathcal{G}_{F}}&\min_{t\in \DC (G) } \mincut_G(s,t),
\end{align}
where $\mathcal{G}_F$ is the collection of IFGs of an FHS scheme $F$. That is, when $\alpha=d\beta$, the additional step of partitioning nodes into sub-groups in the family-plus scheme will monotonically improve the performance when compared to the original FHS scheme without partitioning. Therefore, the family-plus scheme is no worse than the FHS scheme when $\alpha=d\beta$. The proof of \eqref{eq:fpr_vs_fr} is relegated to Appendix~\ref{app:weak_proof}.

\par Equation~\eqref{eq:fpr_vs_fr} can now be used to prove \eqref{eq:weak}. If neither (i) nor (ii) of Proposition~\ref{prop:comparison} is true, one can verify by exhaustively considering all scenarios that one of the following three cases must hold: (a) $d\geq 2$ and $k> \left\lceil \frac{n}{n-d}\right\rceil$; (b) $d=1$, $k> 2$, and even $n$; and (c) $d=1$, $k>3$, and odd $n$. 

\par For case (a), we first note that since $k> \left\lceil \frac{n}{n-d}\right\rceil$, we must also have $d\leq n-2$. Otherwise we will have $k>n$, which contradicts \eqref{eq:ccw1}. We then observe\footnote{A detailed proof of this simple algebraic observation can be found in the proof of Corollary~\ref{cor:low_b} around \eqref{eq:corr1_1} in Appendix~\ref{app:optimal_proof}.} that whenever $2\leq d\leq n-2$ we must have $d> \left\lceil\frac{n}{n-d}\right\rceil-1$. As a result, in case (a) we have that $\min(d+1,k)> \left\lceil\frac{n}{n-d}\right\rceil$. We now apply the FHS scheme to case (a), not the family-plus scheme. Since there are exactly $\left\lceil\frac{n}{n-d}\right\rceil$ families in FHS, among the first $\min(d+1,k)$ indices of a family index permutation $\pi_f$ there is at least one family index that is repeated. Jointly, this observation, Proposition~\ref{prop:low_b}, and the MBR point formula in \eqref{eq:gamma} imply that for the MBR point that has $\alpha=d\beta$, the min-cut value of the FHS scheme is strictly larger than the min-cut value of the BHS scheme. Since \eqref{eq:fpr_vs_fr} shows that the family-plus scheme is no worse than the FHS scheme in the MBR point, we have proved Proposition~\ref{prop:weak} for case (a). 

\par  For both cases (b) and (c), since $n>k$ by \eqref{eq:ccw1}, we have $n\geq 4$. Since $d=1$ in both cases (b) and (c), the construction of the family-plus scheme thus will generate at least 2 groups. That is, the value of $B$ in Proposition~\ref{prop:low_b_plus} must satisfy $B\geq 2$. Moreover, in case (b), we have no remaining group since $n$ is even. Therefore, since $k>2$, for any $\mathbf{k}\in K$ defined in Proposition~\ref{prop:low_b_plus}, there are at least two distinct $b$ values with $k_b\geq 1$. In case (c), we have $k>3=n_{\text{remain}}$ (note that $n_{\text{remain}}=3$ since we have that $2d+1\leq n_{\text{remain}}\leq 4d-1$ by construction). Therefore, similarly, for any $\mathbf{k}\in K$ defined in Proposition~\ref{prop:low_b_plus}, there are at least two distinct $b$ values with $k_b\geq 1$.

\par Using the above observation (at least two distinct $b$ values having $k_b\geq 1$) and \eqref{eq:low_b_plus} in Proposition~\ref{prop:low_b_plus}, we have that in both cases (b) and (c)
\begin{align}
\min_{G\in\gfp} \min_{t\in\DC(G)} \mincut_G(s,t)\geq 2\min(d\beta,\alpha)> \min(\beta,\alpha),
\end{align}
where the first inequality follows from: (i) considering only those $b$ values with $k_b\geq 1$; (ii) plugging in the min-cut formula in Proposition~\ref{prop:low_b}; and (iii) only counting the first term ``$i=1$'' when summing up for all $i=1$ to $k_b$. The second inequality follows from the assumption that $d=1$ in both cases (b) and (c) and the fact that both $\beta$ and $\alpha$ must be strictly positive. By noticing that for cases (b) and (c) the RHS of \eqref{eq:weak} is indeed $\min(\beta,\alpha)$, the proof is complete for cases (b) and (c) as well.b
\end{IEEEproof}


\subsection{The Optimality of the FHS and the Family-plus Schemes} \label{sec:optimality_result}
In the following, we prove that the FHS scheme is indeed optimal for some $(n,k,d)$ values.

\begin{proposition} \label{prop:optimal} For the $(n,k,d)$ values satisfying simultaneously the following three conditions (i) $d$ is even, (ii) $n=d+2$, and (iii) $k=\frac{n}{2}+1$; we have
\begin{align} \label{eq:tight}
\min_{G\in \mathcal{G}_F}\min_{t\in \DC(G)}\mincut_G(s,t) \geq \min_{G\in \mathcal{G}_A}\min_{t\in \DC(G)} \mincut_G(s,t)
\end{align}
for any arbitrary DHS scheme $A$ and any arbitrary $(\alpha,\beta)$ values.
\end{proposition}

\par The proof of Proposition~\ref{prop:optimal} is relegated to Appendix~\ref{app:optimal_proof}.
\par Proposition~\ref{prop:optimal} is the formal version of Proposition~\ref{prop:optimal_main} in Section~\ref{sec:results}. Note that for any $(n,k,d)$ values satisfying conditions (i) to (iii) in Proposition~\ref{prop:optimal}, they must also satisfy neither (i) nor (ii) in Proposition~\ref{prop:comparison}. As a result, by Proposition~\ref{prop:comparison}, there exists some helper selection scheme that strictly outperforms the BHS scheme. Proposition~\ref{prop:optimal} further establishes that among all those schemes strictly better than the BHS scheme, the FHS scheme is indeed optimal. To illustrate that, consider the example of $(n,k,d)=(6,4,4)$. Using Proposition~\ref{prop:comparison}, we know that for this combination of parameters there exists a scheme that can do better than the BHS scheme. Now, it is not hard to check that this combination of parameters also satisfies all the conditions (i), (ii), and (iii) of Proposition~\ref{prop:optimal}. Thus, we know, and as was stated in Result~2 of Section~\ref{sec:preview}, that the FHS scheme is absolutely optimal for $(n,k,d)=(6,4,4)$.

\par We also note that \cite[Theorem~5.4]{hollmann2014minimum} proves that when $k=n-1$ and $\alpha=\beta$, no DHS scheme can protect a file of size $>\frac{nd\alpha}{d+1}$. It was not clear whether such a bound is tight or not. Proposition~\ref{prop:optimal} can be used to prove that the bound in \cite[Theorem~5.4]{hollmann2014minimum} is actually loose for some $(n,k,d)$ combinations. 
\begin{corollary}\label{cor:existing_loose}
When $(n,k,d)=(4,3,2)$ and $\alpha=\beta$, no DHS scheme can protect a file of size $\mathcal{M}>2\alpha$, for which \cite[Theorem~5.4]{hollmann2014minimum} only proves that no scheme can protect a file of size $\mathcal{M}>\frac{8\alpha}{3}$.
\end{corollary}

\begin{IEEEproof}
By Proposition~\ref{prop:low_b}, when $(n,k,d)=(4,3,2)$ and $\alpha=\beta$, the FHS scheme can protect a file of size $2\alpha$. We then notice that $(n,k,d)=(4,3,2)$ satisfies Proposition~\ref{prop:optimal} and, therefore, the FHS scheme is absolutely optimal. As a result, no scheme can protect a file of size $\mathcal{M}>2\alpha$.
\end{IEEEproof}

\par Proposition~\ref{prop:optimal} shows that for certain $(n,k,d)$ value combinations, the FHS scheme is optimal for the entire storage-bandwidth tradeoff curve. If we only focus on the MBR point, we can also have the following optimality results.

\begin{proposition} \label{prop:optimal_2} Consider $k=n-1$ and $\alpha=d\beta$. For the $(n,k,d)$ values satisfying $n\bmod(n-d)=0$, we have
\begin{align} \label{eq:tight_2}
\min_{G\in \mathcal{G}_F}\min_{t\in \DC(G)}\mincut_G(s,t)&=\frac{n\alpha}{2} \nonumber\\
&\geq \min_{G\in \mathcal{G}_A}\min_{t\in \DC(G)} \mincut_G(s,t)
\end{align}
for any arbitrary DHS scheme $A$.
\end{proposition}

\par The proof of Proposition~\ref{prop:optimal_2} is relegated to Appendix~\ref{app:optimal_2}.

\par Proposition~\ref{prop:optimal_2} establishes again that the FHS scheme is optimal in the MBR point ($\alpha=d\beta$), among all DHS schemes, whenever $k=n-1$ and $n\bmod(n-d)=0$. Since Proposition~\ref{prop:optimal_2} is based on FHS, we can generalize Proposition~\ref{prop:optimal_2} by considering the family-plus scheme. We then have 

\begin{proposition}\label{prop:family-plus_optimal}
Consider $k=n-1$ and $\alpha=d\beta$ and a family-plus helper selection scheme that divides $n$ nodes into $B$ groups with $n_1$ to $n_B$ nodes. If $n_b\bmod(n_b-d)=0$ for all $b=1$ to $B$, then we have
\begin{align} \label{eq:tight_2_plus}
\min_{G\in \gfp}\min_{t\in \DC(G)}\mincut_G(s,t)&=\frac{n\alpha}{2} \nonumber\\
&\geq \min_{G\in \mathcal{G}_A}\min_{t\in \DC(G)} \mincut_G(s,t)
\end{align}
for any arbitrary DHS scheme $A$.
\end{proposition}
This result is the formal version of Proposition~\ref{prop:optimal_2_main} in Section~\ref{sec:results}. The proof of Proposition~\ref{prop:family-plus_optimal} is relegated to Appendix~\ref{app:family-plus_optimal_proof}.

\begin{remark}
Thus far, our family-plus scheme assumes all but one group have $n_b=2d$ nodes and the remaining group has $n_b=n_{\text{remain}}\geq 2d+1$ nodes. One possibility for further generalization is to allow arbitrary $n_b$ choices. It turns out that Proposition~\ref{prop:family-plus_optimal} holds even for any arbitrary choices of $n_b$ values. For example, for the case of $(n,k,d)=(19,18,4)$ and $\alpha=d\beta$, the generalized family-plus scheme is absolutely optimal if we divide the 19 nodes into 3 groups of $(n_1,n_2,n_3)=(8,6,5)$. 
\par By allowing arbitrary ways of partitioning $n=\sum_b n_b$, the MBR optimality of the family-plus schemes can be proved for a wider range of $(n,k,d)$ values. For example, one can prove that for any $(n,k,d,\alpha,\beta)$ values satisfying $n\neq 5$, $k=n-1$, $d=2$, and $\alpha=d\beta$, we can always find some $(n_1,\cdots,n_B)$ such that the generalized family-plus helper selection scheme is absolutely optimal. See Result~6 in Section~\ref{sec:preview} for some other $(n,k,d)$ value combinations for which the generalized family-plus scheme is optimal.
\end{remark}

\begin{remark}
Compared to the existing results, \cite{dimakis2010network} showed that when $k=d=n-1$, the optimal MBR point satisfies $\frac{n\alpha}{2}=\mathcal{M}$ with repair-bandwidth $\gamma=d\beta=\frac{2\mathcal{M}}{n}$ and an exact-repair scheme achieving this MBR point is provided in \cite{shah2012distributed}. Our results show that for any $(n,k,d)$ satisfying $k=n-1$ but $d\neq n-1$, as long as we also have  $n_b\bmod(n_b-d)=0$ for all $b=1$ to $B$, the optimal MBR point of the family-plus scheme is absolutely optimal and again satisfies $\frac{n\alpha}{2}=\mathcal{M}$ with a repair-bandwidth also of $\frac{2\mathcal{M}}{n}$. An exact-repair scheme that achieves this MBR point for any $k=n-1$ is provided in Part~II \cite{part2}.
\end{remark}

\par Before closing this section, we should mention that a similar scheme to the family-plus helper selection scheme was devised in \cite{papailiopoulos2012locally} for LRCs when $n$ is a multiple of $(d+1)$. In that scheme the nodes are divided into groups of $(d+1)$ nodes. Whenever a node fails, its set of helpers is the set of $d$ remaining nodes in the same group. This can be viewed as a special example of the generalized family-plus helper selection scheme by choosing $n_b=d+1$ for all $b=1$ to $B$. Each group thus has $\frac{n_b}{n_b-d}=n_b=d+1$ complete families and each family contains only $n_b-d=1$ node. Therefore, all our analysis can be applied to the construction in \cite{papailiopoulos2012locally} and used to rederive the MBR characterization results of that scheme. 

\par In summary, our construction and the corresponding tradeoff curve analysis hold for arbitrary ways\footnote{Our construction and analysis, Proposition~\ref{prop:low_b_plus}, work for arbitrary $n_b$ partitions. On the other hand, the optimality guarantee in Proposition~\ref{prop:family-plus_optimal} only holds when $n_b\bmod(n_b-d)=0$ for all $b$.} of partitioning $n$ nodes into separate groups of $n_b$ nodes, $b=1$ to $B$. This thus significantly broadens the scope of application.

\section{Conclusion}\label{sec:conc}
In practice, it is natural that the newcomer should access only those ``good'' helpers. This paper has provided a necessary and sufficient condition under which optimally choosing good helpers improves the storage-bandwidth tradeoff. We have also analyzed a new class of low-complexity solutions termed the \emph{family helper selection scheme}, including its storage-bandwidth tradeoff, the expression of its MBR point, and its (weak) optimality. In Part~II \cite{part2}, we will construct an explicit exact-repair code, the \emph{generalized fractional repetition code}, that can achieve the MBR point of this scheme.

\par The main goal of this work is to characterize, for the first time in the literature, when can DHS improve RCs. We thus considered the scenario of single failures only in a similar way as in the original RC paper \cite{dimakis2010network}. Since a practical system can easily have multiple failures, as ongoing work, we are studying the helper selection problem under the multiple failures scenario. See \cite{arxiv_multiple} for our current results in this direction. 


\appendices

\section{The Information Flow Graph} \label{app:ifg}
\par We provide in this appendix the description of the information flow graph (IFG) that was first introduced in \cite{dimakis2010network}.

\begin{figure}[h!]
\centering
\includegraphics[width=0.45\textwidth]{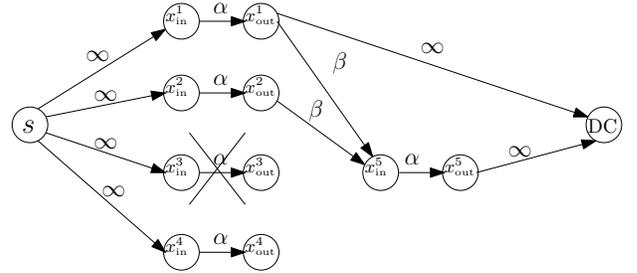} 
\caption{An example of the information flow graph with $(n,k,d)=(4,2,2)$.}
\label{fig:ifg}
\end{figure}

\par As shown in Fig~\ref{fig:ifg}, an IFG has three different kinds of nodes. It has a single \emph{source} node $s$ that represents the source of the data object. It also has nodes $x_{\inp}^i$ and $x_{\out}^i$ that represent storage node $i$ of the IFG. A storage node is split into two nodes so that the IFG can represent the storage capacity of the nodes. We often refer to the pair of nodes $x_{\inp}^i$ and $x_{\out}^i$ simply by storage node $i$. In addition to those nodes, the IFG has \emph{data collector} (DC) nodes. Each data collector node is connected to a set of $k$ active storage nodes, which represents the party that is interested in extracting the original data object initially produced by the source $s$. Fig.~\ref{fig:ifg} illustrates one such data collector, denoted by $t$, which connects to $k=2$ storage nodes. A more detailed description of the IFG is provided as follows.

\par The IFG evolves with time. In the first stage of an information flow graph, the source node $s$ communicates the data object to all the initial nodes of the storage network. We represent this communication by edges of infinite capacity as this stage of the IFG is virtual. See Fig.~\ref{fig:ifg} for illustration. This stage models the encoding of the data object over the storage network. To represent storage capacity, an edge of capacity $\alpha$ connects the input node of storage nodes to the corresponding output node. When a node fails in the storage network, we represent that by a new stage in the IFG where, as shown in Fig.~\ref{fig:ifg}, the newcomer connects to its helpers by edges of capacity $\beta$ resembling the amount of data communicated from each helper. We note that although the failed node still exists in the IFG, it cannot participate in helping future newcomers. Accordingly, we refer to failed nodes by \emph{inactive} nodes and existing nodes by \emph{active} nodes. By the nature of the repair problem, the IFG is always acyclic.

\par Given an IFG $G$, we use $\DC(G)$ to denote the collection of all ${n\choose k}$ {\em data collector nodes} in $G$ \cite{dimakis2010network}. Each data collector $t\in\DC(G)$ represents one unique way of choosing $k$ out of $n$ active nodes when reconstructing the file.

\section{Proof of Proposition~\ref{prop:low_b_gen}} \label{app:low_b_gen}
\par The proof of Proposition~\ref{prop:low_b_gen} below follows the proof of \cite[Lemma 2]{dimakis2010network}.

\par Consider any IFG $G\in \mathcal{G}_A$ where $A$ is a SHS scheme. Consider any data collector $t$ of $G$ and call the set of $k$ active output nodes it connects to $V$. Since all the incoming edges of $t$ have infinite capacity, we can assume without loss of generality that the minimum cut $(U,\overline{U})$ satisfies $s\in U$ and $V\subseteq \overline{U}$.

Let $\mathcal{C}$ denote the set of edges in the minimum cut. Let $x^ i_{\out}$ be the chronologically $i$-th output node in $\overline{U}$, i.e., from the oldest to the youngest. Since $V\subseteq\overline{U}$, there are at least $k$ output nodes in $\overline{U}$. We now consider the oldest $k$ output nodes of $\overline{U}$, i.e., $x^1_{\out}$ to $x^k_{\out}$. For $i=1$ to $k$, let $r_i$ denote the node index of $x^i_{\out}$. Obviously, the vector $\mathbf{r}\stackrel{\Delta}{=}(r_1,\cdots, r_k)$ belongs to $R$.

\par Consider $x^1_{\out}$, we have two cases:
\begin{itemize}
\item If $x^1_{\inp}\in U$, then the edge $(x^1_{\inp},x^1_{\out})$ is in $\mathcal{C}$.
\item If $x^1_{\inp}\in \overline{U}$, since $x^1_{\inp}$ has an in-degree of $d$ and $x^1_\text{out}$ is the oldest node in $\overline{U}$, all the incoming edges of $x^1_{\inp}$ must be in $\mathcal{C}$.
\end{itemize}
From the above discussion, these edges related to $x^1_{\out}$ contribute at least a value of $\min((d-z_1(\mathbf{r}))\beta,\alpha)$ to the min-cut value since by definition $z_1(\mathbf{r})=0$.
Now, consider $x^2_{\out}$, we have three cases:
\begin{itemize}
\item If $x^2_{\inp}\in U$, then the edge $(x^2_{\inp},x^2_{\out})$ is in $\mathcal{C}$.
\item If $x^2_{\inp}\in \overline{U}$ and  $r_1\in D_{r_2}$, since one of the incoming edges of $x^2_{\inp}$ can be from $x^1_{\out}$, then at least $(d-1)$ incoming edges of $x^2_{\inp}$ are in $\mathcal{C}$.
\item If $x^2_{\inp}\in \overline{U}$ and  $r_1\notin D_{r_2}$, since no incoming edges of $x^2_{\inp}$ are from $x^1_{\out}$, then all $d$ incoming edges of  $x^2_{\inp}$ are in $\mathcal{C}$.
\end{itemize}
Therefore, these edges related to $x^2_{\out}$ contribute a value of at least $\min ((d-z_2(\mathbf{r}))\beta,\alpha)$ to the min-cut value, where the definition of $z_2(\mathbf{r})$ takes care of the second and the third cases. Consider $x^3_{\out}$, we have five cases:
\begin{itemize}
\item If $x^3_{\inp}\in U$, then the edge $(x^3_{\inp},x^3_{\out})$ is in $\mathcal{C}$.
\item If $x^3_{\inp}\in \overline{U}$ and  $r_1=r_2\in D_{r_3}$, since one of the incoming edges of $x^3_{\inp}$ can be from $x^2_{\out}$, then at least $(d-1)$ incoming edges of $x^3_{\inp}$ are in $\mathcal{C}$. Note that there cannot be an incoming edge of $x^3_{\inp}$ from $x^1_{\out}$ since $x^3_{\inp}$ only connects to active output nodes at the time of repair and $x^1_\text{out}$ is no longer active since $x^2_{\out}$ (of the same node index $r_2=r_1$) has been repaired after $x^1_{\out}$.
\item If $x^3_{\inp}\in \overline{U}$;  $r_1,r_2\in D_{r_3}$; and $r_1\neq r_2$; since one of the incoming edges of $x^3_{\inp}$ can be from $x ^ 1_{\out}$ and another edge can be from $x^2_{\out}$ , then at least $(d-2)$ incoming edges of $x^3_{\inp}$ are in $\mathcal{C}$.
\item If $x^3_{\inp}\in \overline{U}$ and only one of $r_1$ or $r_2$ is in $D_{r_3}$, since one of the incoming edges of $x^3_{\inp}$ is from either $x^1_{\out}$ or $x^2_{\out}$, then at least $(d-1)$ incoming edges of $x^3_{\inp}$ are in $\mathcal{C}$.
\item If $x^3_{\inp}\in \overline{U}$ and $r_1,r_2 \notin D_{r_3}$, then at least $d$ incoming edges of $x^3_{\inp}$ are in $\mathcal{C}$.
\end{itemize}
Therefore, these edges related to $x^3_{\out}$ contribute a value of at least $\min ((d-z_3(\mathbf{r}))\beta,\alpha)$ to the min-cut value, where the definition of $z_3(\mathbf{r})$ takes care of the second to the fifth cases.

In the same manner, we can prove that the chronologically $i$-th output node in $\overline{U}$ contributes at least a value of $\min ((d-z_i(\mathbf{r}))\beta,\alpha)$ to the min-cut value. If we sum all the contributions of the oldest $k$ output nodes of $\overline{U}$ we get \eqref{eq:low_b_gen}, a lower bound on the min-cut value.

\section{Proof of Proposition~\ref{prop:low_b}} \label{app:low_b_proof}
The outline of the proof is as follows.

\par Phase~I: We will first show that
\begin{align}\label{eq:new2}
\min_{G\in\mathcal{G}_F} \min_{t\in\DC(G)}&\mincut_G(s,t) \leq  \nonumber\\
&\min_{\forall \pi_f} \sum_{i=1}^{k}\min \left(\left(d-y_i(\pi_f)\right)\beta,\alpha\right).
\end{align}

\par Phase~II: By definition, the FHS scheme is a SHS scheme. Thus, \eqref{eq:low_b_gen} is also a lower bound on all IFGs in $\gf$ and we quickly have
\begin{align}\label{eq:upper_lower}
\min_{\mathbf{r}\in R}\sum_{i=1}^{k}\min ((d-z_i(\mathbf{r}))\beta,\alpha)\leq& \nonumber\\
\min_{G\in\mathcal{G}_F} \min_{t\in\DC(G)}\mincut_G(s,t)& \leq  \nonumber\\
\min_{\forall \pi_f} \sum_{i=1}^{k} \min &\left(\left(d-y_i(\pi_f)\right)\beta,\alpha\right).
\end{align}
The remaining step is to prove that
\begin{align}\label{new3}
\min_{\mathbf{r}\in R}\sum_{i=1}^{k}\min ((d-z_i(\mathbf{r}))\beta,\alpha)=\nonumber&\\
\min_{\forall \pi_f} \sum_{i=1}^{k}\min &\left(\left(d-y_i(\pi_f)\right)\beta,\alpha\right).
\end{align}
Once we prove \eqref{new3}, we have \eqref{eq:low_b} since \eqref{eq:upper_lower} is true. The proof is then complete.

\par The proof of Phase~I is as follows. Denote the smallest IFG in $\gf(n,k,d,\alpha,\beta)$ by $G_0$. Specifically, all its nodes are intact, i.e., none of its nodes has failed before. Denote its active nodes arbitrarily by $1,2,\cdots,n$. Consider the family index permutation of the FHS scheme $F$ that attains the minimization of the RHS of \eqref{eq:new2} and call it $\tilde{\pi}_f$. Fail each active node in $\{1,2,\cdots,n\}$ of $G_0$ exactly once in a way that the sequence of the family indices of the failed nodes is $\tilde{\pi}_f$. Along this failing process, we repair the failed nodes according to the FHS scheme $F$. For example, let $(n,d)=(8,5)$ and suppose the minimizing family index permutation is $\tilde{\pi}_f=(1,2,1,-2,0,0,1,2)$. Then, if we fail nodes 1, 4, 2, 6, 7, 8, 3, and 5 in this sequence, the corresponding family index sequence will be $(1,2,1,-2,0,0,1,2)$, which matches the given $\tilde{\pi}_f$. Note that the node failing sequence is not unique in our construction. For example, if we fail nodes 3, 5, 2, 6, 8, 7, 1, and 4 in this sequence, the corresponding family index vector is still $(1,2,1,-2,0,0,1,2)$. Any node failing sequence that matches the given $\tilde{\pi}_f$ will suffice in our construction. We call the resulting new IFG, $G'$.

\par Consider a data collector $t$ in $G'$ that connects to the oldest $k$ newcomers. (Recall that in our construction, $G'$ has exactly $n$ newcomers.)  Now, by the same arguments as in \cite[Lemma 2]{dimakis2010network}, we will prove that $\mincut_{G'}(s,t)=\sum_{i=1}^{k}\min \left(\left(d-y_i(\tilde{\pi}_f)\right)\beta,\alpha\right)$ for the specifically constructed $G'$ and $t$. Number the storage nodes (input-output pair) of the $k$ nodes $t$ is connected to by $1,2,\dots,k$. Define cut $(U,\overline{U})$ between $t$ and $s$ as the following: for each $i\in \{1,\dots,k\}$, if $\alpha \leq (d-y_i(\tilde{\pi}_f))\beta$ then we include $x_{\out}^{i}$ in $\overline{U}$; otherwise, we include both $x_{\out}^{i}$ and $x_{\inp}^{i}$ in $\overline{U}$. It is not hard to see that the cut-value of the cut $(U,\overline{U})$ is equal to $\sum_{i=1}^{k}\min \left(\left(d-y_i(\tilde{\pi}_f)\right)\beta,\alpha\right)$.

Since the LHS of \eqref{eq:new2} further takes the minimum over $\mathcal{G}_F$ and all data collectors $t$, we have proved the inequality \eqref{eq:new2}.

\par Now, we give the proof of Phase~II (i.e., \eqref{new3}). To that end, we first prove that with the helper sets $D_1$ to $D_n$ specified in a FHS scheme, we have
\begin{align}\label{new4}
\text{LHS of \eqref{eq:new2}}=\min_{\mathbf{r}\in R_2}\sum_{i=1}^{k}\min ((d-z_i(\mathbf{r}))\beta,\alpha)
\end{align}
where $R_2=\{(r_1,r_2,\cdots,r_k): \forall i,j\in\{1,\cdots,k\},1 \leq r_i\leq n, r_i\neq r_j \text{ if } i\neq j\}$. That is, when evaluating the LHS of \eqref{new4}, we can minimize over $R_2$ instead of over $R=\{1,\cdots, n\}^k$. We prove \eqref{new4} by proving that for any $\mathbf{r}\in R$ we can always find a vector $\mathbf{r'} \in R_2$ such that
\begin{align}
\sum_{i=1}^{k}\min ((d-z_i(\mathbf{r}))\beta,\alpha) \geq   \sum_{i=1}^{k}\min ((d-z_i(\mathbf{r'}))\beta,\alpha).\label{new5}
\end{align}
Assuming \eqref{new5} is correct, then we have that at least one of the minimizing $\mathbf{r}^*\in R$ of the LHS of \eqref{new3} is also in $R_2$. We thus have \eqref{new4}. The proof of \eqref{new5} is provided in the end of this Appendix~\ref{app:low_b_proof}.

\par We now notice that any $\mathbf{r} \in R_2$ corresponds to the first $k$ coordinates of a permutation of the node indices $(1,2,3,\cdots, n)$. For easier reference, we use $\overline{\mathbf{r}} $ to represent an $n$-dimensional permutation vector such that the first $k$ coordinates of $\overline{\mathbf{r}}$ match $\mathbf{r}$. One can view $\overline{\mathbf{r}}$ as the extended version of $\mathbf{r}$ from a partial $k$-dimensional permutation to a complete $n$-dimensional permutation vector.  Obviously, the choice of $\overline{\mathbf{r}}$ is not unique.  The following discussion holds for any $\overline{\mathbf{r}}$.

\par For any $\mathbf{r}\in R_2$, we first find its extended version $\overline{\mathbf{r}}$. We then construct $\pi_f$ from $\overline{\mathbf{r}}$ by transcribing the permutation of the node indices $\overline{\mathbf{r}}$ to the corresponding family indices. For example, consider the parameter values $(n,k,d)=(8,4,5)$. Then, one possible choice of $\mathbf{r}\in R_2$ is $\mathbf{r}=(3,5,2,4)$ and a corresponding $\overline{\mathbf{r}}$ is $(3,5,2,4,1,6,7,8)$. The transcribed family index vector is $\pi_f=(1,2,1,2,1,-2,0,0)$. We now argue that $z_i(\mathbf{r})=y_i(\pi_f)$ for all $i=1$ to $k$. The reason is that the definition of $y_i(\pi_f)$ is simply a transcribed version of the original definition of $z_i(\mathbf{r})$ under the node-index to family-index translation. In sum, the above argument proves that for any $\mathbf{r}\in R_2$, there exists a $\pi_f$ satisfying

\begin{align}
\sum_{i=1}^{k}\min((d-&z_i(\mathbf{r}))\beta,\alpha)=\sum_{i=1}^{k}\min \left(\left(d-y_i(\pi_f)\right)\beta,\alpha\right).\nonumber
\end{align}
As a result, we have 
\begin{align}
\min_{\mathbf{r}\in R_2}\sum_{i=1}^{k}\min((d-&z_i(\mathbf{r}))\beta,\alpha)\geq \nonumber\\
&\min_{\forall \pi_f} \sum_{i=1}^{k}\min \left(\left(d-y_i(\pi_f)\right)\beta,\alpha\right).\label{eq:new-CCW-001}
\end{align}
Jointly, \eqref{eq:new-CCW-001}, \eqref{new4}, and \eqref{eq:upper_lower} imply \eqref{new3}. The proof of Proposition~\ref{prop:low_b} is thus complete. The remainder of this appendix section is dedicated to proving  \eqref{new5}, which is unfortunately quite long and delicate. 

\par \emph{The proof of \eqref{new5}:}

\par We prove \eqref{new5} by explicit construction. For any vector $\mathbf{r}\in R$, we will use the following procedure, {\sc Modify}, to gradually modify $\mathbf{r}$ in 4 major steps until the end result is the desired $\mathbf{r'}\in R_2$ that satisfies \eqref{new5}. A detailed example illustrating procedure {\sc Modify} is provided in Appendix~\ref{app:modify_ex} to complement the following algorithmic description of {\sc Modify}.

\par \emph{Step~1:} If there are $i, j\in\{1,\cdots,k\}$ such that $i<j$ and   the $i$-th and the $j$-th coordinates of $\mathbf{r}$ are equal, i.e., $r_i=r_j$, then we can do the following modification. For convenience, we denote the value of $r_i=r_j$ by $h$. Suppose that node $h$ belongs to the $Q$-th family. We now check whether there is any value $\gamma$ satisfying simultaneously (i) $\gamma\in\{1,2,\cdots, n\}\backslash h$; (ii) node $\gamma$ is also in the $Q$-th family; and (iii) $\gamma$ is not equal to any of the coordinates of $\mathbf{r}$. If such $\gamma$ exists, we replace the $j$-th coordinate of $\mathbf{r}$ by $\gamma$. Specifically, after this modification, we will have $r_i=h$ and $r_j=\gamma$.

\par Repeat this step until either there is no repeated $r_i=r_j$, or until no such $\gamma$ can be found.

\par \emph{Step~2:} After finishing Step~1, we perform the following modification. If there still are distinct $i,j\in\{1,\cdots,k\}$ such that $r_i=r_j$ and $i<j$, then we again denote the value of $r_i=r_j$ by $h$. Suppose node $h$ belongs to the $Q$-th family. Consider the following two cases. If the $Q$-th family is the incomplete family, then no further modification will be made.

\par If the $Q$-th family is a complete family, then do the following modification.

Find the largest $j_1\in\{1,\cdots, n\}$ such that node $r_{j_1}=h$ and find the largest $j_2\in \{1,\cdots, n\}$ such that $r_{j_2}$ belongs to the $Q$-th family (the same family of node $h$). If $j_1=j_2$, then we set $\mathbf{r'}=\mathbf{r}$. If $j_1\neq j_2$, then we swap the values of $r_{j_1}$ and $r_{j_2}$ to construct $\mathbf{r'}$. That is, we first set $\mathbf{r'}=\mathbf{r}$ for all coordinates except for the $j_1$-th and the $j_2$-th coordinates, and then set $r'_{j_1}=r_{j_2}$ and $r'_  {j_2}=r_{j_1}$. After we have constructed new $\mathbf{r'}$ depending on whether $j_1=j_2$ or not, we now check whether there is any value $\gamma\in\{1,\cdots, n\}$ satisfying simultaneously (i) node $\gamma$ belongs to a complete family (not necessarily the Q-th family); and (ii) $\gamma$ is not equal to any of the coordinates of $\mathbf{r'}$. If such $\gamma$ exists, we replace the $j_2$-th coordinate of $\mathbf{r'}$ by $\gamma$, i.e., set $r'_{j_2}=\gamma$.

\par Repeat this step until the above process does not change the value of any of the coordinates of $\mathbf{r'}$.

After finishing the above two steps, the current vector $\mathbf{r}$ must be in one of the following cases. Case~1: No two coordinates are equal, i.e., $r_i\neq r_j$ for all pairs $i<j$; Case~2: there exist a pair $i<j$ such that $r_i=r_j$. We have two sub-cases for Case~2. Case~2.1: All such $(i,j)$ pairs must satisfy that node $r_i$ belongs to a complete family. Case~2.2: All such $(i,j)$ pairs must satisfy that node $r_i$ belongs to the incomplete family. Specifically, the above construction (Steps~1 and 2) has eliminated the sub-case that some $(i,j)$ pair has $r_i=r_j$ belonging to a complete family and some other $(i,j)$ pair has $r_i=r_j$ belonging to the incomplete family. The reason is as follows. Suppose some $(i,j)$ pair has $r_i$ belonging to a complete family. Since we have finished Step~2, it means that any node $\gamma$ that belongs to a complete family must appear in one of the coordinates of $\mathbf{r}$. Since there are $(n-d)\left\lfloor\frac{n}{n-d}\right\rfloor$ number of nodes belonging to complete families, at least $(n-d)\left\lfloor\frac{n}{n-d}\right\rfloor+1$ number of coordinates of $\mathbf{r}$ must refer to a node in a complete family (since $r_i$ and $r_j$ have the same value). Therefore, there are at most $n-\left((n-d)\left\lfloor\frac{n}{n-d}\right\rfloor+1\right) =(n\bmod (n-d))-1$ number of coordinates of $\mathbf{r}$ referring to a node in the incomplete family. However, if we have another $(i',j')$  pair has $r_{i'}=r_{j'}$ belonging to the incomplete family, then it means that the coordinates of $\mathbf{r}$ can refer to at most $(n\bmod(n-d))-2$ distinct nodes of the incomplete family (since $r_{i'}$ and $r_{j'}$  are equal). Since there are $n\bmod(n-d)$ distinct nodes in the incomplete family, there must exist a $\gamma$ value such that node $\gamma$ belongs to the incomplete family and $\gamma$ does not appear in any one of the coordinates of $\mathbf{r}$. This contradicts the fact that we have exhausted Step~1 before moving on to Step~2.

\par We now consider Cases~1, 2.1, and 2.2, separately. If the $\mathbf{r}$ vector is in Case~1, then such $\mathbf{r}$ belongs to $R_2$ and our construction is complete. If $\mathbf{r}$ belongs to Case~2.2, then do Step~3. If $\mathbf{r}$ belongs to Case~2.1, do Step~4.

\emph{Step~3:}  We use $(i,j)$ to denote the pair of values such that $r_i=r_j$ and $i<j$. Denote the value of $r_i=r_j$ by $h$. Since we are in Case~2.2, node $h$ belongs to the incomplete family. Find the largest $j_1\in\{1,\cdots, n\}$ such that node $r_{j_1}=h$ and find the largest $j_2\in \{1,\cdots, n\}$ such that $r_{j_2}$ belongs to the incomplete family. If $j_1=j_2$, then we keep $\mathbf{r}$ as is. If $j_1\neq j_2$, then we swap the values of $r_{j_1}$ and $r_{j_2}$. Recall that we use $c\stackrel{\Delta}{=}\left\lfloor \frac{n}{n-d}\right\rfloor$ to denote the family index of the last complete family. We now choose arbitrarily a $\gamma$ value from $\{(n-d)\left(c-1\right)+ 1,\dots,(n-d)c\}$. Namely, $\gamma$ is the index of a node of the last complete family. Fix the $\gamma$ value. We then replace $r_{j_2}$ by the arbitrarily chosen $\gamma$.

\par If the value of one of the coordinates of $\mathbf{r}$ (before setting $r_{j_2}=\gamma$) is $\gamma$, then after setting $r_{j_2}=\gamma$ we will have some $i\neq j_2$ satisfying $r_i=r_{j_2}=\gamma$. In this case, we start over from Step~1. If none of the coordinates of $\mathbf{r}$ (before setting $r_{j_2}=\gamma$) has value $\gamma$, then one can easily see that after setting $r_{j_2}=\gamma$ there exists no $i<j$ satisfying ``$r_i=r_j$ belong to a complete family'' since we are in Case~2.2 to begin with. In this case, we are thus either in Case~1 or Case~2.2. If the new $\mathbf{r}$ is now in Case~1, then we stop the modification process. If the new $\mathbf{r}$ is still in Case~2.2, we will then repeat this step (Step~3).

\emph{Step~4:} We use $(i,j)$ to denote the pair of values such that $r_i=r_j$ and $i<j$. Denote the value of $r_i=r_j$ by $h$. Since we are in Case~2.1, node $h$ belongs to a complete family. Suppose $h$ is in the $Q$-th complete family. Find the largest $j_1\in\{1,\cdots, n\}$ such that node $r_{j_1}=h$ and find the largest $j_2\in \{1,\cdots, n\}$ such that $r_{j_2}$ belongs to the $Q$-th complete family. If $j_1=j_2$, then we keep $\mathbf{r}$ as is. If $j_1\neq j_2$, then we swap the values of $r_{j_1}$ and $r_{j_2}$. We now find a $\gamma$ value such that (i) node $\gamma$ belongs to the incomplete family; and (ii) $\gamma$ is not equal to any of the coordinates of $\mathbf{r}$. Note that such $\gamma$ value always exists. The reason is that since we are now in Case~2.1 and we have finished Step~2, it means that any node $\gamma$ that belongs to a complete family must appear in one of the coordinates of $\mathbf{r}$. Therefore, there are at least $(n-d)\left\lfloor \frac{n}{n-d}\right\rfloor+1$ number of coordinates of $\mathbf{r}$ referring to a node in one of the complete families. This in turn implies that there are at most $n-\left((n-d)\left\lfloor\frac{n}{n-d}\right\rfloor+1\right)=(n\bmod (n-d))-1$ number of coordinates of $\mathbf{r}$ referring to a node in the incomplete family. Since there are $n\bmod (n-d)$ distinct nodes in the incomplete family, there must exist a $\gamma$ value such that node $\gamma$ belongs to the incomplete family and $\gamma$ does not appear in any one of the coordinates of $\mathbf{r}$.

\par Once the $\gamma$ value is found, we replace the $j_2$-th coordinate of $\mathbf{r}$ by $\gamma$, i.e., $r_{j_2}=\gamma$. If the new $\mathbf{r}$ is now in Case~1, then we stop the modification process. Otherwise, $\mathbf{r}$ must still be in Case~2.1 since we replace $r_{j_2}$ by a $\gamma$ that does not appear in $\mathbf{r}$ before. In this scenario, we will then repeat this step (Step~4).

\par An example demonstrating the above iterative process is provided in Appendix~\ref{app:modify_ex}.

\par To prove that this construction is legitimate, we need to prove that the iterative process ends in a finite number of time. To that end, for any vector $\mathbf{r}$, define a non-negative function $T(\mathbf{r})$ by
\begin{align}
T(\mathbf{r})&=|\{(i,j):i<j,r_i=r_j\text{ is a complete family node}\}|+\nonumber\\
&2|\{(i,j):i<j,r_i=r_j\text{ is an incomplete family node}\}|.\nonumber
\end{align}

\par One can then notice that in this iterative construction, every time we create a new $\mathbf{r}'$ vector that is different from the input vector $\mathbf{r}$, the value of $T(\mathbf{r})$ decreases by at least 1. As a result, we cannot repeat this iterative process indefinitely. When the process stops, the final vector $\mathbf{r}'$ must be in Case~1. Therefore, the procedure {\sc Modify} converts any vector $\mathbf{r}\in R$ to a new vector $\mathbf{r'}\in R_2$ such that all coordinate values of $\mathbf{r'}$ are distinct. What remains to be proved is that along the above 4-step procedure, the inequality \eqref{new5} always holds. That is, the value of $\sum_{i=1}^{k}\min ((d-z_i(\mathbf{r}))\beta,\alpha)$ is non-increasing along the process. The detailed proof of the non-increasing $\sum_{i=1}^{k}\min ((d-z_i(\mathbf{r}))\beta,\alpha)$ will be provided shortly. From the above discussion, we have proved \eqref{new5}.

In the rest of this appendix, we prove the correctness of {\sc Modify}. For each step of {\sc Modify}, we use $\mathbf{r}$ to denote the input (original) vector and $\mathbf{w}$ to denote the output (modified) vector. In what follows, we will prove that the $\mathbf{r}$ and $\mathbf{w}$ vectors always satisfy
\begin{align} \label{eq:modified}
\sum_{i=1}^{k}\min ((d-z_i(\mathbf{w}))\beta,\alpha) \leq \sum_{i=1}^{k}\min ((d-z_i(\mathbf{r}))\beta,\alpha).
\end{align}
\par In Step~1 of the procedure, suppose that we found such $\gamma$. Denote the vector after we replaced the $j$-th coordinate with $\gamma$ by $\mathbf{w}$. We observe that for $1\leq m\leq j$, we will have $z_m(\mathbf{r})=z_m(\mathbf{w})$ since $r_m=w_m$ over $1\leq m\leq j-1$ and the new $w_j=\gamma$ belongs to the $Q$-th family, the same family as node $r_j$. For $j+1\leq m\leq k$, we will have $z_m(\mathbf{w})\geq z_m(\mathbf{r})$. The reason is that by our construction, we have $w_j=\gamma \neq r_j= r_i=w_i$. For any $m>j$, $z_m(\mathbf{r})$ only counts the repeated $r_i=r_j$ once. Therefore, $z_m(\mathbf{w})$ will count the same $w_i$ as well. On the other hand, $z_m(\mathbf{w})$ may sometimes be larger than $z_m(\mathbf{r})$, depending on whether the new $w_j \in D_{w_m}$ or not. The fact that $z_m(\mathbf{w})\geq z_m(\mathbf{r})$ for all $m=1$ to $k$ implies  \eqref{eq:modified}.

\par In Step~2, if $j_1=j_2$, then we will not swap the values of $r_{j_1}$ and $r_{j_2}$. On the other hand, $j_1=j_2$ also means that $r_{j_1}=r_{j_2}=h$. In this case, $\mathbf{w}$ is modified from $\mathbf{r}$ such that $w_{j_2}=\gamma$ if such a $\gamma$ is found. For $1\leq m\leq j_2-1$, $z_m(\mathbf{w})=z_m(\mathbf{r})$ since $r_m=w_m$ over this range of $m$. We now consider the case of  $m=j_ 2$. Suppose node $\gamma$ belongs to the $Q_{\gamma}$-th family. We first notice that by the definition of $z_m(\cdot)$ and the definition of the FHS scheme, $(z_m(\mathbf{w})-z_m(\mathbf{r}))$ is equal to the number of distinct nodes in the $Q$-th family that appear in the first $(j_2-1)$ coordinates of $\mathbf{r}$ minus the number of distinct nodes in the $Q_{\gamma}$-th family that appear in the first $(j_2-1)$ coordinates of $\mathbf{w}$.  For easier reference, we call the former ${\mathsf{term1}}$ and the latter ${\mathsf{term2}}$ and we will quantify these two terms separately.

\par Since we start Step~2 only after Step~1 cannot proceed any further, it implies that all distinct $(n-d)$ nodes of family $Q$ must appear in $\mathbf{r}$ otherwise we should continue Step~1 rather than go to Step 2. Then by our specific construction of $j_2$, all distinct $(n-d)$ nodes of family $Q$ must appear in the  first $(j_2-1)$-th coordinates of $\mathbf{r}$. Therefore ${\mathsf{term1}}=(n-d)$. Since there are exactly $(n-d)$ distinct nodes in the $Q_{\gamma}$-th family, by the definition of ${\mathsf{term2}}$, we must have ${\mathsf{term2}}\leq (n-d)$. The above arguments show that ${\mathsf{term2}}\leq {\mathsf{term1}}=(n-d)$, which implies the desired inequality $z_m(\mathbf{w})-z_m(\mathbf{r})\geq 0$ when $m=j_2$.

\par
We now consider the case when $m>j_2$. In this case, we still have $z_m(\mathbf{w})\geq z_m(\mathbf{r})$. The reason is that by our construction, we have $w_{j_2}=\gamma \neq r_{j_2}= r_i=w_i$. For any $m>j_2$, $z_m(\mathbf{r})$ only counts the repeated $r_i=r_{j_2}$ once. Therefore, $z_m(\mathbf{w})$ will count the same $w_i$ as well. On the other hand, $z_m(\mathbf{w})$ may sometimes be larger than $z_m(\mathbf{r})$, depending on whether the new $w_{j_2} \in D_{w_m}$ or not. The fact that $z_m(\mathbf{w})\geq z_m(\mathbf{r})$ for all $1\leq m\leq k$ implies \eqref{eq:modified}.

\par Now, we consider the case when $j_1\neq j_2$, which implies that $r_{j_1}=h\neq r_{j_2}$ and Step~2 swaps the $j_1$-th and the $j_2$-th coordinates of $\mathbf{r}$. Note that after swapping, we can see that if we apply the same $j_1$ and $j_2$ construction to the {\em new} swapped vector, then we will have $j_1=j_2$. By the discussion in the case of $j_1=j_2$, we know that replacing the value of $r_{j_2}$ by $\gamma$ will not decrease the value $z_m(\mathbf{w})$ for any $m=1$ to $k$ and \eqref{eq:modified} still holds. As a result, we only need to prove that swapping the $j_1$-th and the $j_2$-th coordinates of $\mathbf{r}$ does not decrease the value of $z_m(\mathbf{r})$.

\par To that end, we slightly abuse the notation and use $\mathbf{w}$ to denote the resulting vector after swapping the $j_1$-th and the $j_2$-th coordinates of $\mathbf{r}$ (but before replacing $r_{j_2}$ by $\gamma$). For the case of $1\leq m\leq j_1$, we have $z_m(\mathbf{w})=z_m(\mathbf{r})$ since for $1\leq m\leq j_1-1$, $r_m=w_m$, and both $r_{j_1}$ and $w_{j_1}=r_{j_2}$ are from the same family $Q$. For $j_1+1\leq m \leq j_2-1$, we have $z_m(\mathbf{w})\geq z_m(\mathbf{r})$. The reason is as follows. We first observe that  $w_{j_1}=r_{j_2} \neq r_{j_1}= r_i=w_i$. For any $j_1+1\leq m \leq j_2-1$, $z_m(\mathbf{r})$ only counts the repeated $r_i=r_{j_1}$ once (since by our construction of $j_1$ we naturally have $j_1> i$). Therefore, $z_m(\mathbf{w})$ will count the same $w_i$ as well. On the other hand, $z_m(\mathbf{w})$ may sometimes be larger than $z_m(\mathbf{r})$, depending on whether the new $w_{j_1} \in D_{w_m}$ or not. We thus have $z_m(\mathbf{w})\geq z_m(\mathbf{r})$ for $j_1+1\leq m\leq j_2-1$.

\par For the case of $m=j_2$, we notice that $w_{j_2}=r_{j_1}$ and $r_{j_2}$ are from the same $Q$-th family. Therefore, we have $z_m(\mathbf{w})= z_m(\mathbf{r})$. For the case of $j_2+1\leq m\leq k$, we argue that $z_m(\mathbf{w})=z_m(\mathbf{r})$. This is true because of the definition of $z_m(\cdot)$ and the fact that both $j_1<m$ and $j_2<m$. In summary, we have proved $z_m(\mathbf{w})\geq z_m(\mathbf{r})$ for $m=1$ to $k$, which implies \eqref{eq:modified}.

\par In Step~3, we first consider the case of $j_1=j_2$, which means that $r_{j_1}=r_{j_2}$ is replaced with $\gamma$, a node from the last complete family. For $1\leq m\leq j_1-1$, since we have $r_m=w_m$ for all $1\leq m\leq j_1-1$, we must have $z_m(\mathbf{r})=z_m(\mathbf{w})$. We now consider the case of $m=j_1$. By the definition of $z_m(\cdot)$ and the definition of the FHS scheme, $(z_m(\mathbf{w})-z_m(\mathbf{r}))$ is equal to the number of distinct nodes in the incomplete family that appear in the first $(j_1-1)$ coordinates of $\mathbf{r}$ minus the number of distinct nodes in the last complete family that simultaneously (i) belong to the helper set of the incomplete family and (ii) appear in the first $(j_1-1)$ coordinates of $\mathbf{w}$. For easier reference, we call the former $\mathsf{term1}$ and the latter $\mathsf{term2}$ and we will quantify these two terms separately. 

\par Since we have finished executing Step~1, it means that all $n\bmod(n-d)$ nodes in the incomplete family appear in the vector $\mathbf{r}$. By our construction of $j_1$,  all  $n\bmod(n-d)$ nodes in the incomplete family must appear in the first $(j_1-1)$ coordinates of $\mathbf{r}$. Therefore, $\mathsf{term1}=n\bmod(n-d)$. Since there are exactly $n\bmod(n-d)$ distinct nodes in the last complete family that belong to the helper set of the incomplete family, by the definition of ${\mathsf{term2}}$, we must have ${\mathsf{term2}}\leq n\bmod(n-d)$. The above arguments show that ${\mathsf{term2}}\leq {\mathsf{term1}}=n\bmod(n-d)$, which implies the desired inequality $z_m(\mathbf{w})-z_m(\mathbf{r})\geq 0$.

\par For the case of  $j_1+1=j_2+1\leq m$, we also have  $z_m(\mathbf{w})\geq z_m(\mathbf{r})$.  The reason is that by our construction, we have $w_{j_2}=\gamma \neq r_{j_2}= r_i=w_i$. For any $m>j_2$, $z_m(\mathbf{r})$ only counts the repeated $r_i=r_{j_2}$ once. Therefore, $z_m(\mathbf{w})$ will count the same $w_i$ as well. On the other hand, $z_m(\mathbf{w})$ may sometimes be larger than $z_m(\mathbf{r})$, depending on whether the new $w_{j_2} \in D_{w_m}$ or not. We have thus proved that $z_m(\mathbf{w})\geq z_m(\mathbf{r})$ for all $m=1$ to $k$, which implies \eqref{eq:modified}.

\par We now consider the case of $j_1\neq j_2$. Namely, we swap the $j_1$-th and the $j_2$-th coordinates of $\mathbf{r}$ before executing the rest of Step~3. We can use the same arguments as used in proving the swapping step of Step~2 to show that after swapping, we still have $z_m(\mathbf{w})\geq z_m(\mathbf{r})$ for all $m=1$ to $k$, which implies \eqref{eq:modified}. The proof of Step~3 is complete.

\par In Step~4, we again consider the case of $j_1=j_2$ first. In this case, $r_{j_1}=h$ is replaced with $\gamma$, a node of the incomplete family. For $1\leq m \leq j_1-1$, $z_m(\mathbf{w})=z_m(\mathbf{r})$ since $w_m=r_m$ over this range of $m$. For $m=j_1$, we have to consider two cases. If the $Q$-th family is the last complete family, then $(z_m(\mathbf{w})-z_m(\mathbf{r}))$ is equal to the number of distinct nodes in the $Q$-th family that simultaneously (i) belong to the helper set of the incomplete family and (ii) appear in the first $(j_1-1)$ coordinates of $\mathbf{r}$, minus the number of distinct nodes in the incomplete family that appear in the first $(j_1-1)$ coordinates of $\mathbf{w}$. For easier reference, we call the former $\mathsf{term1}$ and the latter $\mathsf{term 2}$. If, however, the $Q$-th family is not the last complete family, then $(z_m(\mathbf{w})-z_m(\mathbf{r}))$ is equal to the difference of another two terms. We slightly abuse the notation and refer again to the two terms as $\mathsf{term1}$ and $\mathsf{term2}$ where $\mathsf{term1}$ is the number of distinct nodes in the $Q$-th family that appear in the first $(j_1-1)$ coordinates of $\mathbf{r}$ and $\mathsf{term2}$ is the number of distinct nodes in the last complete family that simultaneously (i) does not belong to the helper set of the incomplete family and (ii) appear in the first $(j_1-1)$ coordinates of $\mathbf{w}$ plus the number of distinct nodes in the incomplete family that appear in the first $(j_1-1)$ coordinates of $\mathbf{w}$. 

\par We will now quantify these two terms separately. Since we have finished executing Step~1 and by the construction of $j_1$, all  $(n-d)$ nodes in the $Q$-th family must appear in the first $(j_1-1)$ coordinates of $\mathbf{r}$, which are the same as the first $(j_1-1)$ coordinates of $\mathbf{w}$. Therefore, the value of $\mathsf{term1}$ is $n\bmod(n-d)$ if the $Q$-th family is the last complete family or $(n-d)$ if it is one of the first $c-1$ complete families. We now quantify $\mathsf{term2}$. For when the $Q$-th family is the last complete family, since there are exactly $n\bmod(n-d)$ distinct nodes in the incomplete family, by the definition of ${\mathsf{term2}}$, we must have ${\mathsf{term2}}\leq n\bmod(n-d)$. When the $Q$-th family is not the last complete family, ${\mathsf{term2}}\leq (n-d)$ since the number of distinct nodes in the incomplete family is $n\bmod(n-d)$ and the number of distinct nodes in the last complete family that do not belong to the helper set of the incomplete family is $(n-d-n\bmod(n-d))$ and their summation is $\leq n-d$. The above arguments show that ${\mathsf{term2}}\leq {\mathsf{term1}}$ for both cases, which implies the desired inequality $z_m(\mathbf{w})-z_m(\mathbf{r})\geq 0$ for $m=j_1$.

\par For $j_1+1\leq m\leq k$, since $r_{j_1}=h=r_i$ was a repeated node, then it was already not contributing to $z_m(\mathbf{r})$ for all $m>j_1$. Thus, $z_{m}(\mathbf{w})\geq z_m(\mathbf{r})$ for all $m=j_1+1$ to $k$. (Please refer to the $j_1+1\leq m$ case in Step~3 for detailed elaboration.) In summary, after Step~4, assuming $j_1=j_2$, we have $z_{m}(\mathbf{w})\geq z_m(\mathbf{r})$ for all $m=1$ to $k$, which implies \eqref{eq:modified}.

\par Finally, we consider the case of $j_1\neq j_2$. Namely, we swap the $j_1$-th and the $j_2$-th coordinates of $\mathbf{r}$ before executing the rest of Step~4. We can use the same arguments as used in proving the swapping step of Step~2 to show that the inequality \eqref{eq:modified} holds after swapping. The proof of Step~4 is thus complete.

\section{An Illustrative Example for the {\sc Modify} Procedure} \label{app:modify_ex}

\par For illustration, we apply the procedure {\sc Modify} to the following example with $(n,d)=(8,5)$ and some arbitrary $k$. Recall that family 1 contains nodes $\{1,2,3\}$, family 2 (last complete family) contains  nodes $\{4,5,6\}$, and the incomplete family, family 0, contains nodes $\{7,8\}$. Suppose the initial $\mathbf{r}$ vector is $\mathbf{r}=(1,2,2,2,4,7,7,7)$. We will use {\sc Modify} to convert $\mathbf{r}$ to a vector $\mathbf{r}'\in R_2$

\par We first enter Step~1 of the procedure. We observe\footnote{We also observe that $r_2=r_3=2$ and we can choose $i=2$ and $j=3$ instead. Namely, the choice of $(i,j)$ is not unique. In {\sc Modify}, any choice satisfying our algorithmic description will work. } that $r_3=r_4=2$ ($i=3$ and $j=4$) and node 2 belongs to the first family. Since node 3 is also in family 1 and it is not present in $\mathbf{r}$,  we can choose $\gamma=3$. After replacing $r_4$ by 3, the resulting vector is $\mathbf{r}=(1,2,2,3,4,7,7,7)$. Next, we enter Step~1 for the second time. We observe that $r_7=r_8=7$. Since node 8 is in family 0 and it is not present in $\mathbf{r}$, we can choose $\gamma=8$. The resulting vector is $\mathbf{r}=(1,2,2,3,4,7,7,8)$. Next, we enter Step~1 for the third time. For the new $\mathbf{r}$, we have $r_2=r_3=2$ and $r_6=r_7=7$, but for both cases we cannot find the desired $\gamma$ value. As a result, we cannot proceed any further by Step~1. For that reason, we enter Step 2. 

\par We observe that for $r_2=r_3=2$, we find $j_1=3$, the last coordinate of $\mathbf{r}$ equal to $2$, and $j_2=4$, the last coordinate of $\mathbf{r}$ that belongs to family~1.  By Step~2, we swap $r_{3}$ and $r_{4}$, and the resultant vector is $\mathbf{r}=(1,2,3,2,4,7,7,8)$. Now, since node 5 belongs to family 2, a complete family, and it is not present in $\mathbf{r}$, we can choose $\gamma=5$. After replacing $r_{j_2}$ by $\gamma$, the resultant vector is $\mathbf{r}=(1,2,3,5,4,7,7,8)$. Next, we enter Step~2 for the second time. Although $r_6=r_7=7$, we notice that node 7 is in family 0. Therefore, we do nothing in Step~2.

\par After Step~2, the latest $\mathbf{r}$ vector is $\mathbf{r}=(1,2,3,5,4,7,7,8)$, which belongs to Case~2.2. Consequently, we enter Step~3. In Step~3, we observe that $j_1=7$, the last coordinate of $\mathbf{r}$ being 7, and $j_2=8$, the last coordinate of $\mathbf{r}$  that belongs to the incomplete family, family 0. Thus, we swap $r_{7}$ and $r_{8}$, and the resultant vector is $\mathbf{r}=(1,2,3,5,4,7,8,7)$. Now, we choose arbitrarily a $\gamma$ value from $\{4,5,6\}$, the last complete family. Suppose we choose\footnote{We can also choose $\gamma=4$ or $5$. For those choices, the iterative process will continue a bit longer but will terminate eventually.} $\gamma =6$. The resultant vector is $\mathbf{r}=(1,2,3,5,4,7,8,6)$. Since we have no other repeated nodes of family 0, the procedure finishes at this point. Indeed, we can see that the final vector $\mathbf{r'}=(1,2,3,5,4,7,8,6)\in R_2$, which has no repeated nodes and is the result expected.

\section{Proof of Proposition~\ref{prop:mbr}}\label{app:mbr_proof}
For fixed $(n,k,d)$ values, define function $g$ as
\begin{align}
g(\alpha,\beta)=\min_{G\in\mathcal{G}_F} \min_{t\in\DC(G)}\mincut_G(s,t).
\end{align}
We first note that by \eqref{eq:low_b}, we must have $g(d\beta,\beta)=m\beta$ for some integer $m$. The value of $m$ depends on the $(n,k,d)$ values and the minimizing family index permutation $\pi_f$, but does not depend on $\beta$. We then define $\beta^*$ as the $\beta$ value such that $g(d\beta,\beta)= \mathcal{M}$. We will first prove that $\beta_{\MBR}=\beta^*$ by contradiction. Suppose $\beta_{\MBR}\neq \beta^*$. Since $(\alpha,\beta)=(d\beta^*,\beta^*)$ is one way that can satisfy $g(\alpha,\beta)=\mathcal{M}$, the minimum-bandwidth consumption $\beta_{\MBR}$ must satisfy $\beta_{\MBR}\leq \beta^*$. Therefore, we must have $\beta_{\MBR}<\beta^*$. However, we then have the following contradiction.
\begin{align}
\mathcal{M}\leq g(\alpha_{\MBR},\beta_{\MBR})\leq g(\infty, \beta_{\MBR})&=\nonumber \\
g(d \beta_{\MBR}, \beta_{\MBR})&<g(d\beta^*,\beta^*)= \mathcal{M}, \label{new:MBR}
\end{align}
where the first inequality is by knowing that $(\alpha_{\MBR},\beta_{\MBR})$ satisfies the reliability requirement; the second inequality is by the definition of $g(\alpha,\beta)$; the first equality is by \eqref{eq:low_b}; and the third inequality (the only strict inequality) is by the fact that $g(d\beta,\beta)=m\beta$ for all $\beta$ and by the assumption of $\beta_{\MBR}<\beta^*$; and the last equality is by the construction of $\beta^*$.

\par The above arguments show that $\beta_{\MBR}=\beta^*$. To prove that $\alpha_{\MBR}=d\beta^*$, we first prove
\begin{align}
g(\alpha,\beta)<g(d\beta,\beta), \mbox{ if } \alpha<d\beta. \label{eq:g1}
\end{align}
The reason behind \eqref{eq:g1} is that (i) $k\geq 1$ and we thus have at least one summand in the RHS of \eqref{eq:low_b}; and (ii) the first summand is always $\min(d\beta,\alpha)$ since $y_1(\pi_f)=0$ for any family index permutation $\pi_f$. Suppose $\alpha_{\MBR}\neq d\beta^*$. Obviously, we have $\alpha_{\MBR}\leq d\beta^*$ by the construction of $\beta^*$. Therefore, we must have $\alpha_{\MBR}<d\beta^*$. However, we then have the following contradiction
\begin{align}
\mathcal{M}\leq g(\alpha_{\MBR},\beta_{\MBR}) < g(d\beta^*,\beta^*)= \mathcal{M}, \label{new:MBR2}
\end{align}
where the first inequality is by knowing that $(\alpha_{\MBR},\beta_{\MBR})$ satisfies the reliability requirement, the second inequality is by \eqref{eq:g1}, and the equality is by the construction of $\beta^*$.

The above arguments prove that $\alpha_{\MBR}=d\beta_{\MBR}$.  This also implies that when considering the MBR point, instead of finding a $\pi_f$ that minimizes \eqref{eq:low_b}, we can focus on finding a $\pi_f$ that minimizes 
\begin{align}
\sum_{i=1}^k (d-y_i(\pi_f)) \label{eq:new15}
\end{align}
instead, i.e., we remove the minimum operation of \eqref{eq:low_b} and ignore the constant $\beta$, which does not depend on $\pi_f$. We are now set to show that $\pi_f^*$ is the minimizing family index permutation at the MBR point. 
\par First, define
\begin{align} \label{eq:y_def}
y_{\text{offset}}(\pi_f)=\sum_{i=1}^k (i-1-y_i(\pi_f)).
\end{align}
Notice that a family index permutation that minimizes $y_{\text{offset}}(\cdot)$ also minimizes \eqref{eq:new15}. Therefore, any minimizing family index permutation for \eqref{eq:new15}, call it $\pi_f^{\min}$, must satisfy
\begin{align} \label{eqn:y}
y_{\text{offset}}(\pi_f^{\min})=\min_{\forall \pi_f} y_{\text{offset}}(\pi_f).
\end{align}

\par Consider the following two cases:

\underline{Case 1:} $n\bmod(n-d)=0$, i.e., we do not have an incomplete family.

Consider any family index permutation $\pi_f$ and let $l_j$ be the number of the first $k$  coordinates of $\pi_f$ that have value $j$. Recall that there is no incomplete family in this case. Suppose the $i$-th coordinate of $\pi_f$ is $m$. Then, we notice that the expression ``$(i-1)-y_i(\pi_f)$'' counts the number of appearances of the value $m$ in the first $i-1$ coordinates of $\pi_f$ (recall that there is no incomplete family in this case). Therefore, we can rewrite \eqref{eq:y_def} by
\begin{align} \label{eq:y_families}
y_{\text{offset}}(\pi_f)=\sum_{i=1}^{l_1}(i-1) + \sum_{i=1}^{l_2}(i-1) + \dots + \sum_{i=1}^{l_{\frac{n}{n-d}}}(i-1).
\end{align}
We now prove the following claim.

\begin{claim} \label{clm:diverse} The above equation implies that a family index permutation is a minimizing permutation $\pi_f^{\min}$ if and only if
\begin{align} \label{eq:diverse}
|l_i-l_j|\leq1 \text{ for all $i,j$ satisfying }1\leq i,j\leq \frac{n}{n-d}.
\end{align}
\end{claim}

\begin{IEEEproof}
We first prove the only if direction by contradiction. The reason is as follows. If $l_i>l_j+1$ for some $1\leq i,j\leq \frac{n}{n-d}$, then we consider another family permutation $\pi_f'$ and denote its corresponding $l$ values by $l'$, such that $l'_i=l_i-1$, $l'_j=l_j+1$, and all other $l$s remain the same. Clearly from \eqref{eq:y_families}, such $\pi_f'$ will result in strictly smaller $y_{\text{offset}}(\pi_f')<y_\text{offset}(\pi_f)$. Note that such $\pi_f'$ with the new $l'_i=l_i-1$, $l'_j=l_j+1$ always exists. The reason is the following. By the definition of $l_j$ and the fact that $\pi_f$ is a family index permutation, we have $0\leq l_j\leq (n-d)$ for all $j=1,\cdots, \frac{n}{n-d}$. The inequality $l_i>l_j+1$ then implies $l_i\geq 1$ and $l_j\leq (n-d)-1$. Therefore, out of the first $k$ coordinates of $\pi_f$, at least one of them will have value $i$; and out of the last $(n-k)$ coordinates of $\pi_f$, at least one of them will have value $j$. We can thus swap arbitrarily one of the family indices $i$ from the first $k$ coordinates with another family index $j$ from the last $(n-k)$ coordinates and the resulting $\pi_f'$ will have the desired $l_i'$ and $l_j'$.

\par We now prove the if direction. To that end, we first observe that the equality
$\sum_{i=1}^{\frac{n}{n-d}}l_i=k$ always holds because of our construction of $l_i$. Then  \eqref{eq:diverse} implies that we can uniquely decide the {\em distribution} of $\{l_i:i=1,\cdots,\frac{n}{n-d}\}$ even though we do not know what is the minimizing permutation $\pi_f^{\min}$ yet. For example, if $\frac{n}{n-d}=3$, $k=5$, $l_1$ to $l_3$ satisfy \eqref{eq:diverse}, and the summation $l_1+l_2+l_3$ is $k=5$, then among $l_1$, $l_2$, and $l_3$, two of them must be 2 and the other one must be 1. On the other hand, we observe that the value of $y_{\text{offset}}(\cdot)$ depends only on the distribution of $\{l_i\}$, see \eqref{eq:y_families}. As a result, the above arguments prove that any $\pi_f$ satisfying \eqref{eq:diverse} is a minimizing $\pi_f^{\min}$.

\end{IEEEproof}

\par Finally, by the construction of the RFIP $\pi_f^*$, it is easy to verify that the RFIP $\pi_f^*$ satisfies \eqref{eq:diverse}. Therefore, the RFIP $\pi_f^*$ is a minimizing permutation for this case.

\underline{Case 2:} $n\bmod(n-d)\neq 0$, i.e., when we do have an incomplete family.
In this case, we are again interested in minimizing \eqref{eq:new15}, and equivalently minimizing \eqref{eq:y_def}. To that end, we first prove the following claim.

\begin{claim}\label{clm:rfip_incomplete} Find the largest $1\leq j_1\leq k$ such that the $j_1$-th coordinate of $\pi_f$ is 0. If no such $j_1$ can be found, we set $j_1=0$.  Find the smallest $1\leq j_2\leq k$ such that the $j_2$-th coordinate of $\pi_f$ is a negative number if no such $j_2$ can be found, we set $j_2=k+1$. We claim that if we construct $j_1$ and $j_2$ based on a $\pi_f$ that minimizes $\sum_{i=1}^k (d-y_i(\pi_f))$, we must have $j_1<j_2$.
\end{claim}

\begin{IEEEproof}
We prove this claim by contradiction. Consider a minimizing family index permutation $\pi_f$ and assume $j_2<j_1$. This means, by our construction, that $1\leq j_2<j_1\leq k$.  Since the $j_2$-th coordinate of $\pi_f$ is a negative number by construction, $y_{j_2}(\pi_f)$ counts all coordinates before the $j_2$-th coordinate of $\pi_f$ with values in $\{1,2,\cdots,c-1,0\}$, i.e., it counts all the values before the $j_2$-th coordinate except for the values $c$ and $-c$, where $c$ is the family index of the last complete family. Thus, knowing that there are no $-c$ values before the $j_2$-th coordinate of $\pi_f$, we have that
\begin{align} \label{eq:y_j2_pi}
y_{j_2}(\pi_f)=j_2-1-\lambda^{[1,j_2)}_{\{c\}},
\end{align}
where $\lambda^{[1,j_2)}_{\{c\}}$ is the number of $c$ values before the $j_2$-th coordinate. Similarly, since the $j_1$-th coordinate is 0, we have that $y_{j_1}(\pi_f)$ counts all coordinates before the $j_1$-th coordinate of $\pi_f$ with values in $\{1,2,\cdots,c\}$, i.e., it counts all the values before the $j_1$-th coordinate except for the values $-c$ and $0$. Thus, we have that 
\begin{align}
y_{j_1}(\pi_f)&=j_1-1-\lambda^{[1,j_1)}_{\{0\}}-\lambda^{[1,j_1)}_{\{-c\}}
\end{align}
where $\lambda^{[1,j_1)}_{\{0\}}$ is the number of 0 values preceding the $j_1$-th coordinate in $\pi_f$ and $\lambda^{[1,j_1)}_{\{-c\}}$ is the number of $-c$ values preceding the $j_1$-th coordinate in $\pi_f$. Now, swap the $j_2$-th coordinate and the $j_1$-th coordinate of $\pi_f$, and call the new family index permutation $\pi_f'$. Specifically, $\pi_f'$ has the same values as $\pi_f$ on all its coordinates except at the $j_2$-th coordinate it has the value 0 and at the $j_1$-th coordinate it has the value $-c$. For $1\leq m\leq j_2-1$, we have that $y_m(\pi_f')=y_m(\pi_f)$ since the first $j_2-1$ coordinates of the two family index permutations are equal. Moreover, since there are no negative values before the $j_2$-th coordinate of $\pi_f'$, we have that
\begin{align}\label{eq:y_j2}
y_{j_2}(\pi_f')=j_2-1-\phi^{[1,j_2)}_{\{0\}},
\end{align}
where $\phi^{[1,j_2)}_{\{0\}}$ is the number of 0 values in $\pi_f'$ preceding the $j_2$-th coordinate. 

\par For $j_2+1\leq m\leq j_1-1$, if the $m$-th coordinate of $\pi_f'$ is either $c$ or $-c$, then $y_m(\pi_f')=y_m(\pi_f)+1$; otherwise, $y_m(\pi_f')=y_m(\pi_f)$. The reason behind this is that the function $y_m(\pi_f')$ now has to take into account the new 0 at the $j_2$-th coordinate when the $m$-th coordinate is either $c$ or $-c$. When the value of the $m$-th coordinate is in $\{1,\cdots, c-1\}$, then by the definition of $y_m(\cdot)$, we have $y_m(\pi_f')=y_m(\pi_f)$. The last situation to consider is when the value of the $m$-th coordinate is $0$. In this case, we still have $y_m(\pi_f')=y_m(\pi_f)$  since $y_m(\pi_f)$ already does not count the value on the $j_2$-th coordinate of $\pi_f$ since it is a negative value. 

\par Denote the number of $c$ and $-c$ values from the $(j_2+1)$-th coordinate to the $(j_1-1)$-th coordinate of $\pi_f'$ by $\phi^{(j_2,j_1)}_{\{c,-c\}}$. We have that
\begin{align}\label{eq:y_j1_pip}
y_{j_1}(\pi_f')=j_1-1-\lambda^{[1,j_2)}_{\{c\}} -\phi^{(j_2,j_1)}_{\{c,-c\}},
\end{align}
since the $j_1$-th coordinate of $\pi_f'$ has a $-c$ value. Finally, for $j_1+1\leq m\leq n$, we have that $y_m(\pi_f')=y_m(\pi_f)$ since the order of the values preceding the $m$-th coordinate in a permutation does not matter for $y_m(\cdot)$. By the above, we can now compute the following difference
\begin{align}
\sum_{i=1}^k&(d-y_i(\pi_f))-\sum_{i=1}^k(d-y_i(\pi_f')) \nonumber\\
&= \sum_{i=1}^k (y_i(\pi_f')-y_i(\pi_f))\nonumber\\
&= \sum_{i=j_2}^{j_1} (y_i(\pi_f')-y_i(\pi_f))\label{eq:CCWnew3}\\
&=(y_{j_2}(\pi_f')-y_{j_2}(\pi_f))+ \phi^{(j_2,j_1)}_{\{c,-c\}}+(y_{j_1}(\pi_f')-y_{j_1}(\pi_f)) \label{eq:diff_1}\\
&=\left(\lambda^{[1,j_2)}_{\{c\}}-\phi^{[1,j_2)}_{\{0\}}\right)+\phi^{(j_2,j_1)}_{\{c,-c\}}+\nonumber\\
&\qquad\qquad\qquad\left(\lambda^{[1,j_1)}_{\{0\}}+\lambda^{[1,j_1)}_{\{-c\}}-\lambda^{[1,j_2)}_{\{c\}} -\phi^{(j_2,j_1)}_{\{c,-c\}}\right) \label{eq:CCWnew4}\\
&=\lambda^{[1,j_1)}_{\{0\}}+\lambda^{[1,j_1)}_{\{-c\}}-\phi^{[1,j_2)}_{\{0\}}\nonumber\\
&>0\label{eq:diff_2},
\end{align}
where \eqref{eq:CCWnew3} follows from $y_i(\pi_f')=y_i(\pi_f)$ for all $i<j_2$ and for all $i>j_1$; \eqref{eq:diff_1} follows from our analysis about $y_i(\pi_f')=y_i(\pi_f)+1$ when the $i$-th coordinate of $\pi_f$ belongs to $\{-c,c\}$ and $y_i(\pi_f')=y_i(\pi_f)$ otherwise, and there are thus $\phi^{(j_2,j_1)}_{\{c,-c\}}$ coordinates between the $(j_2+1)$-th coordinate and the $(j_1-1)$-th coordinate of $\pi_f'$ that satisfy $y_i(\pi_f')=y_i(\pi_f)+1$; \eqref{eq:CCWnew4} follows from \eqref{eq:y_j2_pi} to \eqref{eq:y_j1_pip}; and \eqref{eq:diff_2} follows from the facts that $\lambda^{[1,j_1)}_{\{0\}} \geq \lambda^{[1,j_2)}_{\{0\}}=\phi^{[1,j_2)}_{\{0\}}$ and that $\lambda^{[1,j_1)}_{\{-c\}} \geq 1$ since we have a $-c$ value at the $j_2$-th coordinate of $\pi_f$. By \eqref{eq:diff_2}, we have that $\pi_f'$ has a strictly smaller ``$\sum_{i=1}^k(d-y_i(\cdot))$''. As a result, the case of $j_1>j_2$ is impossible. 

\par By the construction of $j_1$ and $j_2$, it is obvious that $j_1\neq j_2$. Hence, we must have $j_1<j_2$. The proof of this claim is complete.
\end{IEEEproof}

\par Claim~\ref{clm:rfip_incomplete} provides a necessary condition on a minimizing permutation vector. We thus only need to consider permutations for which $j_1<j_2$. That is, instead of taking the minimum over all $\pi_f$, we now take the minimum over only those $\pi_f$ satisfying $j_1<j_2$.

This observation is critical to our following derivation. The reason is that if we consider a permutation $\pi_f$ that has $1\leq j_2<j_1\leq k$, then the expression ``$(j_1 - 1)- y_{j_1}(\pi_f )$'' is not equal to the number of appearances of the value $0$ in the first $j_1-1$ coordinates of $\pi_f$ (recall that by our construction the $j_1$-th coordinate of $\pi_f$ is 0). Instead, by the definition of $y_{i}(\cdot)$, $(j_1-1)-y_{j_1}(\pi_f)$ is the number of appearances of the values 0 {\em and} $-c$ in the first $(j_1-1)$ coordinates of $\pi_f$. Therefore, we cannot rewrite \eqref{eq:y_def} as \eqref{eq:y_families} if $1\leq j_2<j_1\leq k$. 

\par On the other hand, Claim~\ref{clm:rfip_incomplete} implies that we only need to consider those $\pi_f$ satisfying $j_1<j_2$. We now argue that given any $\pi_f$ satisfying $j_1<j_2$, for all $i=1$ to $k$, the expression $(i-1)-y_{i}(\pi_f)$ is now representing the number of appearances of $m$ and $-m$ in the first $(i-1)$ coordinates of $\pi_f$, where $m$ is the {\em absolute value} of the $i$-th coordinate of $\pi_f$. The reason is as follows. Let $m$ denote the absolute value of the $i$-th coordinate of $\pi_f$. If $m\neq 0$, then by the definition of $y_i(\pi_f)$, we have that $(i-1)-y_{i}(\pi_f)$ represents the number of appearances of $m$ in the first $(i-1)$ coordinates of $\pi_f$. If $m=0$, then by the definition of $y_i(\pi_f)$, we have that $(i-1)-y_{i}(\pi_f)$ represents the number of appearances of 0 and $-c$ in the first $(i-1)$ coordinates of $\pi_f$. However, by the construction of $j_1$, we have $i\leq j_1$. Since $j_1<j_2$, we have $i<j_2$. This implies that in the first $(i-1)$ coordinates of $\pi_f$, none of them is of value $-c$. As a result, we have that $(i-1)-y_{i}(\pi_f)$ again represents the number of appearances of 0 in the first $(i-1)$ coordinates of $\pi_f$. 

\par
We now proceed with our analysis while only considering those $\pi_f$ satisfying $j_1<j_2$ as constructed in Claim~\ref{clm:rfip_incomplete}. Let $l_j$ be the number of the first $k$  coordinates of $\pi_f$ that have values $j$ or $-j$. We can then rewrite \eqref{eq:y_def} by
\begin{align} \label{eq:new-y_families}
y_{\text{offset}}(\pi_f)= \sum_{i=1}^{l_0}&(i-1)+\sum_{i=1}^{l_1}(i-1)+ \nonumber\\
& \sum_{i=1}^{l_2}(i-1) + \dots + \sum_{i=1}^{l_{\left\lfloor\frac{n}{n-d}\right\rfloor}}(i-1).
\end{align}
The above equation implies that a family index permutation is a minimizing permutation $\pi_f^{\min}$ if and only if either
\begin{align}\label{eq:diverse2_1}
\begin{cases}l_0=n\bmod(n-d),\\
|l_i-l_j|\leq1 \text{ for all $i,j$ satisfying}~1\leq i,j\leq c,\\
l_i\geq l_0 \text{ for all $i$ satisfying}~1\leq i\leq c.
\end{cases}
\end{align}
or
\begin{align} \label{eq:diverse2_2}
|l_i-l_j|\leq1, \text{for all $i,j$ satisfying}~0\leq i,j\leq c. 
\end{align}
If we compare \eqref{eq:diverse2_1} and \eqref{eq:diverse2_2} with \eqref{eq:diverse} in Claim~\ref{clm:diverse}, we can see that \eqref{eq:diverse2_2} is similar to \eqref{eq:diverse}. The reason we need to consider the situation described in \eqref{eq:diverse2_1} is that the range of $l_0$ is from 0 to $n\bmod(n-d)$ while the range of all other $l_i$s is from 0 to $(n-d)$. Therefore, we may not be able to make $l_0$ as close to other $l_i$s (within a distance of 1) as we would have hoped for due to this range discrepancy. For some cases, the largest $l_0$ we can choose is $n\bmod(n-d)$, which gives us the first scenario when all the remaining $l_i$s are no less than this largest possible $l_0$ value. If $l_0$ can also be made as close to the rest of $l_i$s, then we have the second scenario.

The proof that \eqref{eq:diverse2_1} and \eqref{eq:diverse2_2} are the if-and-only-if condition on $\pi_f^{\min}$ can be completed using the same arguments as in the proof of Claim~\ref{clm:diverse}. Finally, notice that the RFIP $\pi_f^*$ satisfies \eqref{eq:diverse2_1} or \eqref{eq:diverse2_2} and has $j_1<j_2$. As a result, $\pi_f^*$ must be one of the minimizing permutations $\pi_f^\text{min}$. The proof of Proposition~\ref{prop:mbr} is hence complete.

\section{Proof of Proposition~\ref{prop:msr}} \label{app:msr_proof}
We first consider the case when $d\geq k$. We have $\alpha_{\MSR}\geq\frac{\mathcal{M}}{k}$ since otherwise the MSR point cannot satisfy \eqref{eq:condition} even when plugging in $\beta=\infty$ in \eqref{eq:low_b}. Define
\begin{align}\label{eq:max-y-msr}
y_{\max}\stackrel{\Delta}{=}\max_{\forall \pi_f} \max_{1\leq i\leq k} y_i(\pi_f).
\end{align}
By \eqref{eq:low_b}, we have that the $(\alpha,\beta)$ pair
\begin{align}
(\alpha,\beta)=\left(\frac{\mathcal{M}}{k}, \frac{\mathcal{M}}{k(d-y_{\max})}\right)
\end{align}
satisfies \eqref{eq:condition} since $(d-y_i(\pi_f))\beta\geq (d-y_{\max})\beta= \frac{\mathcal{M}}{k}=\alpha$. Therefore, $\frac{\mathcal{M}}{k}$ is not only a lower bound of $\alpha_{\MSR}$ but is also achievable, i.e.,  $\alpha_{\MSR}=\frac{\mathcal{M}}{k}$. Now, for any $(\alpha,\beta)$ pair satisfying
\begin{align}
(\alpha,\beta)=\left(\frac{\mathcal{M}}{k},\beta\right)
\end{align}
for some $\beta<\frac{\mathcal{M}}{k(d-y_{\max})}$, we argue that \eqref{eq:condition} does not hold anymore. The reason is the following. When $\alpha=\frac{\mathcal{M}}{k}$ and $\beta<\frac{\mathcal{M}}{k(d-y_{\max})}$, we plug in the $\pi_f^{\circ}$  vector that maximizes \eqref{eq:max-y-msr} into \eqref{eq:low_b}. Therefore, for at least one $i^{\circ}\leq k$, we will have $(d-y_{i^{\circ}}(\pi_f^{\circ}))\beta<\alpha=\frac{\mathcal{M}}{k}$. This implies ``$\eqref{eq:low_b}< \mathcal{M}$'' when evaluated using $\pi_f^{\circ}$.  By taking the minimum over all $\pi_f$, we still have ``$\eqref{eq:low_b}< \mathcal{M}$''.  Therefore, the above choice of $(\alpha,\beta)$ cannot meet the reliability requirement at the MSR point. As a result, we have $\beta_{\MSR}=\frac{\mathcal{M}}{k(d-y_{\max})}$.

\par We now argue that $y_{\max}=k-1$. According to the definition of function $y_i(\cdot)$, $y_i\leq k-1$. Recall that the size of a helper set is $d$, which is strictly larger than $k-1$. We can thus simply set the values of the first $(k-1)$ coordinates of $\pi_f$ to be the family indices of the $(k-1)$ distinct helpers (out of $d$ distinct helpers) of a node and place the family index of this node on the $k$-th coordinate. Such a permutation $\pi_f$ will have $y_k(\pi_f)=k-1$. Therefore, we have proved that $\beta_{\MSR}=\frac{\mathcal{M}}{k(d-k+1)}$.

We now consider the remaining case in which $d<k$. To that end, we first notice that for any $(n,k,d)$ values we have $\left\lfloor \frac{n}{n-d}\right\rfloor \geq 1$ number of complete families. Also recall that family 1 is a complete family and all families $\neq 1$ are the helpers of family 1, and there are thus $d$ number of nodes in total of family index $\neq 1$. We now consider a permutation $\pi_f^{\circ}$ in which all its first $d$ coordinates are family indices not equal to 1 and its last $(n-d)$ coordinates are of family index 1. Observe that if we evaluate the objective function of the RHS of \eqref{eq:low_b} using $\pi_f^{\circ}$, out of the $k$ summands, of $i=1$ to $k$, we will have exactly $d$ non-zero terms since (i) by the definition of $y_i(\cdot)$, we always have $y_i(\pi_f^{\circ})\leq (i-1)$ and, therefore, when $i\leq d$, we always have $(d-y_i(\pi_f^{\circ}))\geq 1$; (ii) whenever $i>d$, the corresponding term $y_i(\pi_f^{\circ})=d$ due to the special construction of the $\pi_f^{\circ}$. As a result, when a sufficiently large $\beta$ is used, we have 
\begin{align}
\sum_{i=1}^k \min((d-y_i(\pi_f^{\circ}))\beta,\alpha) = d \alpha.
\end{align}

The above equality implies $\alpha_{\MSR}\geq \frac{\mathcal{M}}{d}$. Otherwise if $\alpha_{\MSR}<\frac{\mathcal{M}}{d}$, then we will have ``$\eqref{eq:low_b} <\mathcal{M}$'' when using the aforementioned $\pi_f^{\circ}$, which implies that  ``$\eqref{eq:low_b} <\mathcal{M}$''  holds still when minimizing over all $\pi_f$. This contradicts the definition that $\alpha_{\MSR}$ and $\beta_{\MSR}$ satisfy the reliability requirement.

\par On the other hand, we know that $\alpha_{\MSR}=\frac{\mathcal{M}}{d}$ and $\beta_{\MSR}=\frac{\mathcal{M}}{d}$ for the BHS scheme when $d<k$ \cite{dimakis2010network}. Since the performance of the FHS scheme is not worse than that of the BHS scheme, we have $\alpha_{\MSR}=\frac{\mathcal{M}}{d}$ and $\beta_{\MSR}\leq \frac{\mathcal{M}}{d}$ for the FHS scheme. Hence, the proof is complete. 

\section{Proof of Corollary~\ref{cor:mbr_plus}}\label{app:mbr_plus_proof}
Consider first the case when $n\bmod(2d)\neq 0$. Without loss of generality, assume that $n_B=n_{\text{remain}}$ and $n_b=2d$ for $b=1$ to $B-1$, i.e., the indices $b=1$ to $B-1$ correspond to the regular groups and the index $b=B$ corresponds to the remaining group. Now, applying the same reasoning as in the proof of Proposition~\ref{prop:mbr} to \eqref{eq:low_b_plus}, we have that $\alpha_{\MBR}=\gamma_{\MBR}=d\beta_{\MBR}$ for the family-plus helper selection scheme as well. In the following, we will prove that (i) if $k\leq 2d$, then one minimizing $\mathbf{k}$ vector can be constructed by setting $k_b=0$ for $b=1$ to $B-1$ and $k_B=k$; (ii) if $k>2d$, then we can construct a minimizing $\mathbf{k}$ vector by setting $k_B=\min(n_{\text{remain}},k)$ and among all $b=1$ to $B-1$, at most one $k_b$ satisfies $0<k_b<2d$.


\par To prove this claim, we first notice that since we are focusing on the MBR point, we can assume $\alpha$ is sufficiently large. Therefore, we can replace the minimizing permutation for each summand of \eqref{eq:low_b_plus} by the RFIP (of $(n,d)=(2d,d)$ for the summand $b=1$ to $B-1$ and of $(n,d)=(n_{\text{remain}},d)$ for summand $b=B$) using the arguments in the proof of Proposition~\ref{prop:mbr}. Therefore, we can rewrite \eqref{eq:low_b_plus} by
\begin{align}\label{eq:inter5}
\eqref{eq:low_b_plus}=
 \min_{\mathbf{k}\in K} \sum_{b=1}^{B} \sum_{i=1}^{k_b}(d-y_i(\pi_b))\beta
\end{align}
where $\pi_b$ is the RFIP of $(n,d)=(2d,d)$ for $b=1$ to $B-1$ and the RFIP of $(n,d)=(n_{\text{remain}},d)$ for $b=B$. Note that for $(n,d)=(2d,d)$, in the FHS scheme we have 2 complete families and no incomplete family and the RFIP in this case is $\pi_1^*=(1,2,1,2,\cdots,1,2)$. As a result, $\pi_b=\pi_1^*$ for all $b=1$ to $B-1$. For $(n,d)=(n_{\text{remain}},d)$, we have one complete family and one incomplete family and the RFIP in this case is 
\begin{align}
\pi_2^*=(\overbrace{1,0,1,0,\cdots,1,0}^{2d \text{ coordinates}},\overbrace{-1,-1,\cdots,-1}^{(n_{\text{remain}}-2d) \text{ coordinates}}).
\end{align} 
We thus have $\pi_B=\pi_2^*$. We now argue that a vector $\mathbf{k^*}$ satisfying conditions (i) and (ii) stated above minimizes \eqref{eq:inter5}. Note first that both $y_i(\pi_1^*)$ and $y_i(\pi_2^*)$ are non-decreasing with respect to $i$ according to our construction of the RFIP. Also, we always have $y_i(\pi_1^*)=y_i(\pi_2^*)$ for all $1\leq i\leq 2d$. 

\par We are now ready to discuss the structure of the optimal $\mathbf{k}$ vector. Since for each $b=1$ to $B$, we are summing up the first $(d-y_i(\pi_b))$ from $i=1$ to $k_b$ and in total there are $\sum_b k_b=k$ such terms, \eqref{eq:inter5} implies that to minimize \eqref{eq:low_b_plus} we would like to have as many terms corresponding to ``large $i$'' as possible in the summation $\sum_b k_b=k$ terms. If $k\leq2d$, this can be done if and only if we set all $k_b$ to 0 except for one $k_b$ value to be $k$, which is our construction (i). If $k>2d$, this can be done if and only if we set $k_B=\min(n_{\text{remain}},k)$ and, for $b=1$ to $B-1$, we set all $k_b$ to either $2d$ or $0$ except for one $k_b$. 

\par Knowing that $\mathbf{k^*}$ is of this special form, we can compute the RHS of \eqref{eq:low_b_plus} by
\begin{align}
\text{RHS of \eqref{eq:low_b_plus}}&=\left\lfloor \frac{k-\min(n_{\text{remain}},k)}{2d} \right\rfloor \text{sum}^{(1)} \nonumber\\
&+ \text{sum} ^{(2)}+\text{sum} ^{(3)},
\end{align}
where $\left\lfloor \frac{k-\min(n_{\text{remain}},k)}{2d} \right\rfloor$ is the number of $b$ from 1 to $B-1$ with $k_b=2d$ in the minimizing vector $\mathbf{k^*}$; $\text{sum}^{(1)}$ is the contribution to the min-cut value from those groups with $k_b=2d$, which is equal to $\sum_{i=1}^{2d}(d-y_i(\pi_1^*))\beta$; $\text{sum}^{(2)}$ is the contribution to the min-cut value from the single regular group with $k_b=(k-\min(n_{\text{remain}},k))\bmod (2d)$, which is equal to $\sum_{i=1}^{k_b}(d-y_i(\pi_1^*))\beta$; and $\text{sum}^{(3)}$ is the contribution to the min-cut value from the remaining group (group $B$), which is equal to 
\begin{align}
\text{sum}^{(3)}=\sum_{i=1}^{\min(n_{\text{remain}},k)}(d-y_i(\pi_2^*))\beta.
\end{align}
By plugging in the expressions of the RFIPs $\pi_1^*$ and $\pi_2^*$, we have
\begin{align}
\text{sum}^{(1)}&= \sum_{i=0}^{2d-2}\left(d-i+\left\lfloor \frac{i}{2}\right\rfloor\right)\beta=d^2\beta, \nonumber\\
\text{sum}^{(2)}&=\sum_{i=0}^{q}\left(d-i+\left\lfloor\frac{i}{2}\right\rfloor\right) \beta, \text{ and}\nonumber\\
\text{sum}^{(3)}&=\sum_{i=0}^{\min(k,2d-1)-1}\left(d-i+\left\lfloor\frac{i}{2}\right\rfloor\right)\beta,\label{eq:sum3}
\end{align}
where $q=((k-\min(n_{\text{remain}},k))\bmod(2d))-1=((k-n_{\text{remain}})^+\bmod(2d))-1$ and \eqref{eq:sum3} follows from the fact that $y_j(\pi_2^*)=d$ when $j\geq 2d$ and $n_{\text{remain}}\geq 2d+1$. The minimum repair-bandwidth $\beta_{\MBR}$ thus satisfies \eqref{eq:gamma_plus}.

\par Now, for the case when $n\bmod(2d)=0$, in a similar fashion, we can prove that a $\mathbf{k}$ vector minimizes the RHS of \eqref{eq:low_b_plus} at the MBR point if and only if there is at most one $b\in\{1,\cdots,B\}$ such that $0<k_b<2d$. By setting $\pi_b=\pi_1^*$ for all $b$ in \eqref{eq:inter5}, recall that $\pi_1^*$ is the RFIP for $(n,d)=(2d,d)$, we get  
\begin{align}
\text{RHS of \eqref{eq:low_b_plus}}&=d^2\left\lfloor \frac{k}{2d}\right\rfloor\beta+\sum_{i=0}^{(k\bmod(2d))-1}\left(d-i+\left\lfloor\frac{i}{2}\right\rfloor\right)\beta,
\end{align}
and thus $\beta_{\MBR}$ satisfies \eqref{eq:gamma_plus} for this case too.
The proof is hence complete.

\section{Proof of \eqref{eq:fpr_vs_fr}} \label{app:weak_proof}
\par To prove \eqref{eq:fpr_vs_fr}, we first notice that when $n<4d$, the family-plus helper selection scheme collapses to the FHS scheme since each group of the family-plus scheme needs to have at least $2d$ nodes and when $n<4d$ we can have at most 1 group. Thus, trivially, we have \eqref{eq:fpr_vs_fr} when $n<4d$. Now, we consider the case when $n\geq 4d$.

\par We first consider the original FHS scheme (the RHS of \eqref{eq:fpr_vs_fr}). In this case, the FHS scheme has $\left\lfloor \frac{n}{n-d}\right\rfloor=1$ complete family and one incomplete family. The corresponding RFIP $\pi_f^*$ is thus
\begin{align}
\pi_f^*=(\overbrace{1,0,1,0,\cdots, 1,0}^{2d \text{ coordinates}}, \overbrace{-1,-1,\cdots, -1}^{(n-2d) \text{ coordinates}})\nonumber.
\end{align}
By Proposition~\ref{prop:mbr}, we have

\begin{align}
\min_{G\in\mathcal{G}_{F}}\min_{t\in \DC (G)}\mincut_G(s,t)&=\nonumber\\ \sum_{i=0}^{\min(k,2d-1)-1}&\left(d-i+\left\lfloor\frac{i}{2}\right\rfloor\right)\beta,\label{eq:optimality_fr}
\end{align}
where \eqref{eq:optimality_fr} from the fact that $y_j(\pi_f^*)=d$ when $j\geq 2d$.

\par We now turn our focus to the family-plus helper selection scheme. Consider first the case when $n\bmod(2d)=0$. If $k<2d$, we have by \eqref{eq:gamma_plus} and \eqref{eq:optimality_fr} that \eqref{eq:fpr_vs_fr} is true since the third term on the LHS of \eqref{eq:gamma_plus} is the RHS of \eqref{eq:optimality_fr}. If $k\geq 2d$, we again have by \eqref{eq:gamma_plus} and \eqref{eq:optimality_fr} that \eqref{eq:fpr_vs_fr} is true since the second term on the LHS of \eqref{eq:gamma_plus} is no less than the RHS of \eqref{eq:optimality_fr}. Now, consider the case when $n\bmod(2d)\neq 0$. Similarly, we have by \eqref{eq:gamma_plus} and \eqref{eq:optimality_fr} that \eqref{eq:fpr_vs_fr} is true since the first term on the LHS of \eqref{eq:gamma_plus} is the RHS of \eqref{eq:optimality_fr}.

\section{Proof of Proposition~\ref{prop:optimal}}\label{app:optimal_proof}
We first introduce the following corollary that will be used shortly to prove Proposition~\ref{prop:optimal}.

\begin{corollary} \label{cor:low_b} For any $(n,k,d)$ values satisfying $d\geq 2$ and $k=\left\lceil \frac{n}{n-d}\right\rceil + 1$, we consider the corresponding IFGs $\gf (n,k,d,\alpha,\beta)$ generated by the FHS scheme $F$. We then have that
\begin{align}
\min_{G\in\gf}&\min_{t\in\DC(G)}\mincut(s,t) = \nonumber\\
&\sum_{i=2}^{k-1}\min ((d-i)\beta,\alpha) + 2\min(d\beta,\alpha) \label{eq:low_b_spec}.
\end{align}
\end{corollary}

\begin{IEEEproof}
First consider the case when $d\geq k-1=\left\lceil \frac{n}{n-d}\right\rceil$. Since there are $\left\lceil \frac{n}{n-d}\right\rceil$ number of families (complete plus incomplete families) and $k=\left\lceil \frac{n}{n-d}\right\rceil+1$, any family index permutation has at least one pair of indices of the same family in its first $k$ coordinates. Using \eqref{eq:low_b}, this observation implies that
\begin{align}
\min_{G\in\gf}&\min_{t\in\DC(G)}\mincut(s,t)\nonumber\\
&=\min_{\forall \pi_f} \sum_{i=1}^{k}\min \left(\left(d-y_i(\pi_f)\right)\beta,\alpha\right) \geq \min_{2\leq m\leq k} C_m \label{eq:another-new-prop-11}
\end{align}
where $C_m=\sum_{i=0}^{k-1}\min ((d-i)\beta,\alpha)1_{\{i\neq m-1\}} +  \min((d-m+2)\beta,\alpha)$ for $2\leq m\leq k$. Namely, $C_m$ is a lower bound of the following sum 
\begin{align}
\sum_{i=1}^{k}\min \left(\left(d-y_i(\pi_f)\right)\beta,\alpha\right)
\end{align}
conditioning on that the $m$-th oldest nodes in the family index permutation $\pi_f$ turns out to be a repeated one.

We now prove that the inequality \eqref{eq:another-new-prop-11} is actually an equality. To that end, we first define $\pi_f^{[m]}$ as a family index permutation such that its first $k$ coordinates, in this order, are $1,2,\cdots,m-1,1,m+1,\cdots,c,0$ if $n\bmod(n-d)\neq 0$ and define $\pi_f^{[m]}$ as $1,2,\cdots,m-1,1,m+1,\cdots,c$ if $n\bmod(n-d)=0$. Since all the $k$ coordinates have different values except the first coordinate and the $m$-th coordinate have equal value $1$, and since they have no $-c$ value, we have
\begin{align}
\sum_{i=1}^k \min\left(\left(d-y_i\left(\pi_f^{[m]}\right)\right)\beta,\alpha\right)=C_m,
\end{align}
Thus, we get that
\begin{align}
\min_{G\in\gf}\min_{t\in\DC(G)}\mincut(s,t) = \min_{2\leq m\leq k} C_m.
\end{align}

\par By observing that the RHS of \eqref{eq:low_b_spec} is identical to $C_2$, what remains to be proved is to show now that $\min_{2\leq m\leq k} C_m=C_2$. First, notice that we have
\begin{align}
C_m-C_2&=\min((d-1)\beta,\alpha)-\min(d\beta,\alpha)+\nonumber\\
&\min((d-m+2)\beta,\alpha)-\min((d-m+1)\beta,\alpha)\label{eq:cm_exp}.
\end{align}
Since we always have $C_m-C_2=0$ when $m=2$, we only consider the $m$ values satisfying $3\leq m\leq k$. We then observe that the $\alpha$ value in \eqref{eq:cm_exp} is compared to four different values: $(d-m+1)\beta$, $(d-m+2)\beta$, $(d-1)\beta$, and $d\beta$, listed from the smallest to the largest. Depending on the relative order between $\alpha$ and these 4 values, we have 5 cases:
\begin{itemize}
\item If $\alpha\leq (d-m+1)\beta$, then $C_m-C_2=\alpha-\alpha+\alpha-\alpha=0$.
\item If $(d-m+1)\beta \leq \alpha \leq (d-m+2)\beta$, then $C_m-C_2=\alpha-\alpha+\alpha-(d-m+1)\beta\geq \alpha-\alpha+\alpha-\alpha =0$.
\item If $(d-m+2)\beta <\alpha \leq (d-1)\beta$ (this case does not exist for $m=3$), then $C_m-C_2=\alpha-\alpha+(d-m+2)\beta-(d-m+1)\beta=\beta\geq 0$.
\item If $(d-1)\beta <\alpha \leq d\beta$, then $C_m-C_2=\alpha-(d-1)\beta+\beta\geq \alpha-\alpha +\beta \geq 0$.
\item If $\alpha \geq d\beta$, then $C_m-C_2=(d-1)\beta-d\beta+\beta=0$.
\end{itemize}

\par We have shown by the above that $C_m\geq C_2$ for all $3\leq m \leq k$. Therefore, we have proved that $\min_{2\leq m\leq k} C_m=C_2=\sum_{i=2}^{k-1}\min ((d-i)\beta,\alpha) + 2\min(d\beta,\alpha)$ and we get the equality in \eqref{eq:low_b_spec}.

\par We now consider the case when $d<k-1=\left\lceil \frac{n}{n-d}\right\rceil$. Before proceeding, we first argue that among all $(n,k,d)$ values satisfying \eqref{eq:ccw1}, the only possible cases of having $d\leq \left\lceil \frac{n}{n-d}\right\rceil-1$ are either $d=1$ or $d=n-1$. The reason behind this is the following. Suppose $d\leq \left\lceil \frac{n}{n-d}\right\rceil-1$ and $2\leq d\leq n-2$. For any $2\leq d\leq n-2$, we have
\begin{align}
0\leq \left\lceil\frac{n}{n-d}\right\rceil-1-d&=\left\lceil1+\frac{d}{n-d}\right\rceil-1-d\nonumber\\
&=\left\lceil\frac{d}{n-d}\right\rceil-d\nonumber\\
&\leq \left\lceil\frac{d}{2}\right\rceil-d \label{eq:corr1_1}\\
&=\begin{cases} -\frac{d}{2}, & \mbox{if } d\mbox{ is even} \\ \frac{1-d}{2}, & \mbox{if } d\mbox{ is odd} \end{cases}\nonumber\\
&<0\label{eq:corr1_2},
\end{align}
where we get \eqref{eq:corr1_1} by our assumption that $d\leq n-2$ and \eqref{eq:corr1_2} follows from the assumption that $d\geq 2$. The above contradiction implies that when $d\leq \left\lceil \frac{n}{n-d}\right\rceil-1$  we have either $d=1$ or $d=n-1$.
Since Corollary~\ref{cor:low_b} requires $d\geq 2$, the only remaining possibility in this case of $d\leq \left\lceil \frac{n}{n-d}\right\rceil-1$ is when $d=n-1$. However, $k$ will not have a valid value since in this case we have $d=n-1<k-1$, which implies $k>n$, an impossible parameter value violating \eqref{eq:ccw1}. Hence, the proof is complete.
\end{IEEEproof}

\par We now prove Proposition~\ref{prop:optimal} by proving the following. Consider any fixed $(n,k,d)$ values that satisfy the three conditions of Proposition~\ref{prop:optimal} and any $G\in \mathcal{G}(n,k,d,\alpha,\beta)$ where all the active nodes of $G$ have been repaired at least once. We will prove the statement that such $G$ satisfies that there exists a data collector, denoted by $t_2 \in \DC(G)$, such that
\begin{align} \label{eq:tight_proof}
\mincut_G(s,t_2)\leq \sum_{i=2}^{k-1}\min ((d-i)\beta,\alpha) + 2\min(d\beta,\alpha).
\end{align}
Note that the above statement plus Corollary~\ref{cor:low_b} immediately prove Proposition~\ref{prop:optimal} since it says that no matter how we design the helper selection scheme $A$, the resulting $G$ (still belongs to $\mathcal{G}(n,k,d,\alpha,\beta)$) will have $\min_{t\in \DC(G)}\mincut_G(s,t)\leq \sum_{i=2}^{k-1}\min ((d-i)\beta,\alpha) + 2\min(d\beta,\alpha)$.

\par We now prove the above statement. We start with the following definition.

\begin{definition}\label{def:mp_set}A set of $m$ active storage nodes (input-output pairs) of an IFG is called an $(m,2)$-set if the following conditions are satisfied simultaneously. (i) Each of the $m$ active nodes has been repaired at least once; (ii) for easier reference, we use $x_1$ and $x_2$ to denote the oldest and the second-oldest nodes, respectively, among the $m$ nodes of interest. If we temporarily add an edge connecting $x_{2,\text{in}}$ and $x_{1,\text{out}}$, then we require that the $m$ nodes of interest form an $m$-set as defined in Definition~\ref{def:m_set}. Specifically, in an $(m,2)$-set, the only possible ``disconnect'' among the $m$ nodes is between $x_{2,\text{in}}$ and $x_{1,\text{out}}$ and every other node pairs must be connected. Note that whether $x_{2,\text{in}}$ and $x_{1,\text{out}}$ are actually connected or not is of no significance in this definition. 
\end{definition}	

We now prove the following claim, which will later be used to prove the desired statement.
\begin{claim}
Consider any $G\in\mathcal{G}(n,k,d,\alpha,\beta)$ where $(n,k,d)$ satisfy the three conditions of Proposition~\ref{prop:optimal} and all the active nodes of $G$ have been repaired at least once. In any $l$ active nodes of $G$, where $l$ is an even integer value satisfying $4\leq l\leq n$, there exists a $(\frac{l}{2}+1,2)$-set.
\end{claim}

\begin{IEEEproof}
We prove this claim by induction on $l$. We first prove that the claim holds for $l=4$. Consider any set $H_1$ of 4 active nodes of $G$. We will now prove the existence of a $(3,2)$-set. First, call the chronologically fourth active node of $G$, $u$. Since $d=n-2$, $u$ can avoid at most 1 active node during repair and $u$ is thus connected to at least $3-1=2$ older active nodes in $H_1$. Pick two nodes that $u$ is connected to and call this set of two nodes $V$. Then, we claim that $\{u\}\cup V$ forms a $(3,2)$-set. The reason is the following. Let $v_1$ and $v_2$ denote the two nodes in $V$ and, without loss of generality, we assume $v_1$ is older than $v_2$. We have that $u$ is connected to $v_1$ and $v_2$. One can verify that $\{v_1,v_2,u\}$ satisfy the properties (i) and  (ii) of Definition~\ref{def:mp_set} since the first and the second oldest nodes are $V=\{v_1,v_2\}$. Therefore, $\{v_1,v_2,u\}$ form a $(3,2)$-set. Note that $v_2$ may or may not be connected to $v_1$.

\par Now, assume that the claim holds for $l\leq l_0-2$. Consider any set of $l_0$ active nodes of $G$ and call it $H_2$. Since $d=n-2$, each node can avoid connecting to at most 1 active node. Therefore, the youngest node in $H_2$, call it $x$, is connected to $l_0-2$ older nodes in $H_2$. Call this set of $(l_0-2)$ nodes, $V_2$. We assumed that the claim holds for $l\leq l_0-2$, this tells us that in $V_2$ there exists an $(\frac{l_0}{2},2)$-set. Moreover, for any $(\frac{l_0}{2},2)$-set in $V_2$, denoted by $V_3$, we argue that the set $V_3\cup \{x\}$ is a $(\frac{l_0}{2}+1,2)$-set in $H_2$. The reason is that the first and the second oldest nodes in $V_3\cup\{x\}$ are also the first and the second oldest nodes in $V_3$. Since node $x$ is connected to all nodes in $V_2\supseteq V_3$, $V_3\cup\{x\}$ satisfies properties (i) and (ii) in Definition~\ref{def:mp_set} and thus form a $(\frac{l_0}{2}+1,2)$-set. Hence, the proof is complete.
\end{IEEEproof}

By the above claim, we have that for any $G\in\mathcal{G}(n,k,d,\alpha,\beta)$ where all the active nodes of $G$ have been repaired at least once there exist a $(\frac{n}{2}+1,2)$-set. We then consider a data collector that connects to this $(\frac{n}{2}+1,2)$-set and we denote it by $t_2$.

\par We now apply a similar analysis as in the proof of \cite[Lemma 2]{dimakis2010network} to prove \eqref{eq:tight_proof}. We need to prove that \eqref{eq:tight_proof} is true for the $t_2$ we are considering. Denote the storage nodes (input-output pair) of this $(\frac{n}{2}+1,2)$-set by $1,2,\dots,\frac{n}{2}+1$. Define cut $(U,\overline{U})$ between $t_2$ and $s$ as the following: for each $i\in \{0,2,3,4,\dots,\frac{n}{2}\}$, if $\alpha \leq (d-i)\beta$ then we include $x_{\out}^{i+1}$ in $\overline{U}$; otherwise, we include both $x_{\out}^{i+1}$ and $x_{\inp}^{i+1}$ in $\overline{U}$. For $i=1$, if $\alpha \leq d\beta$, then we include $x_{\out}^2$ in $\overline{U}$; otherwise, we include both $x_{\out}^2$ and $x_{\inp}^2$ in $\overline{U}$. It is not hard to see that the cut-value of the cut $(U,\overline{U})$ is no larger than $\sum_{i=2}^{k-1}\min ((d-i)\beta,\alpha) + 2\min(d\beta,\alpha)$. Therefore, we get \eqref{eq:tight_proof} and the proof is complete.

\section{Proof of Proposition~\ref{prop:optimal_2}}\label{app:optimal_2}
\cite[Theorem~5.2]{hollmann2014minimum} proved that for $k=n-1$ and $\alpha=d\beta$,
\begin{align}\label{eq:optimal_2_proof_4}
\min_{G\in \mathcal{G}_A}\min_{t\in \DC(G)} \mincut_G(s,t)\leq \frac{nd\beta}{2}
\end{align}
for any arbitrary DHS scheme $A$. As a result, we only need to prove that when $n\bmod(n-d)=0$, the min-cut of the FHS scheme equals $\frac{nd\beta}{2}$.

Since $\alpha=d\beta$, we know by Proposition~\ref{prop:mbr} that
\begin{align} \label{eq:optimal_2_proof_1}
\min_{G\in \mathcal{G}_F}\min_{t\in \DC(G)}\mincut_G(s,t)= \sum_{i=1}^{n-1} (d-y_i(\pi_f^*)) \beta.
\end{align}
Now, when $n\bmod(n-d)=0$, we have no incomplete family in the FHS scheme and the RFIP has the following form
\begin{align}
\pi_f^*=(1,2,\cdots,c,1,2,\cdots,c,\cdots,1,2,\cdots,c)\label{eq:rfip_optimal_2},
\end{align}
where recall that $c=\left\lfloor\frac{n}{n-d}\right\rfloor=\frac{n}{n-d}$. Using \eqref{eq:rfip_optimal_2}, we get that
\begin{align}\label{eq:yi_optimal_2}
y_i(\pi_f^*)=i-1-\left\lfloor \frac{i-1}{c}\right\rfloor.
\end{align}
The reason behind \eqref{eq:yi_optimal_2} is the following. Examining the definition of $y_i(\cdot)$, we can see that $y_i(\cdot)$ counts all the coordinates $j<i$ of $\pi_f^*$ that have a family index different than the family index at the $i$-th coordinate. For each coordinate $i$, with the aid of \eqref{eq:rfip_optimal_2}, there are $\left\lfloor \frac{i-1}{c}\right\rfloor$ coordinates in $\pi_f^*$ preceding it with the same family index. Therefore, in total there are $i-1-\left\lfloor \frac{i-1}{c}\right\rfloor$ coordinates in $\pi_f^*$ preceding the $i$-th coordinate with a different family index, thus, we get \eqref{eq:yi_optimal_2}.

\par By \eqref{eq:optimal_2_proof_1} and \eqref{eq:yi_optimal_2}, we get 
\begin{align}
\min_{G\in \mathcal{G}_F}\min_{t\in \DC(G)}&\mincut_G(s,t)=\sum_{i=0}^{n-2}\left(d-i+\left\lfloor\frac{i}{\frac{n}{n-d}}\right\rfloor\right)\beta \nonumber\\
&=\sum_{i=0}^{n-1}\left(d-i+\left\lfloor\frac{i}{\frac{n}{n-d}}\right\rfloor\right)\beta\label{eq:n-1_n-2}\\
&=\left(nd-\frac{(n-1)n}{2}+\sum_{i=0}^{n-1}\left\lfloor\frac{i}{\frac{n}{n-d}}\right\rfloor\right)\beta\nonumber\\
&=\left(nd-\frac{(n-1)n}{2}+\frac{n}{n-d}\sum_{i=0}^{n-d-1}i\right) \beta\nonumber\\
&=\left(nd-\frac{(n-1)n}{2}+\frac{n(n-d-1)}{2}\right)\beta\nonumber\\
&=\frac{nd\beta}{2}, \nonumber
\end{align}
where we get \eqref{eq:n-1_n-2} by the fact that $d-(n-1)+\left\lceil \frac{n-1}{c}\right\rceil =d-(n-1)+(n-d-1)=0$. The proof is thus complete

\section{Proof of Proposition~\ref{prop:family-plus_optimal}} \label{app:family-plus_optimal_proof}
\par By Proposition~\ref{prop:low_b_plus} and the fact that $k=n-1$, we must have all but one $k_b=n_b$ and the remaining one $k_b=n_b-1$. Without loss of generality, we assume $k_1=n_1-1$ and all other $k_b=n_b$ for $b=2$ to $B$ for the minimizing $\mathbf{k}$ vector in \eqref{eq:low_b_plus}. Since $n_1\bmod(n_1-d)=0$, by Proposition~\ref{prop:optimal_2}, the first summand of \eqref{eq:low_b_plus} must be equal to $\frac{n_1\alpha}{2}$.

\par For the case of $b=2$ to $B$, we have $k_b=n_b$ instead of $k_1=n_1-1$. However, if we examine the proof of Proposition~\ref{prop:optimal_2},  we can see that Proposition~\ref{prop:optimal_2} holds even for the case of $k=n$ since (i) when compared to the case of $k=n-1$, the case of $k=n$ involves one additional summand $(d-y_n(\pi_f^*))\beta$ in \eqref{eq:optimal_2_proof_1} and (ii) $(d-y_n(\pi_f^*))=0$. By applying Proposition~\ref{prop:optimal_2} again, the $b$-th summand of \eqref{eq:low_b_plus}, $b=2$ to $B$, must be $\frac{n_b\alpha}{2}$ as well.

\par Finally, by Proposition~\ref{prop:low_b_plus}, we have the equality in \eqref{eq:tight_2_plus}

\begin{align}
\min_{G\in \gfp}\min_{t\in \DC(G)}\mincut_G(s,t)=\sum_{b=1}^B\frac{n_b\alpha}{2}=\frac{n\alpha}{2}\label{eq:family-plus_optimal2}.
\end{align}
The inequality in \eqref{eq:tight_2_plus} is by \cite[Theorem~5.4]{hollmann2014minimum}. The proof is thus complete.

\bibliography{paper}
\bibliographystyle{IEEEtranS}

\end{document}